\documentclass[11pt, oneside]{article}   	
\usepackage{jheppub}
\usepackage{amsmath}
\usepackage{amssymb}
\usepackage{amsfonts}
\usepackage{amsthm}
\usepackage{amsbsy}
\usepackage{array}
\usepackage{mathtools}
\usepackage[usenames,dvipsnames]{xcolor}
\usepackage{subcaption}
\usepackage{dsdshorthand}
\usepackage{graphicx,subcaption}
\usepackage[vcentermath]{youngtab}
\usepackage{tikz}
\usepackage{booktabs,multirow}
\usepackage{changepage}
\usetikzlibrary{decorations.pathreplacing}
\usetikzlibrary{decorations.markings}
\usetikzlibrary{arrows.meta}
\usetikzlibrary{calc}
\makeatletter
\def\@fpheader{\ }
\makeatother

\title{A genus-2 crossing equation in $d\geq2$}
\author{David Simmons-Duffin, Yixin Xu}
\affiliation{Walter Burke Institute for Theoretical Physics, Caltech, Pasadena, California 91125, USA}
\emailAdd{dsd@caltech.edu}
\emailAdd{yixinxu@caltech.edu}

\date{}
\abstract{We explore a ``genus-2" crossing equation obeyed by CFTs in general dimensions $d\geq 2$.
This crossing equation relates two different decompositions of the ``genus-2 partition function" --- namely the partition function on the connected sum $M_2=(S^1\times S^{d-1})\sharp (S^1\times S^{d-1})$. The ``sunrise" channel decomposition expresses $M_2$ as a pair of three-punctured spheres glued together with cylinders, while the ``dumbbell" channel decomposition expresses $M_2$ as a gluing of two one-point functions on $S^1\times S^{d-1}$. We introduce coordinates to describe each channel, and write down Casimir equations obeyed by the corresponding blocks. We also explain why equality between the two channels guarantees mapping class group invariance of the genus-2 partition function in 3d CFTs. As an application of the genus-2 crossing equation, we derive a novel relation between asymptotics of ``heavy-heavy-heavier" OPE coefficients and squares of thermal one-point coefficients in 3d CFTs. Along the way, we demonstrate how expectation values of conformal generators can help locate saddle points in large quantum number limits.}
\preprint{}

\begin{document}

\maketitle
\pagenumbering{roman}
\setcounter{page}{2}
\newpage
\pagenumbering{arabic}
\setcounter{page}{1}

\section{Introduction}
\label{sec:intro}
        
Requiring a CFT to be consistently defined on different geometries places strong constraints on its dynamical data. In 2d, modular invariance of the torus partition function is a well-studied example. In higher dimensions, the best-studied probe of these constraints is the four-punctured sphere. The analysis (both analytical and numerical) of four-point crossing has proved extremely fruitful. However, the information in a given finite set of four-point correlation functions may not provide a full picture of a theory. 
This observation has motivated the investigation of larger, more intricate systems of four-point correlators, see e.g.\ \cite{Chang_2025}, as well as recent investigations of higher-point correlators \cite{Buric:2020dyz,Buric:2021kgy,Buric:2021ywo,Antunes:2023kyz,Kaviraj:2022wbw,Poland:2023vpn,Poland:2023bny,Harris:2024nmr,Antunes:2025vvl,Harris:2025cxx,Poland:2025ide}.

\begin{figure}[h]
    \centering
\includegraphics[width=.8\linewidth]{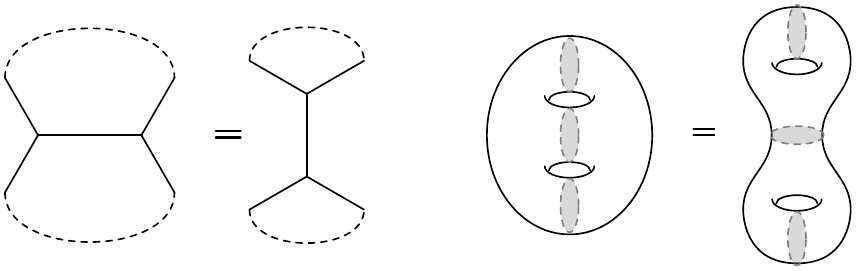}
\caption{\label{fig:dumbbell_sunrise_as_contracted_4pt_X} 
Left: the genus-2 crossing equation obtained by starting with the four-point crossing equation (solid lines), and contracting the external operators (indicated by dashed lines). The left-hand side of the resulting equation is the ``sunrise" channel (since the diagram looks like a sunrise Feynman diagram), and the right-hand side is the ``dumbbell" channel. Right:  the resulting crossing equation in terms of geometry, drawn in 2d for simplicity. A resolution of unity is inserted at each gray circle,
giving a sum over genus-2 blocks in the sunrise channel and a sum over squared torus one-point functions in the
dumbbell channel.
}
\end{figure}

In this paper, we explore a crossing equation that encapsulates infinitely 
many four-point functions in a different way. 
Our main focus is the \textit{genus-2 partition function}, 
which was explored from a bootstrap perspective in
2d CFTs in \cite{Keller:2017iql,Cho:2017fzo,Cardy:2017qhl}, and more recently in $d>2$ in \cite{Benjamin:2023qsc}.
In $d>2$, we define the \textit{genus-2 manifold} $M_2$ as the connected sum of two 
copies of $S^1\times S^{d-1}$. The manifold $M_2$ can be built by gluing punctured spheres in multiple ways. Each gluing yields different decompositions for its partition function into sums over states.
We will focus on two such decompositions: the 
sunrise channel, where we sum over OPE coefficients multiplied by
the \textit{genus-2 block} from \cite{Benjamin:2023qsc},
and the dumbbell channel, where we sum over squares of 
one-point functions on $S^1\times S^{d-1}$.  Schematically, equality between these two channels takes the form:
\begin{equation}
\label{eq:ourequation}
    \sum_{123}c_{123}^2 B_{123}=\sum_{1'2'3'} c_{1'1'2'}c_{2'3'3'} B'_{1'2'3'},
\end{equation}
where $B_{123}$ is the sunrise block, and $B'_{1'2'3'}$ is the dumbbell block.
This equation can also be understood as a consequence of taking the usual four-point crossing equation and summing over external operators, with weights that depend on the moduli of $M_2$, see figure~\ref{fig:dumbbell_sunrise_as_contracted_4pt_X}.

In section \ref{sec:the genus-2 geometry}, we review the moduli space of flat conformal structures on $M_2$, introducing coordinates
suited for each channel. The dimension of the moduli space is $(d+1)(d+2)/2$, much larger than that 
of the four-punctured sphere. This counting reflects the fact that even though the genus-2 crossing equation follows ultimately from the usual four-point crossing,
it encodes more information than any finite set of  four-point functions.
 After explaining the matching between dumbbell and sunrise coordinates,  we comment on why the genus-2 crossing equation (\ref{eq:ourequation}) ensures mapping class group invariance of the partition function in 
3d. 

\begin{figure}
    \centering
\includegraphics[width=0.8\linewidth]{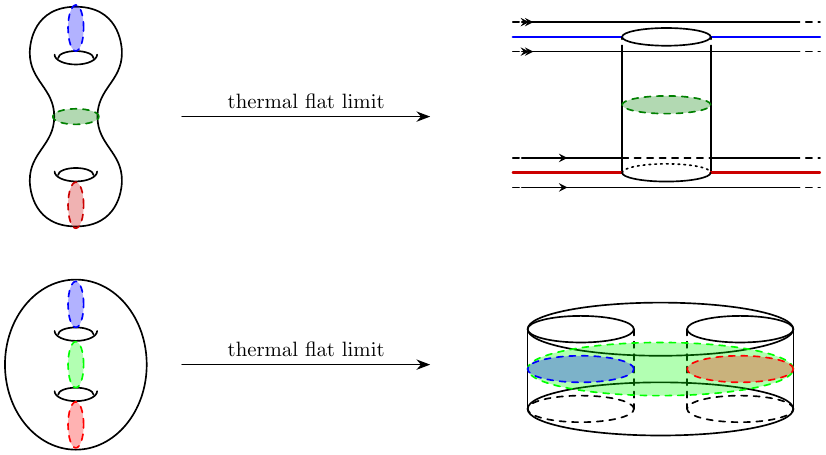}
\caption{\label{fig:thermalflatlimit}Cartoon illustrating the thermal flat limit of the genus-2 manifold with colored circles and lines representing 
different  $d-$dimensional balls. In the upper graph,
the two strips in the right diagram 
represent two copies of $S^1\times \mathbb{R}^{d-1}$ where the lines in the front and back are identified as suggested by the arrows.
The lower graph represents the dumbbell channel geometry. In the thermal flat limit, 
the two smaller balls are tangential to the larger one but they are not tangent to each other.
}
\end{figure}

To further understand the implications of the crossing equation (\ref{eq:ourequation}), we restrict to an interesting one-dimensional locus within the aforementioned moduli space. Starting from section \ref{sec:thermal flat limit},
we focus on the 
\textit{thermal flat limit}, where the $S^1\times S^{d-1}$'s blow up into copies of thermal flat 
space $S^1\times \mathbb{R}^{d-1}$. The genus 2 manifold $M_2$ then becomes two copies of thermal flat space $S^1 \x \R^{d-1}$, connected by a cylinder, see figure~\ref{fig:thermalflatlimit}. A special case within the thermal flat limit was investigated in \cite{Benjamin:2023qsc},
producing an asymptotic expression for thermal one-point coefficients. 
Relatedly, in the crossing equation considered in this work, the following function naturally arises in the dumbbell channel decomposition:
\begin{equation}
\label{eq:thefunctionh}
    h(z)=\sum_{\mathcal{O}} b^2_\mathcal{O} q_J z^{\Delta_\mathcal{O}},
\end{equation}
where  $z$ is the coordinate along the thermal flat locus, 
$q_J$ is a $J$-dependent constant, 
and $b_\mathcal{O}$ is the thermal one-point function coefficient of $\cO$, see e.g.\ \cite{Iliesiu:2018fao}. This function simultaneously encodes all the thermal one-point functions of a CFT, and may have interesting applications to the thermal bootstrap.

As a concrete application of the thermal flat limit of the genus 2 crossing equation, in section \ref{sec:OPE asymptotic} we derive an asymptotic relation between (averaged) heavy-heavy-heavy OPE coefficients 
and thermal one-point coefficients.
Analogous relations between thermal one-point data and heavy-heavy-light OPE coefficients $c_{\phi\mathcal{O}\mathcal{O}}$  are known
\cite{Gobeil:2018fzy,Buric:2025uqt}, obtained by inverting the block decomposition of 
one-point functions on $S^1 \times S^{d-1}$ \cite{Buric:2024kxo}. The OPE asymptotics we present in this work, however, hold in a qualitatively different regime. 
For example, for scalars in 3d, we find that, schematically: 
\be
\label{eqn:main_formula_intro}
    \rho_1 \rho_2 \rho_3 (c_{123})^2 &\sim 
    h(z_*)
    \exp \left(3 (f\pi)^{1/3}(\Delta_1^{2/3}+\Delta_2^{2/3})-16 \pi c_1\right) \times \textrm{``one-loop factors"},
\ee
where $\rho_i$ are the density of states for representations $\pi_i=(\Delta_i,\lambda_i)$, $h$ is given in (\ref{eq:thefunctionh}), 
\be
 z_* &= \left(\frac{\Delta _1^{2} \Delta _2^{2}}{\pi f  \Delta _3^3}\right)^{\frac{2}{3}},
\ee
(The ``one-loop factors" are also important at large $\Delta$, but we have not included them in (\ref{eqn:main_formula_intro}) for brevity. The precise formula is given below in 
 (\ref{eqn:main_formula}).) The quantities $f$ and $c_1$ are theory-dependent Wilson coefficients in the thermal effective action \cite{Bhattacharyya:2007vs,Banerjee:2012iz,Kang:2022orq,Benjamin:2023qsc,Benjamin:2024kdg,Jensen:2012jh}.  The formula (\ref{eqn:main_formula_intro}) holds for large $\De_1,\De_2,\De_3$, with $z_*$ held fixed --- in other words, for $\De_3$ parametrically larger than $\De_1$ and $\De_2$. (This hierarchy reflects an asymmetry among the three cutting spheres in the sunrise channel in the thermal flat limit.) We sometimes refer to this regime as the ``heavy-heavy-heavier" limit. A similar result holds for spinning operators,
 as long as their spins are held fixed as $\Delta_i \rightarrow \infty $, see (\ref{eqn:main_formula}) below. Unlike the OPE coefficient asymptotics in \cite{Benjamin:2023qsc}, our results do not depend on the ``hot spot conjecture" of \cite{Benjamin:2023qsc}, although they are fully consistent with it.

Equation (\ref{eqn:main_formula_intro}) follows from taking an inverse Laplace transform of the genus-2
partition function in the thermal flat limit.
 A crucial step in this calculation is to determine 
the quantum numbers dominating the sunrise channel decomposition. This can be done by computing expectation values of various 
Casimir operators, as we explain below. The expectation values of Casimirs tell us which scaling limit to take of the genus-2 block. We can then compute the genus-2 block in this scaling limit using the shadow formalism and saddle point approximations. 
The integral of interest has multiple saddle points, corresponding to 
the genus-2 block and seven different shadow blocks. The identification of the correct saddle point can be non-trivial --- for example in \cite{Benjamin:2023qsc} it was done by
numerically following the correct low-temperature saddle to a high temperature limit.
Here, we find a more direct method to identify the correct saddle using expectation values of conformal generators.

Finally, we devote space in appendices to further discussion of genus 2 blocks. 
Similarly to four-point conformal blocks, the genus-2 blocks satisfy a set of Dolan-Osborn-like Casimir equations \cite{DO1,DO2,DO3}. 
We derive these equations for genus-2 sunrise and dumbbell 
blocks in appendix \ref{appendix:Casimir}.

\section{The genus-2 geometry}\label{sec:the genus-2 geometry}
In this section, we review the parameterization of flat conformal structures on $M_2$. We introduce two sets of 
coordinates: dumbbell coordinates and sunrise coordinates, and discuss the translation between them.
After writing down the genus-2 crossing equation, we comment on why mapping class group invariance in 3d is trivialized by this 
crossing equation. 

\subsection{Genus-2 geometry in the sunrise channel}
\label{sec:genus2geomtrysunrise}

The genus-2 manifold $M_2$ can be built by taking two copies of the plane $\mathbb{R}^d$, drilling out three balls $(B_1,B_2,B_3)$ and 
$(B'_1,B'_2,B'_3)$ from each plane and connecting their boundaries by three cylinders $C_1,C_2,C_3$.
Using the Weyl-equivalence between a cylinder and an annulus, one can map each cylinder $C_i$ either to 
the interior of $\partial B_i$ or the interior of $\partial B'_i$, introducing two sets of coordinates $x$ and $x'$ inside $C_i$.
A flat conformal structure on $M_2$  can be specified by giving transition maps between these coordinates in each cylinder:
\begin{equation}
    x = g_i x',\quad g_i \in G^-,\quad (\text{inside } \ C_i),
\end{equation}
where $G^-$ denotes the orientation reversing\footnote{Even though we are gluing by orientation-reversing conformal transformations, the manifold itself is orientable: along any closed loop
on $M_2$, an even number of orientation reversals are applied, creating no obstruction for the existence of a global orientation.} component of $O(d+1,1)$ . Since conformal transformations performed on each plane $\R^d$ 
alter the coordinates $x,x'$ without changing the conformal structure of $M_2$, the moduli (covering)\footnote{The actual moduli space is the quotient of $\mathcal{M}$ by the action of the mapping class group.} space 
of the flat conformal structure is given by the following double quotient:
\begin{equation}
    \mathcal{M} = G\backslash (G^-)^3/G.
\end{equation}

More concretely, let us enumerate the geometric parameters relevant to this construction and understand how they are related by conformal transformations.
We have four types of parameters to start with:
\begin{itemize}
    \item The positions of the punctures $(x_1,x_2,x_3)$ and $(x_1',x_2',x_3')$ on both planes.
    \item The radii of the six removed balls $r_i$ and $r_i'$.
    \item The lengths $\beta_i$ of the three cylinders.
    \item The angular twists $h_i$ along the three cylinders.
\end{itemize}

By performing individual conformal transformations on each plane, we can bring the punctures to standard locations, say, $0,e,\infty$ on each plane, 
leaving an $\SO(d-1)\times \SO(d-1)$ subgroup that preserves this configuration.
 The radii of the removed balls can be converted into the length of the cylinders: to glue an $S^{d-1}$ of radius $r_i'$ with 
 an $S^{d-1}$ with radius $r_i$ via a cylinder of length $\beta_i$, the transition map we write down is:
 \begin{equation}
    x'=e^{-(\beta_i-\log r_i-\log r_i')D} I\cdot  x.
 \end{equation} 
 This is equivalent to the map which glues two unit sphere with a cylinder of length $\beta_i-\log r_i-\log r_i'$.
Intuitively, one can use the plane-cylinder map to ``flatten" part of the cylinder onto the plane or ``squeeze" part of the plane onto the cylinder.
Therefore, the triplet $(r_i,r_i',\beta_i)$ contributes collectively as a single modulus. 
Because of the remaining stabilizer subgroup after fixing the centers  of 
 the removed balls, the angular twists $(h_1,h_2,h_3)$ are subject to the following equivalence relation:
 \begin{equation}
    (h_1,h_2,h_3)\sim (k h_1 k',k h_2 k',k h_3 k'),\quad k,k'\in \SO(d-1).
 \end{equation}
Together, the number of independent moduli is:
\begin{equation}
3+3\dim \SO(d)-2\dim \SO(d-1)=\frac{(d+1)(d+2)}{2}.
\end{equation}

For doing calculations, it is often useful to choose a \textit{conformal frame}, that is, to fix a subset of geometric
parameters so that variations of the rest lead to a truly different flat conformal structure.
However, completely fixing the conformal frame comes at the cost of breaking certain symmetries. Therefore, 
we will ``partially gauge fix''  and leave the $\SO(d-1)\times \SO(d-1)$ redundancy untouched. 

One possible conformal frame for describing the sunrise channel is described in \cite{Benjamin:2023qsc}. That conformal frame captures the ``high-temperature" limit described in that work, but it is not well-suited to the ``thermal flat limit" that will be our eventual focus. Instead, we will work with the following frame:
\begin{equation}
    \begin{split}
    x_1&=(-a_1-1,0,\cdots,0), \quad x_2=(a_2+1,0,\cdots,0),\quad x_3=\infty,\\
    r_1&=r_1'=1-a_1,\quad r_2=r_2'=1-a_2,\quad r_3=r_3'=2,\quad \beta_1=\beta_2=0,
    \end{split}
\end{equation}
 where $0<a_1<1,0<a_2<1$. The corresponding gluing maps are:
\begin{equation}
    \begin{split}
    g_1&= e^{-(1+a_1)P^1} e^{2\log(1-a_1)D}h_1  I e^{(1+a_1)P^1},\\
    g_2&= e^{(1+a_2)P^1} e^{2\log(1-a_2)D}h_2  I e^{-(1+a_2)P^1},\\ 
    g_3&=e^{(\beta_3+2\log(2))D} h_3 I,
    \end{split}
\end{equation}
and the angular twists $h_i$ are given by:
\begin{equation}\label{eqn:sunrise_channel_angular_twist}
    h_i = \exp\left(\sum_{a}i\alpha_i^{a} M_{1,a}\right)\exp\left(\sum_{a,b=2}^{d}i\Phi^{a,b}_{i} M_{a,b}\right),\quad i=1,2,3,
\end{equation}
where $M_{i,j}$ are (hermitian) rotation generators.
In this parameterization of the angular twists we've chosen to separate:
\begin{itemize}
\item The generators $M_{a,b}$ with $2\leq a,b\leq d$ that generate the subgroup $\SO(d-1)$ preserving the $x^1$-axis
an $x^1$-preserving $\SO(d-1)$, from
\item The generators $M_{1,a}$ with $2\leq a \leq d$ that transform in the vector representation under the $\SO(d-1)$.
\end{itemize}

The essential difference between this frame and the one in \cite{Benjamin:2023qsc} is as follows. In \cite{Benjamin:2023qsc}, the three spheres $\ptl B_i$ (and similarly $\ptl B_i'$) were mutually tangent. This made it easy to probe the limit where three ``hot spots" form by tuning the lengths of all three cylinders to zero. By contrast, the thermal flat limit includes only two hot spots. Our new conformal frame makes it easier to probe this regime by choosing $\ptl B_1$ and $\ptl B_2$ to be tangent to $\ptl B_3$, but not to each other (unless $a_1=a_2=0$).

Note that our choice of conformal frame gives rise to a piecewise differentiable metric on $M_2$, given by the flat metric on the planes $\R^d$ and the conventional metric on the cylinders $C_i$ connecting them. We will refer to this as the ``sunrise metric."

\subsection{Sunrise channel decomposition of the partition function}
The above construction of $M_2$ leads to the decomposition of $Z(M_2)$ as a sum of squared OPE coefficients, weighted 
by a purely kinematic function --- the genus-2 sunrise block: 
\be
&\left.Z(M_2)\right|_\mathrm{sunrise}
\nn\\
&=\left( \frac{|Z_{\text{glue}}(2)|}{|Z_{\text{glue}}(1-a_1)||Z_{\text{glue}}(1-a_2)| }\right)^2\times e^{-\varepsilon_0 \beta_3}
\nn\\ 
&\quad\times  \sum_{\mathcal{O}_1,\mathcal{O}_2,\mathcal{O}_3} \Big( e^{-\beta_3 \Delta_3}
\nn\\ 
&\quad\quad \langle
 \mathcal{O}_1^{a'}\left(-(a_1+1)e\right) \mathcal{O}_2^{b'}((a_2+1)e) \mathcal{O}_3^{c'}(\infty e) \rangle^*
\langle h_1\cdot \mathcal{O}_1^{a}\left(-(a_1+1)e\right)h_2\cdot  \mathcal{O}_2^{b}((a_2+1)e) h_3\cdot \mathcal{O}_3^{c}(\infty e) \rangle
\nn\\
&\quad\quad \times  \left( _{1-a_1}\langle \mathcal{O}_1^a|\mathcal{O}_1^{a'}\rangle_{1-a_1}\right)^{-1} 
\left( _{1-a_2}\langle \mathcal{O}_2^b|\mathcal{O}_2^{b'}\rangle_{1-a_2}\right)^{-1}
\left( _{2}\langle \mathcal{O}_1^a|\mathcal{O}_1^{a'}\rangle_{2}\right)^{-1}
\nn\\
&\quad\quad + \text{descendants}
\Big).
\label{eq:sunrisesumoverstates}
\ee
In odd dimensions, where there is no Weyl anomaly, $Z(M_2)$ will depend only in the flat conformal structure on $M_2$. In the presence of a Weyl anomaly, there is some residual dependence on the metric. The expression (\ref{eq:sunrisesumoverstates}) for $Z(M_2)$ holds in the sunrise metric described in section~\ref{sec:genus2geomtrysunrise}.

A detailed derivation of (\ref{eq:sunrisesumoverstates}) can be found in \cite{Benjamin:2023qsc}.  For now, we briefly unpack the notation:
\begin{itemize}
\item The inner product $ _{r}\langle \mathcal{O}|  \mathcal{O}'\rangle_r$ is taken over the Hilbert space associated with a sphere $S^{d-1}_r$ with radius $r$.
It is related the usual inner product over $\mathcal{H}_{S^{d-1}_1}$ by a numerical factor:
\begin{equation}
    _{r}\langle \mathcal{O}|\mathcal{O}'\rangle_r = r^{2\Delta_{\mathcal{O}}} \langle \mathcal{O}|\mathcal{O}' \rangle.
\end{equation}
\item The three-point functions can be further expanded with respect to a basis of conformally-invariant three-point structures:
\begin{equation}
    \langle \mathcal{O}_1^{a'}(-(a_1+1)e) \mathcal{O}_2^{b'}((a_2+1)e) \mathcal{O}_3^{c'}(\infty e) \rangle
    =(a_1+a_2+2)^{\Delta_3-\Delta_1-\Delta_2} c_{123}^s V^{s;abc}(0,e,\infty).
\end{equation}
Separating the OPE coefficients from the kinematic structures as shown above, we can write each term in the 
summand as :
\begin{equation}
    (c_{123}^{s'})^{*}c_{123}^s B_{123}^{s',s},
\end{equation}
where the function $B_{123}^{s',s}$ encapsulates the sum of contracted three-point structures over a triplet 
of conformal multiplets.
\item Whenever a cylinder is glued onto a plane, the metric has delta-function-type curvature localized at the 
junction. To account for this curvature, a ``junction factor'' is associated 
with the spheres $\partial B_i,\partial B_i'$. The exact value of the junction factor depends on
the radius of the sphere and the   ``type'' of the junction. There are two types of junctions in the 
geometry discussed above: an open junction where the exterior  of the sphere $\partial B_i$
 is filled and a closed junction where the interior of the sphere $\partial B_i$ is filled. An open junction 
 factor is associated with $\partial B_1,\partial B_2,\partial B'_1,\partial B'_2$ and a closed junction factor 
 is associated with $\partial B_3$ and $\partial B'_3$.
 
 To compute these junction factors,
 first note that the open junction factor should be the inverse of the closed function factor,
 since the plane is Weyl equivalent to an open junction placed infinitely close to a closed junction. This relation
 holds exactly even in the presence of a nontrivial Weyl anomaly, since the cylinder connecting an open and closed junction can 
 be made infinitesimally short. One can then use the Weyl equivalence between a sphere $S^{d}$
 and a capped cylinder to determine the closed junction factor. We denote the closed junction factor by $Z_{\text{glue}}$
 and it is given by:
 \begin{equation}
    Z_{\text{glue}}(r)=Z(S^{d})^{1/2}\times f_{\text{Weyl}}(r,\beta)\times e^{\varepsilon_0\beta/2},
 \end{equation}
where $\varepsilon_0$ is the Casimir energy in $\mathcal{H}_{S^{d-1}}$, $\beta$ is the length of the capped cylinder 
and $f_{\text{Weyl}}$ accounts for the Weyl anomaly associated with the transformation between a unit sphere and the 
capped cylinder. A detailed computation of $f_{\text{Weyl}}$ can be found in \cite{Benjamin:2023qsc}.

Our eventual focus will be on 3d, where Weyl factors and Casimir energies don't contribute, and the gluing factor is just $Z(S^d)^{1/2}$.
\end{itemize} 

\subsection{Genus-2 geometry in the dumbbell channel}
\label{subsec: genus-2 geometry in the dumbbell channel}

In the dumbbell channel, we use a different cutting and gluing scheme to construct $M_2$. We start by 
building the two ``dumbbells" as  mapping tori of conformal group elements $g_L,g_R\in \SO(d+1,1)$. We then remove
a ball from each of the $S^1\times S^{d-1}$'s and glue their boundary components via an orientation reversing 
conformal map $g_M\in G^{-}$ (``$M$" for ``middle"). The conformal structure is determined by $(g_L,g_M,g_R)\in G\times G^{-}\times G$,
up to the gauge redundancy:
\begin{equation}
    (g_L,g_M,g_R)\sim (g g_L g^{-1},g g_M g'^{-1},g' g_R g'^{-1}),\quad g,g'\in SO(d+1,1).
\end{equation}

\begin{figure}[htbp]
  \centering
  \includegraphics[trim=50 120 70 50, clip,width=0.5\textwidth]{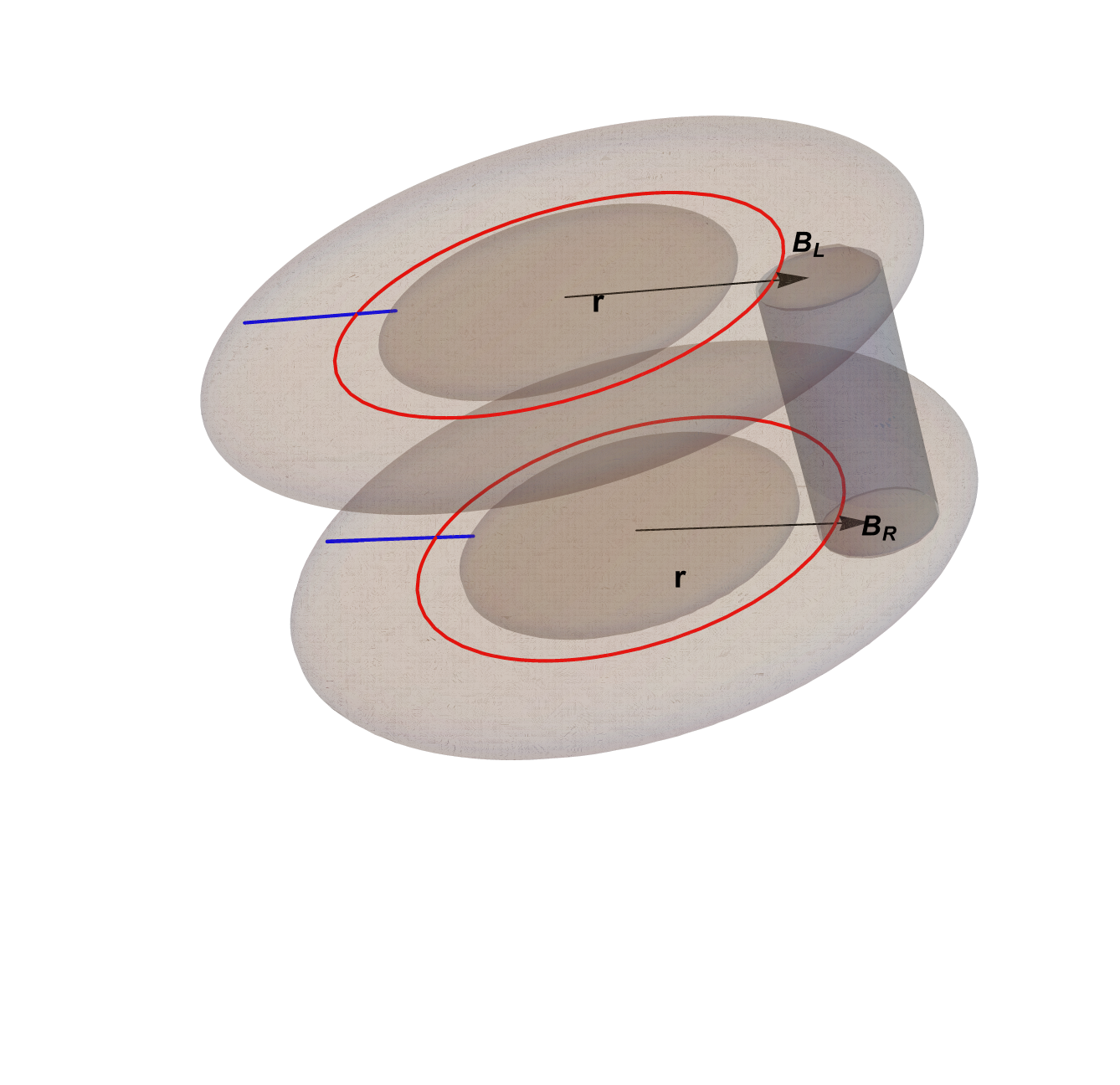}

  \caption{In the dumbbell channel, we obtain $M_2$ by removing two unit balls
  $B_L$ and $B_R$ from two pieces of $S^1\times S^{d-1}$ and glue their boundaries with a cylinder of 
  inverse temperature $\beta$. The two blue lines represent two non-contractable loops
  with length $2r\sinh(\beta_L/2r)$ and  $2r\sinh(\beta_R/2r)$ 
  . In $d>2$, the two red loops can be shrunk towards poles of $S^{d-1}$.
  }
  \label{fig:myimage}
\end{figure}

We choose the following representative geometry (see figure~\ref{fig:myimage}): 
take two pieces of $\mathbb{R}^d$ with flat coordinates $x_L$ and $x_R$.
We can build the left dumbbell $A_L$ by gluing an $S^{d-1}$ of radius $e^{-\beta_L/2 r}r$  with another $S^{d-1}$ with radius $e^{\beta_L/2 r} r$ after applying an
angular twist $h_L\in \SO(d)$. The right dumbbell $A_R$ can be built similarly via a gluing map $g_R \equiv e^{-(\beta_R/r_R) D}h_R$.

From the two annuli $A_L$ and $A_R$, we remove unit balls $B_L$ and $B_R$ centered at $x_L=(-r,0,...,0)$ and $x_R=(-r,0,...,0)$,
respectively\footnote{For this operation to make sense geometrically, we need $\beta_L,\beta_R > 2 r\log(\frac{r}{r-1})$.}. Let $x_M$,$x_M'$ be coordinate systems with origins at the centers of $B_L$, $B_R$ and let $\partial B_L$ and $\partial B_R$ be located at $|x_M|=1$ and $|x_M'|=1$. Using 
Weyl equivalence between the cylinder and the annulus, we can extend the coordinates 
$x_M$ and $x_M'$ to cover a cylinder connecting $\partial B_L$ and $\partial B_R$. At any point on such a 
cylinder, the transition map between $x_M$ and $x_M'$ is given by an orientation reversing conformal group element:
\begin{equation}
   x_M=  e^{-\beta D}h_M I x'_M,\quad h_M\in \SO(d).
\end{equation}
 Combined with the transition map between $x_L, x_M$ and $x_R,x_M'$:
 \begin{equation}
    x_M=e^{r P^1} x_L,\quad x_M'=e^{r P^1} x_R,
 \end{equation}
this leads to the transition map between $x_L$ and $x_R$:
\begin{equation}
    x_L=e^{-r P^1}e^{-\beta D}h_M I e^{r P^1} x_R.
\end{equation}

In this conformal frame, the flat conformal structure on $M_2$ is therefore specified by the following 
gluing maps:
\begin{equation}
    \begin{split}
    g_L&=e^{-(\beta_L/r) D}h_L,\\
    g_M&= e^{-r P^1} e^{-\beta D}h_M I e^{r P^1},\\
    g_R&=e^{-(\beta_R/r) D}h_R,
    \end{split}
\end{equation}
subject to the equivalence relation:
\begin{equation}\label{eqn:redundancy_dumbbell_1}
    (h_L,h_M,h_R)\sim (k_L h_L k_L^{-1},k_L h_M k_R^{-1},k_R h_R k_R^{-1}),\quad k_L,k_R\in \SO(d-1),
\end{equation}
and 
\begin{equation}\label{eqn:redundancy_dumbbell_2}
    (r,\beta)\sim (e^{-\delta}r,\beta+2\delta),\quad \delta\in \mathbb{R}.
\end{equation}
The number of independent parameters is again $(d+1)(d+2)/2$.
One can in principle  fix the gauge redundancy in  (\ref{eqn:redundancy_dumbbell_2}) but we will
keep $r$ and $\beta$ both as free parameters since it makes the definition of the thermal flat limit more convenient. 

\subsection{Dumbbell channel decomposition of the partition function}
\label{sec:dumbbellpartition}

Now we are ready to derive the dumbbell channel decomposition of $Z(M_2)$ in detail. The basic idea is to insert 
resolutions of the identity at $\partial B_L$ and $\partial B_R$. More precisely, we have:
\be
&\left.Z(M_2)\right|_\mathrm{dumbbell}
\nn\\
&= \frac{1}{|Z_{\text{glue}}(1)|^2}  \sum_{\mathcal{O}_L,\mathcal{O}_R} Z(A_L\backslash B_L)(|\mathcal{O}_L^{a}\rangle)\langle \mathcal{O}_L^a|\mathcal{O}_L^{a'}\rangle^{-1} \langle \mathcal{O}_L^{a'}| e^{-\beta D}h_M| \mathcal{O}_R^{b'}\rangle \langle \mathcal{O}_R^{b}|\mathcal{O}_R^{b'}\rangle^{-1} (Z(A_R \backslash B_R)(|\mathcal{O}_R^b\rangle))^* 
\nn\\
&\quad+ \text{descendants},
\ee
In the equation above, the indices $a,a',b,b'$ are implicitly summed over.  $\langle \mathcal{O}_L^a|\mathcal{O}_L^{a'}\rangle$ 
is the inner product in $\mathcal{H}_{S^{d-1}_{r=1}}$ and it is equivalently computed by the following two point function:
\begin{equation}
    \langle \mathcal{O}_L^{a}(0)[\mathcal{O}_L^{a'}(0)]^{\dagger}]\rangle,
\end{equation}
where $[\cdots]^\dagger$ denotes BPZ conjugation. Since $\partial(A_L\backslash B_L) = - \partial B_L$, we can view 
$Z(A_L\backslash B_L)$ as a map $\mathcal{H}_{S^{d-1}_{r=1}}\rightarrow \mathbb{C}$. When acting on $|\mathcal{O}^a\rangle$,
it evaluates an (unnormalized) one-point function on $A_L$ with $\mathcal{O}^a$ inserted at the center of $B_L$:
\begin{equation}
    Z(A_L\backslash B_L)(|\mathcal{O}^a\rangle)=\langle \mathcal{O}^a(r,\hat{n}) \rangle_{A_L}\times  \langle 1 \rangle_{A_L}.
\end{equation}
Using Weyl equivalence between an annulus and the cylinder, we have: 
\begin{equation}
    \langle 1 \rangle_{A_L} = Z(S^1_{\beta_L}\times S^{d-1}_{r}),\quad 
     \langle \mathcal{O}^a(r,\hat{n}) \rangle_{A_L} = e^{-\beta_L\Delta_\mathcal{O}/2 r}\langle\mathcal{O}^a(\beta_L/2,\hat{n}) \rangle_{S^{1}_{\beta_L}\times S^{d-1}_r}.
\end{equation} 
Combining everything, we get:
\be\label{eqn:ZM2_dumbbell}
&Z(M_2)|_\mathrm{dumbbell}
\nn\\
 &= \frac{1}{|Z_{\text{glue}}(1)|^2} Z(S^1_{\beta_L}\times S^{d-1}_{r})Z(S^1_{\beta_R}\times S^{d-1}_r)
\nn\\
&\quad  \times  e^{-\beta \varepsilon_0}\sum_{\mathcal{O}}\Bigg(
    \langle \mathcal{O}^a(\beta_L/2,\hat{n}) \rangle_{S^1_{\beta_L}\times S^{d-1}_r}
    \langle (h_M\cdot\mathcal{O})^{a'}(\beta_R/2,\hat{n}) \rangle_{S^1_{\beta_R}\times S^{d-1}_r}
    \langle\mathcal{O}^a | \mathcal{O}^{a'} \rangle^{-1} 
    e^{-(\beta+\frac{\beta_L+\beta_R}{2r})\Delta_{\mathcal{O}}}
\nn\\
&\quad \quad + \text{descendants} \Bigg),
\ee
where $\varepsilon_0$ is the Casimir energy on $S^{d-1}$ and for an operator in the conformal representation $(\Delta,\lambda)$, $h_M$ acts by:
\begin{equation}
    h_M \cdot \mathcal{O}^a=\lambda(h_M^{-1})^{a}_{\ b} \mathcal{O}^b.
\end{equation}
This expression for $Z(M_2)$ is valid in the metric associated to the geometry described in section~\ref{subsec: genus-2 geometry in the dumbbell channel}.

Let us review some basic features of the thermal one-point function on $S^{d-1}$ \cite{Buric:2024kxo}, which appears in (\ref{eqn:ZM2_dumbbell}). When the angular twists $h_R$ and $h_L$
    are turned off, any correlation function on $S^1_{\beta}\times S^{d-1}_r$ must be invariant under the $SO(2)\times SO(d)$ isometry group of the background 
    geometry. Therefore, the one-point function $\langle \mathcal{O}^a(\tau,\hat{n})\rangle_{S^1_{\beta}\times S^{d-1}_r}$ cannot have
    $\tau$ or $\hat{n}$ dependence. In this case, one-point functions of descendants vanish, and the sum (\ref{eqn:ZM2_dumbbell}) truncates to only the primary contributions. Furthermore, only even-spin symmetric traceless tensors can have non-vanishing one-point functions. Their form is fixed by symmetry to be
    \begin{equation}
        \langle \mathcal{O}^{\mu_1,\cdots\mu_\ell}(\tau,\hat{n})\rangle_{S^1_{\beta} \times S^{d-1}_{r}}=\frac{b_\mathcal{O}}{\beta^{\Delta_\mathcal{O}}}\times f_{\mathcal{O}}(\beta/r)(e^{\mu_1}\cdots e^{\mu_\ell}-\text{traces}),
        \label{eq:evenspinonepointfunction}
    \end{equation}
where $e^\mu$ denotes the unit vector along the $\tau$-direction. The function 
$f_\mathcal{O}(\beta/r)$
satisfyies the boundary condition $f_{\mathcal{O}}(0)=1$. Odd-spin one-point functions are forbidden by a $\pi$ rotation that fixes the operator insertion, 
flips the direction of $S^1$, and reflects the $S^{d-1}$.

When angular fugacities are turned on, $SO(d)$ invariance is generically broken to the Cartan subgroup, allowing 
new coordinates to enter the one-point function. For example, when $d=3$, in the presence of an angular twist $e^{i\theta M_{12}}$,
a one-point function can also depend on $\hat{n}_1^2+\hat{n}_2^2=1-\hat{n}_3^2$. In this case,
descendant operators can have non-trivial one-point functions.
For more discussion in 3d, see \cite{Buric:2024kxo}.

The large-$r$ asymptotics of the partition functions $Z(S^1_{\beta_L}\times S^{d-1}_{r})$ and $Z(S^1_{\beta_R}\times S^{d-1}_{r})$
are captured by the thermal effective action  \cite{Benjamin:2023qsc}. At leading order:

\begin{equation}
Z(\beta_L)\sim \exp\left( \frac{f r^{d-1}\ \vol S^{d-1}}{\beta_L^{d-1}\prod_{i=1}^{n}(1+\Omega_i^2)}\right),
\quad 
Z(\beta_R)\sim \exp\left( \frac{f r^{d-1} \vol S^{d-1}}{\beta_R^{d-1}\prod_{i=1}^{n}(1+\Omega_i^2)}\right),
\end{equation} 
where $n=\lfloor d/2\rfloor$ is the rank of $SO(d)$ and $\beta_L \Omega_i/r$ and $\beta_R \Omega_i/r$ are angular fugacities in 
 $h_L$ and $h_R$, respectively. Here, $f$ is a theory-dependent coefficient, and is proportional to the central charge in 2d. 
In general dimensions, $f$ is proportional to the free energy density of the CFT at finite
temperature, the Casimir energy of the CFT on $S^1_\beta\times \mathbb{R}^{d-2}$ and the thermal 
one-point function of the stress energy tensor on $S^1\times \mathbb{R}^{d-1}$.

\subsection{Matching between dumbbell and sunrise channels}
\begin{figure}[h]
    \centering
\includegraphics[scale=1.2]{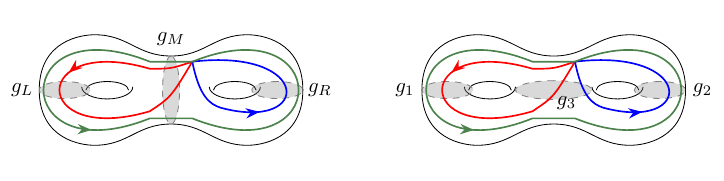}
\caption{\label{fig:figurewithloops}The correspondence of loops and conformal group elements in different channels. In the sunrise channel (right)
the $a$-loop (red) is mapped to the group element $g_1^{-1} g_3$, the loop $b$ (blue) is mapped to 
$g_3^{-1} g_2$ and the loop $a\cdot b$ (green) is mapped to $g_1^{-1} g_2$. In the dumbbell channel 
(left),the loop $a$ is mapped to $g_M^{-1} g_L^{-1} g_M$, the loop $b$ is mapped to $g_R$ and the loop $a\cdot b$ is 
mapped to $g_M^{-1} g_L^{-1} g_M g_R$. 
}
\end{figure}

Now that we've introduced two sets of coordinates on the moduli space of flat conformal structures $\mathcal{M}$, let us discuss how to match them. The basic idea is the following:
parameterizations of flat conformal structures 
in both channels give two homomorphisms from the fundamental group of $M_2$ to the conformal group $\varphi_D,\varphi_S:\pi_1(M_2) \rightarrow G$. 
For example, in the sunrise channel, the elements $a,b$ illustrated in figure~\ref{fig:figurewithloops}
are mapped to
\begin{equation}
    \varphi_S(a)=g_1^{-1}g_3,\quad \varphi_S(b)=g_3^{-1}g_2,
\end{equation}
 while in the dumbbell channel:
\begin{equation}
    \varphi_D(a)=g_M^{-1}g_L^{-1} g_M,\quad \varphi_S(b)=g_R.
\end{equation}

If $(g_L,g_M,g_R)$ and $(g_1,g_2,g_3)$ correspond to the same conformal structure, then
for any element $\ell \in \pi_1(M_2)\simeq \mathbb{Z}*\mathbb{Z}$, the group element $\varphi_D(\ell)$ should be conjugate to $\varphi_S(\ell)$ in $G$. In practice, we will 
choose a set of loops $\{\ell_i\}$ and a particular representation $\rho$ of the conformal group $G$, then match the dumbbell coordinates and the sunrise coordinates 
via the identity:
\begin{equation}
    \chi_\rho(\varphi_D(\ell_i))=\chi_\rho(\varphi_S(\ell_i)),
\end{equation}
where $\chi_\rho(g)$ denotes the character of $g$ in the representation $\rho$.

As a simple example, let us look at the case $d=1$. (The genus-2 manifold $M_2$ does not make sense in this case, but the essential group theory still works. The computations below also apply when the angular twists are turned off, so that all gluing group elements lie in an $\SO(1,2)$ subgroup.) Let us choose $\rho$ to be the vector representation of $SO(1,2)$ and 
$\{\ell_1,\ell_2,\ell_3\}=\{a,b,a\cdot b\}$. We can match the three dumbbell coordinates $\beta_L,\beta_R,\beta$
with the three sunrise coordinates by solving the following equations:
\begin{equation}\label{dumbbell_sunrise_matching_1d_full}
    \begin{split}
    \frac{a_1^2 \left(4 e^{-\beta _3}-1\right)-6 a_1+4 e^{\beta
   _3}-1}{\left(a_1-1\right){}^2} &=e^{-\beta_L/r}+e^{\beta_L/r}+1,\\
   \frac{a_2^2 \left(4 e^{-\beta _3}-1\right)-6 a_2+4 e^{\beta
   _3}-1}{\left(a_2-1\right){}^2}&= e^{-\beta_R/r}+e^{\beta_R/r}+1,\\
  \frac{\left(7 a_2+a_1 \left(a_2+7\right)+1\right) \left(5 a_2+a_1 \left(3
   a_2+5\right)+3\right)}{\left(a_1-1\right){}^2 \left(a_2-1\right){}^2}
   &=e^{-\frac{\beta _L+\beta _R}{r}} 
   \Big[
    e^{2 \beta} r^4
     \left(e^{\beta_L/r}-1\right){}^2 
   \left(e^{\beta_R/r}-1\right){}^2\\
   -2 e^{\beta } r^2
   \left(e^{\beta _L/r}-1\right)
    \left(e^{\beta _R/r}-1\right)&
   \left(e^{\frac{(\beta _L+\beta _R)}{r}}+1\right)
   +e^{\frac{\beta _L+\beta
   _R}{r}}+e^{\frac{2 \left(\beta _L+\beta _R\right)}{r}}+1
   \Big].
    \end{split}
\end{equation}

We will come back to these equations later and solve them in the large-$r$ limit. A priori, it
might seem like this matching method can provide infinitely many ``loop relations.'' However, only finitely many 
loops give rise to independent identities. For example, when $d=1$ and $\dim \rho=3$, the identity coming 
from $a^2$ is redundant since:
\begin{equation}
 \chi_\rho(\varphi_{S,D}(a^2))=\chi_\rho(\varphi_{S,D}(a))^2-2 \chi_\rho(\varphi_{S,D}(a)).
\end{equation}

When working with vector representations of $SO(d+1,1)$ in higher dimension, we find the following set of 
loops to give a sufficient set of independent relations:
\begin{equation}\label{eqn:coordinates}
    \begin{split}
    d=2&: a,a^2,b,b^2,ab,abab\\
    d=3&: a,a^2,b,b^2,ab,abab,aba,bab,ab^{-1},ab^{-1}ab\\
    d=4&: a,a^2,a^3,b,b^2,b^3,ab,(ab)^2,(ab)^3,aba,bab,ab^{-1},ab^{-1}ab,b^{-2}a,(ab)^2a\\
    d=5&: a,a^2,a^3,b,b^2,b^3,ab,(ab)^2,(ab)^3,aba,bab,ab^{-1},ab^{-1}ab,b^{-2}a,(ab)^2a\\
     & (ab)^3a,(ab)^3b,b^{-3}a,a^3b^{-1},a^3b,a^4b\\
    d=6&: a,a^2,a^3,a^4,b,b^2,b^3,b^4,ab,(ab)^2,(ab)^3,(ab)^4,aba,bab,ab^{-1},ab^{-1}ab,b^{-2}a,(ab)^2a\\
     & (ab)^3a,(ab)^3b,b^{-3}a,a^3b^{-1},a^3b,a^4b,b^{-3}a^{-1},b^{-2}(ab)^{-2},b^{-1}a(ab)^{-2},a^2(ab)^{-2}.
    \end{split}
\end{equation}
The number of independent loop relations is always equal to $(d+1)(d+2)/2$. In fact, $\chi_\rho(\ell_i)$
furnishes another set of coordinates on $\mathcal{M}$. In a later section we will write down Casimir operators acting on the genus-2
block using these coordinates.

Given these identifications between sunrise and dumbbell coordinates, the genus-2 crossing equation is
\be
\label{eq:theequation}
Z(M_2)|_\mathrm{sunrise} &= e^{S_\mathrm{Weyl}} Z(M_2)|_\mathrm{dumbbell},
\ee
where $Z(M_2)|_\mathrm{sunrise}$ is given in (\ref{eq:sunrisesumoverstates}), $Z(M_2)|_\mathrm{dumbbell}$ is given in (\ref{eqn:ZM2_dumbbell}), and $e^{S_\mathrm{Weyl}}$ is the contribution from the Weyl anomaly in changing from the the dumbbell metric (section~\ref{subsec: genus-2 geometry in the dumbbell channel}) to the sunrise metric (section~\ref{sec:genus2geomtrysunrise}). We will not attempt to evaluate this contribution in general. However our eventual interest will be in 3d CFTs, where $e^{S_\mathrm{Weyl}}$ is 1.

\subsection{Mapping class group of $(S^1\times S^2)\sharp (S^1\times S^2)$}
The mapping class group MCG$(M_2)$ of a manifold is defined to be the group of isotopy classes of orientation-preserving diffeomorphisms.
Apart from satisfying Ward identities, CFT correlation functions should be invariant under the mapping class group.  In this section, we discuss the mapping class group of 
$(S^1\times S^2)^{\sharp 2}$ and explain why our genus 2 crossing equation trivializes MCG invariance. 
Roughly speaking, the generators of MCG($M_2$) come in two types: the ones that act non-trivially on the fundamental group $\pi(M_2)\simeq \mathbb{Z}*\mathbb{Z}$
and those that act trivially. The mapping class group elements that act non-trivially on $\pi_1(M_2)$ form the group $\text{Out}(F_2).$\footnote{Recall that the outer automorphism group of a group $G$ is defined as $\text{Aut}(G)/\text{Inn}(G)$ where the inner automorphism
group $\text{Inn}(G)$ is given by conjugation action of a group element. The reason $\text{Out}(\pi_1)$ appears is because 
the action of the inner automorphisms $\text{Inn}(\pi_1)$ simply alters the starting points of the loops.} 
Those that act trivially form 
$(\mathbb{Z}/2\mathbb{Z})^2$ --- they consist of sphere twists
around the core $S^2$'s in each $S^1\times S^2$.\footnote{Given a sphere $S^2$ and a tubular neighborhood $S^2 \times [0,1]$ around it, 
to implement a sphere twist, keep everything outside of the tubular neighborhood unchanged
but implement a full turn from $S^2\times \{0\}$ to $S^2\times \{1\}$.
This is very much like a higher dimensional version of Dehn twist in 2d surfaces
but unlike in $2d$, the sphere twist has order (at most) two --- the configuration of 
the sphere inside the cylinder $S^2\times [0,1]$ traces out a loop in the group $SO(3)$
 and $\pi_1(SO(3))\simeq \mathbb{Z}/2\mathbb{Z}$.}
 The precise statement is \cite{Brendle_2023}:
\begin{equation}
    \text{MCG}(M_2) \simeq (\mathbb{Z}/2\mathbb{Z})^2 \rtimes \text{Out}(F_2).
\end{equation}

To analyze MCG invariance of the CFT partition function, we must understand 
how to translate MCG generators to operators acting on Hilbert space. 
Generators of $(\mathbb{Z}/2\mathbb{Z})^2$ correspond to insertions of a $2\pi$ rotation
in $g_2,g_3$ or $g_L,g_R$. For bosonic theories, this has no effect, and so the partition function is invariant under this operation. In fermionic theories, this corresponds to an insertion of $(-1)^F$ and hence relates partition functions with different 
spin-structures. The partition functions with different spin structures can be collected into a ``partition vector," and insertions of $(-1)^F$ have a linear action on this vector --- we can say that the partition vector is MCG covariant.

The $\text{Out}(F_2)$ part of the mapping class group is generated by three generators 
$\sigma,\iota,\rho$ with the following actions on $a,b$:
\begin{equation}
    \begin{split}
    \sigma&: (a,b)\rightarrow (b^{-1},a^{-1}),\\
    \iota&: (a,b)\rightarrow (a,b^{-1}),\\
    \rho&:(a,b)\rightarrow (ab,b^{-1}),
    \end{split}
\end{equation}
The action $\sigma$ and $\rho$ can be easily understood in the sunrise channel: they correspond to 
exchanging $g_1,g_2$ and $g_2,g_3$. From the sunrise channel expansion of $Z(M_2)$, it is clear that the partition function is invariant under 
such exchanges since invariance of three-point functions under permutation implies:
\begin{equation}
\sum_{s,s'} P^{s,s'}_{123} B^{s',s}_{123}(g_1,g_2,g_3)= \sum_{s,s'} P^{s,s'}_{213} B^{s',s}_{213}(g_2,g_1,g_3)=
\sum_{s,s'} P^{s,s'}_{132} B^{s',s}_{132}(g_1,g_3,g_2).
\end{equation}

The action of $\iota$, however, is better understood from the dumbbell channel. 
Under $\iota$, an extra $\pi$-rotation is performed along the ``neck'' of the dumbbell so 
$g_M\rightarrow e^{i\pi M_{12}}g_M$. We call this operation a ``dumbbell flip." In $A_R$, $\iota$ flips the sign of a spatial direction 
and a time direction, therefore:
\begin{equation}
    g_R = e^{-(\beta_R/r)D}h_R I \rightarrow e^{-(\beta_R/r)D} \mathcal{R} h_R^{-1}\mathcal{R} I = (I \mathcal{R}) g_R^{-1}(I \mathcal{R}),
\end{equation} 
where $\mathcal{R}$ is a reflection along, say $x_2$. When viewed as an operator acting on $\mathcal{H}_{S^{d-1}_r}$, the change in $g_R$
can be implemented as conjugation by $J_{\Omega}\equiv\mathsf{CRT}$. Recall that the operator $J_{\Omega}$ is an anti-unitary operator that commutes with the Hamiltonian. Thus:
\begin{equation} 
J_{\Omega} e^{-(\beta_R/r) D} J_{\Omega}^{-1}=e^{-(\beta_R/r) D},\quad  J_{\Omega} e^{i \vec{M}\cdot\vec{\theta}} J_{\Omega}^{-1} = \mathcal{R} e^{-i \vec{M}\cdot\vec{\theta}} \mathcal{R}^{-1},
\end{equation}
Similarly the change in $ (h_M\cdot \mathcal{O})_{a}$ under $\iota$ is implemented as:
\begin{equation}
   (e^{i\pi M_{12}}h_M\cdot \mathcal{O})_{a}(x)=J_{\Omega} (h_M\cdot \mathcal{O})_{a}(x)J_{\Omega}^{-1}.
\end{equation}
Invariance under $\iota$ then follows naturally in the dumbbell channel from cyclicity of the trace.

The argument for invariance under the dumbbell flip $\iota$ is even simpler when angular fugacities are turned off in $A_R$. In that case, as discussed in section~\ref{sec:dumbbellpartition}, only even-spin operators have non-vanishing one-point functions on $S^1 \x S^{d-1}$. The rotation $e^{i\pi M_{12}}$ flips the direction of $e^\mu$ (the tangent vector in the $S^1$ direction), and even-spin one-point functions (\ref{eq:evenspinonepointfunction}) are invariant under this operation. It follows that the partition function is invariant under the dumbbell flip. 

We conclude this section with a side remark. The mapping class group of the $n-$fold 
connected sum $M_n\equiv (S^1\times S^2)^{\sharp n}$ is known to be \cite{Brendle_2023}:
\begin{equation}
    (S^1\times S^2)^{\sharp n} = H^1(M_n,\mathbb{Z}/2\mathbb{Z})\rtimes \text{Out}(F_n),
\end{equation}
where $H^1(M_n,\mathbb{Z}/2\mathbb{Z})$, as a group, is just the group of sphere twists. Restricting to the case $n=1$,
we find that the mapping class group of $(S^1\times S^2)$ is $\mathbb{Z}/2\mathbb{Z}\rtimes\mathbb{Z}/2\mathbb{Z}$. One of the 
generators is a sphere twist along $S^1$. In bosonic theories, the partition function is invariant under this action, while in fermionic theories the spin-structure will be changed. The other $\mathbb{Z}/2\mathbb{Z}$ is generated by the $\pi$-rotation
discussed in section \ref{subsec: genus-2 geometry in the dumbbell channel} and invariance under this subgroup forbids odd spin operators from having non-zero thermal one-point functions.

\section{The thermal flat limit }\label{sec:thermal flat limit}

In this section, we focus on a simplifying limit of the genus-2 crossing equation (\ref{eq:theequation}). 
 After defining this limit in terms of the dumbbell coordinates, we
translate it into sunrise coordinates via the matching procedure described previously. In section \ref{sec:OPE asymptotic}, 
we use our simplified crossing equation to derive asymptotic relations between OPE coefficients and thermal one-point coefficients. As 
a prerequisite, we must pinpoint the dominant quantum numbers in the genus-2 block decomposition of  $Z(M_2)$. We explain how to do this 
by evaluating the expectation value of Casimir operators in subsection \ref{subsubsec:saddles in Delta}. Similar techniques can be used to locate saddle points
in the partial wave integral --- an integral essential for the computation of the genus-2 sunrise block. We discuss this in subsection \ref{subsubsec:saddles in the shadow integral}.
\subsection{Thermal flat limit of the genus-2 geometry}
\subsubsection{Thermal flat limit in dumbbell coordinates}
We define the thermal flat limit as the limit where $r$ scales towards infinity and $\beta_L,\beta_R,\beta$ are all held fixed. Physically, this means that the radii of the annuli $A_L,A_R$ get large, while their temperatures are held fixed. For simplicity, let us simultaneously turn off the angular twists:
\begin{equation}
    \text{thermal flat limit}: h_M= h_L=h_R=1,\quad r\rightarrow \infty.
\end{equation}
As discussed in subsection \ref{subsec: genus-2 geometry in the dumbbell channel}, we work in a conformal frame
where the ``dumbbell'' $A_L$ (respectively $A_R$) is built by gluing a sphere $S^{d-1}$ of radius $e^{-\beta_L/2r} r$  (resp. $e^{-\beta_R/2r} r$) with a larger 
sphere of radius $e^{\beta_L/2r} r$ (respectively $e^{\beta_R/2r} r$). As $r\rightarrow \infty$, the spheres blow up to flat space $\mathbb{R}^{d-1}$
while the distance between them stays fixed.

In the thermal flat limit, the genus-2 partition function $Z(M_2)$ becomes infinite, but this singularity is entirely captured by the factors
$Z(S^1_{\beta_L}\times S_{r}^{d-1})Z(S^1_{\beta_R}\times S_{r}^{d-1})$. The quotient by these factors remains finite. From 
 (\ref{eqn:ZM2_dumbbell}), we find:
\begin{equation}
    \lim_{r\rightarrow \infty}\frac{Z(M_2)}{Z(S^1_{\beta_L}\times S_{r}^{d-1})Z(S^1_{\beta_R}\times S_{r}^{d-1})}
    =\frac{e^{-\beta \varepsilon_0}}{|Z_{\text{glue}}(1)|^2}
    \sum_{\cO_i,\cO_j}
    \langle\mathcal{O}_{i}^a\rangle_{S^1_{\beta_L}\times \mathbb{R}^{d-1}}
    \langle\mathcal{O}_{j}^b\rangle_{S^1_{\beta_R}\times \mathbb{R}^{d-1}}
    \langle \mathcal{O}_{i}^a \mathcal{O}_{j}^b\rangle^{-1} e^{-\beta \Delta_{\mathcal{O}_i}}.
\end{equation}
The sum on the right hand side is over even-spin STT primaries. We used superscripts $a,b$ as
spin indices, while subscripts $i,j$ indicate multiplicities. Using the identity:\footnote{This follows because homogeneity in $x,y$ dictates the contraction takes the form $|x|^J|y|^Jf\left(\frac{x\cdot y }{|x||y|}\right)$. Meanwhile, tracelessness translates to the differential equation:
    \begin{equation}
    \delta^{\mu\nu}\frac{d}{dy^\mu}\frac{d}{dy^\nu}\left(|x|^J|y|^Jf\left(\frac{x\cdot y }{|x||y|}\right)\right)=0
    \end{equation}
    Together with regularity at $\frac{x\cdot y}{|x||y|}=1$, these constraints force $f$ to be a Gegenbauer polynomial.}
\begin{equation}
    (x^{\mu_1}\cdots x^{\mu_J}-\text{traces})(y_{\mu_1}\cdots y_{\mu_J}-\text{traces})=q_J \mathcal{P}_J\left(\frac{x\cdot y}{|x||y|}\right)|x|^J|y|^J,
\end{equation}
where $\mathcal{P}_J(z)={}_2F_1(-J,d-2+J,\frac{d-1}{2};\frac{1-z}{2})$ is a Gegenbauer polynomial and the constant $q_J$ is:
\begin{equation}
    q_J = \frac{\Gamma(\frac{d-2}{2})\Gamma(J+d-2)}{2^J \Gamma(d-2)\Gamma(J+\frac{d-2}{2})},
\end{equation}
the sum above simplifies to:
\begin{equation}
\frac{e^{-\beta \varepsilon_0}}{|Z_{\text{glue}}(1)|^2}
\sum_{\mathcal{O}}b_{\mathcal{O}}^2 c^{-1}_{\mathcal{O}}
     q_{J_{\cO}} \left(\frac{e^{-\beta}}{\beta_L \beta_R}\right)^{\Delta_\mathcal{O}}\equiv \frac{e^{-\beta \varepsilon_0}}{|Z_{\text{glue}}(1)|^2} h\left( \frac{e^{-\beta}}{\beta_L \beta_R}\right),
\end{equation}
where the coefficient $c_{\mathcal{O}}$ is the normalization constant entering the two point function $\langle\mathcal{O}^a \mathcal{O}^b\rangle$.
 The function $h(z)$ is therefore independent of normalization conventions.

Note that the normalized genus-2 partition function now depends on a single combination of the moduli: $z\equiv e^{-\beta}/(\beta_L \beta_R)$.
By sending $r$ to infinity, we've scaled towards a one-dimensional locus in $\mathcal{M}$. This can alternatively be seen by examining the characters $\chi_\rho(\varphi_D(a))$,
$\chi_\rho(\varphi_D(b))$, and $\chi_\rho(\varphi_D(ab))$. 
In 1d, if $\rho = \square$, the characters read:
\be
\lim_{r\rightarrow \infty}\chi_\rho(\varphi_D(a))&=3,
\nn\\
\lim_{r\rightarrow \infty}\chi_\rho(\varphi_D(b))&=3,
\nn\\
\lim_{r\rightarrow \infty}\chi_\rho(\varphi_D(a\cdot b))&= 3+4 \beta_L \beta_R e^{\beta}-(\beta_L \beta_R e^{\beta})^2.
\ee
As expected, the only free parameter that appears is $z\equiv e^{-\beta}/\beta_L \beta_R$. 
This pattern holds in general dimensions. For example, choosing $d=3$ and $\rho=\square$, we similarly have:
\be
        &\lim_{r\rightarrow \infty}\chi_\rho(\varphi_D(a))=\lim_{r\rightarrow \infty}\chi_\rho(\varphi_D(a^2))
        =\lim_{r\rightarrow \infty}\chi_\rho(\varphi_D(b))=\lim_{r\rightarrow \infty}\chi_\rho(\varphi_D(b^2))=5
        \nn\\
        &\lim_{r\rightarrow \infty}\chi_\rho(\varphi_D(ab))=5-4z^{-1}+z^{-2},
        \nn\\ 
        &\lim_{r\rightarrow \infty}\chi_\rho(\varphi_D((ab)^2)) = z^{-4}-8 z^{-3}+20 z^{-2}-16 z^{-1}+5
        \nn\\
        &\lim_{r\rightarrow \infty}\chi_\rho(\varphi_D(aba))=\lim_{r\rightarrow \infty}\chi_\rho(\varphi_D(bab))
        =4 z^{-2}-8 z^{-1}+5,
        \nn\\
        &\lim_{r\rightarrow \infty}\chi_\rho(\varphi_D(ab^{-1}))=z^{-2}+4 z^{-1}+5,
        \nn\\
        &\lim_{r\rightarrow \infty}\chi_\rho(\varphi_D(ab^{-1}a b))=z^{-4}-4 z^{-2}+5.
\ee

Though we will mostly focus on the special case $h_M=h_L=h_R=1$, let us make 
some comments on angular twists. We distinguish different cases:

\begin{itemize}
    \item If $h_L=h_R=1$ and we only turn on $h_M$, then the two thermal one-point functions are still fixed by conformal symmetry. If $h_M$ lies in 
    the $\SO(d-1)$ subgroup that preserves the $x^1$ axis, the geometry is unaffected by $h_M$. (This can also be checked explicitly by computing all the characters and 
    observe that they are not affected.) More generically, suppose $h_M$ takes the following form:
    \begin{equation}\label{eqn:non-trivial hM}
        h_M = \exp(i \alpha^i M_{1i})\exp(i \phi^{ij}M_{ij}), \quad i,j = 2,\dots, d.
    \end{equation}
    We can use the redundancy  (\ref{eqn:redundancy_dumbbell_1}) to rotate the $(d-1)$-dimensional vector  $\vec{\alpha}$
    onto the $x^2$-axis so that the only effective parameter is $\theta\equiv \sqrt{\sum_{i=2}^{d} (\alpha^i)^2}$.
    Thus, the flat locus becomes two-dimensional. We can directly verify this claim by computing the ``loop coordinates''
    in the presence of a non-trivial $h_M$ defined in  (\ref{eqn:non-trivial hM}).
    Again, working with the vector representation in 3d, we find that in the $r\rightarrow \infty$ limit:
    \begin{equation}\label{eqn:3d_loop_coordiantes_with_hM}
        \begin{split}
       &\chi_\rho(\varphi_D(a))=\chi_\rho(\varphi_D(a^2))=\chi_\rho(\varphi_D(b))=\chi_\rho(\varphi_D(b^2))=5\\
       &\chi_\rho(\varphi_D(ab))=5-4 z^{-1}\cos(\theta)+z^{-2},\\
       &\chi_\rho(\varphi_D((ab)^2))=5-16 z^{-1}\cos(\theta)+(16+4 \cos(2\theta))z^{-2}-8\cos(\theta) z^{-3}+z^{-4}\\
       &\chi_\rho(\varphi_D(aba))=\chi_\rho(\varphi_D(bab))=5-8\cos(\theta) z^{-1}+4 z^{-2}\\
       &\chi_\rho(\varphi_D(ab^{-1}))=5+4 z^{-1}\cos(\theta)+z^{-2},\\
       &\chi_\rho(\varphi_D(ab^{-1}a b))=5-4 \cos(2\theta)z^{-2}+z^{-4}.
        \end{split}
    \end{equation} 
The function $h(z)$ gets replaced by
\be
        h(z) &\to \sum_{\mathcal{O}}
            b_{\mathcal{O}}^2
            c_{\mathcal{O}}^{-1}
            q_{J_\cO} 
            \mathcal{P}_{J_\cO}(\cos \theta)
            z^{\Delta_\mathcal{O}}.
\ee

\item If angular fugacities entering $h_L$ and $h_R$ are infinitesimally small, that is, we assume $h_L,h_R$
takes the following form:
\begin{equation}\label{eqn:non-trivial hLR}
    h_{L/R}=\exp\left(\mathrm{i} \left(\frac{s^i_{L/R}}{r}\right)M_{1i}\right)\exp\left(\mathrm{i} \left(\frac{y^{ij}_{L/R}}{r}\right)M_{ij}\right)
    ,\quad i,j=2,\cdots d,
\end{equation}
the flat locus remains two-dimensional with the two moduli being $z_{\text{eff}}$
and $\theta_{\text{eff}}$, where:
\begin{equation}
    z_{\text{eff}}=\frac{e^{-\beta}}{\beta_{L}^{\text{eff}}\beta_{R}^{\text{eff}}},\quad
    \beta^\text{eff}_{L/R}=\sqrt{\beta_{L/R}^2+\vec{s}_{L/R}\cdot \vec{s}_{L/R}},
\end{equation}
and $\theta_{\text{eff}}$ is the angle between the two $d-$dimensional vectors
 $(\beta_L,\vec{s}_L)$ and $h_M\cdot(\beta_R,\vec{s}_R)$. 
 The normalized genus-2 partition function takes the form:
\begin{equation}
    \sum_{\mathcal{O},J\in 2\mathbb{Z}}
            b_{\mathcal{O}_i}
            b_{\mathcal{O}_j}
            c_{\mathcal{O}_i\mathcal{O}_j}^{-1}
            q_J 
            \mathcal{P}_J(\cos \theta_{\text{eff}})
            \left( 
                \frac{e^{-\beta}}{
                                \beta_{L,\text{eff}}
                                \beta_{R,\text{eff}}}
            \right)^{\Delta_\mathcal{O}}.
\end{equation}
In this case, every point on the inner sphere of $A_L$ and $A_R$ is still identified with another point finite distance away, even as $r\rightarrow \infty$. If we zoom in to the neighborhood of the two operator insertions, located at the center of $B_L$ and $B_R$,
the net effect of $h_L$ and $h_R$ is to tilt the unit vector along the Euclidean time direction from $(1,\vec{0})$ 
to $e'_{L/R}=(\cos(|\vec{s}_{L/R}|/r),\hat{s}_{L/R}\sin(|\vec{s}_{L/R}|/r))$. Since the point located at
point $r e^{-\beta_{L/R}/2r}(1,\vec{0})$ is now identified with the point at $r e^{\beta_{L/R}/2r}e'_{L/R}$,
the effective length of the thermal circle becomes $\beta^{\text{eff}}_{L/R}$. 
We checked that in the presence of  (\ref{eqn:non-trivial hLR}), (\ref{eqn:3d_loop_coordiantes_with_hM})
still holds as long as $z,\theta$ is replaced with $z_{\text{eff}}$, $\theta_{\text{eff}}$!
\item When angular fugacities in $h_L$ and $h_R$ are finite, points on $A_L$ and $A_R$ will be identified with 
another point infinitely far away as $r\rightarrow \infty$. For this reason, thermal effective theory breaks down
and in general, it is not clear if the $r\rightarrow \infty$ limit makes sense.\footnote{Even though when the angles are rational, $Z(S^1\times S^{d-1})$ can still be computed in combination with the ``folding trick'' \cite{Benjamin:2024kdg}.}
However, if only $y^{ij}_{L/R}/r$ become $O(1)$ and $s^i_{L/R}/r$ remains small, the operator insertion at the 
center of $B_L$ and $B_R$ are still identified with points finite distance away. Orbifold singularities develop 
on the quotient $S^{d-1}/(h_{L/R}|_{s^i_{L/R}=0})$. We expect that the
$r\rightarrow \infty$ limit still exists and it should only depend on 
the geometry in the vicinity of the two orbifold points. 
 
\end{itemize}

We leave the systematic exploration of angular twists to future work and focus on the zero-angle geometry for the rest of this paper.
\subsubsection{Thermal flat limit in sunrise coordinates }

In this subsection, we match coordinates between the two channels in the thermal flat limit. Solving (\ref{dumbbell_sunrise_matching_1d_full})
order by order in $1/r$, we find:
\begin{equation}
    \begin{split}
    a_1&=\frac{\beta _L-2 \sqrt{z} \beta _R}{\beta _L+2 \sqrt{z} \beta _R}-\frac{\sqrt{z} \left(-3 \beta
   _L^3 \beta _R-\beta _L \beta _R^3+24 z \beta _L^2 \beta _R^2+24 z \beta _L \beta _R^3\right)}{24 r^2
   \left(\beta _L+2 \sqrt{z} \beta _R\right){}^2}+O(r^{-4}),\\
   \\
   a_2&=\frac{\beta _R-2 \sqrt{z} \beta _L}{\beta _R+2 \sqrt{z} \beta _L}-\frac{\sqrt{z} \left(-\beta _L^3
   \beta _R-3 \beta _L \beta _R^3+24 z \beta _L^3 \beta _R+24 z \beta _L^2 \beta _R^2\right)}{24 r^2
   \left(\beta _R+2 \sqrt{z} \beta _L\right){}^2}+O(r^{-4}),\\
   \\
   \beta_3&=\frac{1}{2 r^2}\sqrt{z} \beta _L \beta _R+\frac{1}{192\ r^4} \sqrt{z} \beta _L \beta _R \left(-\beta
   _L^2+24 z \beta _L \beta _R-\beta _R^2\right)+O(r^{-6}).
    \end{split}
\end{equation}
Here, it suffices to use the 1d matching equations, since all gluing elements lie inside the 1d conformal group.

\begin{figure}[htbp]
  \centering
    \includegraphics[scale=1.6]{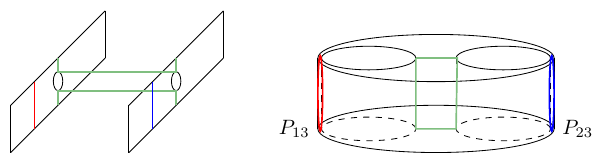}
    \caption{Matching of loops between the two channels. In the sunrise channel geometry  (right picture), the red, blue and green 
    circles have length $\beta_{13},\beta_{23} $ and $\beta_{12}$. The corresponding loops in the dumbbell channel geometry 
    are drawn in the left picture. The two rectangular strips represent two pieces of $S^1\times \mathbb{R}^{d-1}$
    with the vertical directions being the $S^1$ direction. The upper edges of the two strips are all identified with the 
    lower edges, as indicated by the arrow. By carefully tracing the identifications, one can check that the green lines in the left picture form a closed loop.
    The height of the cylinder in the right picture becomes infinitesimally 
    small in the thermal flat limit, but with a generic value of $z$, the two horizontal segments of the green loop will always have a finite length. The 
    blue, red circle will shrink to size zero and two hot spots will develop at $P_{13},P_{23}$. In the dumbbell channel, the two hot spots 
    get blown up to two pieces of flat space.
    }
    \label{fig:matching_of_loops}
\end{figure}

In the sunrise channel, it is useful to define the ``relative inverse temperature" $\beta_{ij}$ and ``relative angles'' $\theta_{ij}$
defined using eigenvalues of the product $g_{i}^{-1} g_{j}$:
\begin{equation}
    g_i^{-1}g_j \xleftrightarrow{\text{conjugate to}}  \exp(-\beta_{ij}D+i\theta_{ij}M_{12}+\cdots).
\end{equation}
In the thermal flat limit, all the relative angles vanish and 
the relative inverse temperatures become:
\begin{equation}
\beta_{13}=\beta_{L}/r,\quad \beta_{23}=\beta_{R}/r, \quad \beta_{12}= \cosh ^{-1}\left(\frac{1-4 z}{2 z^2}+1\right)+O(r^{-2}).
\label{eq:thermaflatbetas}
\end{equation}
The three relative temperatures can be understood as the lengths of the three colored loops indicated in figure \ref{fig:matching_of_loops}. 
For generic $z$, $\beta_{12}$ remains finite as $r\rightarrow \infty$, while $\beta_{13},\beta_{23}\rightarrow 0$.

\begin{figure}[h]
    \centering
    \includegraphics[scale=0.4]{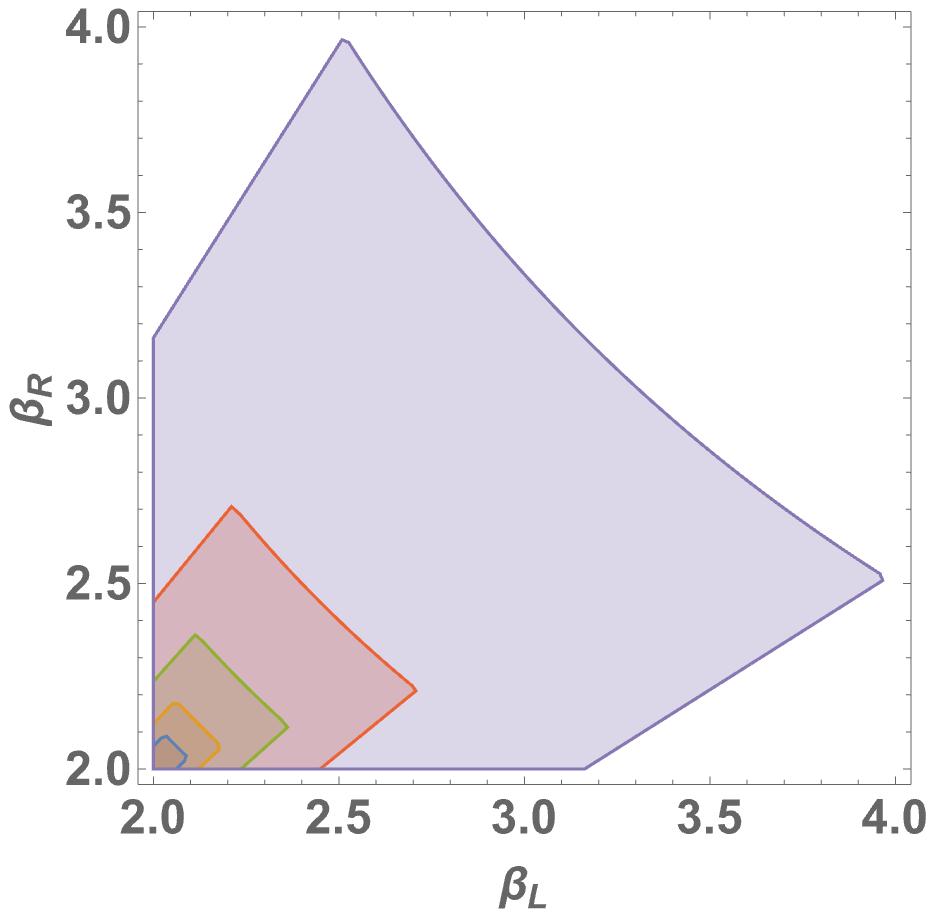}
    \caption{\label{fig:allowedregionofstuff}The allowed region on the  $\beta_L,\beta_R$ plane for different 
    of $z$. We chose $z$ to be $4/17,2/9,1/5,1/6,1/10$. As $z$ approaches $1/4$, the allowed 
    region shrinks to the left corner $(2,2)$. }
\end{figure}

Let us pause and discuss the range of various parameters. 
In order for $\beta_{12}$ to be real, $z$ must satisfy $0<z<1/4$. 
For any fixed $z$ within this interval, to ensure $a_1,a_2,\beta$ are in their appropriate range and that the representative 
geometry discussed in subsection \ref{subsec: genus-2 geometry in the dumbbell channel} makes sense,
$\beta_L,\beta_R$ must satisfy the following conditions, see figure~\ref{fig:allowedregionofstuff}:
\begin{equation}
    \beta_L,\beta_R \geq 2, \quad 2\sqrt{z} \leq\beta_L/\beta_R\leq (2\sqrt{z})^{-1},\quad \beta_L\beta_R\leq z^{-1}.
\end{equation}

When $z=1/4$, the only allowed region in the $\beta_L,\beta_R$ plane is the point $(2,2)$. In the dumbbell channel, 
the ``neck'' of the dumbbell becomes infinitesimally short and $B_L$, $B_R$ become tangent to themselves since they wrap 
around the thermal circle. In the sunrise channel,  
$a_1=a_2=O(r^{-1})$, and therefore $B_1,B_2$ become tangent to each other, and all three relative inverse temperatures become parametrically small. 
This is the geometry studied in \cite{Benjamin:2023qsc} to derive 
an asymptotic formula for thermal one-point coefficients. 

Another important limit is when $z\rightarrow 0$. In this limit, $\beta\rightarrow \infty$ and 
the neck of the dumbbell becomes infinitely long. Up to gluing factors and Casimir energy contributions, the genus-2
partition function factorizes into two $S^1\x S^{d-1}$ partition functions. 
Meanwhile,  $a_1$ and $a_2$ approach $1$, and therefore the radii of 
the two balls $B_1$ and $B_2$ shrink to zero. From the sunrise channel perspective, this is a mixture of an infinite temperature limit 
and a zero temperature limit since $\beta_{12}\rightarrow \infty$ while $\beta_{13},\beta_{23}\rightarrow 0$.

\paragraph{Comment on the hotspot conjecture.}
In \cite{Benjamin:2023qsc}, the ``hot spot conjecture'' is used to estimate $Z(M_2)$ in an appropriate ``high-temperature" limit. 
It states that the leading contribution to the partition function in the limit comes from 
regions where the geometry looks like a circle fibration with a large local temperature.
Contributions from these regions can be well-approximated by the thermal effective action. 
Though  the hot spot conjecture  is not needed in this work, we note that 
the dumbbell channel decomposition is consistent with it: in the thermal flat limit, two 
hot spots develop with local temperatures $1/\beta_{13}$ and $1/\beta_{23}$. 
Since $\beta_{13}=\beta_L/r$ and $\beta_{23}=\beta_R/r$, the contribution from these hot spots
exactly reproduces the product of partition functions $Z(S^1_{L}\times S^{d-1}_r)$$Z(S^1_{R}\times S^{d-1}_r)$! 
Indeed, $Z(S^1_{L}\times S^{d-1}_r)$ and $Z(S^1_{R}\times S^{d-1}_r)$ encompasses all the singular terms as $r\rightarrow \infty$.
\subsubsection{The function $h(z)$}

The function $h(z)$ is the partition function of a CFT on two copies of thermal flat space connected by a cylinder (normalized by the square of the thermal partition function). This is an interesting function about which little has been written in the literature. 

As a first example, it is interesting to compute $h(z)$ in 2d CFTs. One way to access $h(z)$ would be to determine the Weyl rescaling from the sphere to a pair of infinite cylinders connected by a finite cylinder (i.e.\ an infinitely tall ``H"), and then use the Weyl anomaly. This shows in particular that $h(z)$ depends only on the central charge in 2d and takes the form $h(z) = e^{c f(z)}$ for some universal $f(z)$. We will not attempt this route in this work, but instead we will compute $h(z)$ to the first few orders in $z$ by summing over low-lying states.

Due to Weyl equivalence between the cylinder and the plane, only 
quasi-primaries in the conformal family of the identity operator can have non-zero cylinder one-point function. 
For simplicity, we only include the holomorphic sector. The first few quasi-primary states
are given as follows:
\begin{equation}
    \begin{split}
    &|\mathbb{Q}_2 \rangle \equiv L_{-2}|0\rangle,\\
    &|\mathbb{Q}_4\rangle \equiv (L_{-4}-\frac{5}{3}L_{-2}L_{-2})|0\rangle,\\
    &|\mathbb{Q}^{1}_6\rangle \equiv \left(L_{-6}+\frac{21}{4}L_{-3}^2-\frac{14}{3}L_{-2}^3\right)|0\rangle, \quad 
    |\mathbb{Q}^{2}_6\rangle \equiv \left(L_{-4}L_{-2}-\frac{5}{2}L_{-3}^2+\frac{5}{3}L_{-2}^3\right)|0\rangle
    \\ & \cdots
    \end{split}
\end{equation}
Under the plane-cylinder map $z_\text{pl.}=\exp(2\pi w_{\text{cyl.}}/\beta)$, these quasi-primaries admit the following
non-zero one-point functions \cite{Gaberdiel:1994fs}:
\begin{equation}
    \begin{split}
    \langle \mathbb{Q}_2 \rangle_\text{cyl.}=-\frac{\pi^2 c}{6}\frac{1}{\beta^2},& \quad
    \langle \mathbb{Q}_4\rangle_\text{cyl.}=-\frac{\pi ^4 c (5 c+22)}{108 \beta ^4},\quad
    \\
    \langle \mathbb{Q}_6^1\rangle_\text{cyl.}= \frac{\pi ^6 c (7 c (5 c+66)+2062)}{1620 \beta ^6},& \quad
     \langle \mathbb{Q}_6^2\rangle_\text{cyl.}= -\frac{\pi ^6 c (5 c (5 c+66)+1712)}{3240 \beta ^6}.
    \end{split}
\end{equation}
Combining both the holomorphic and the anti-holomorphic sectors, we find the first few terms in $h(z)$ are:
\begin{equation}
h(z)=  1+ \frac{1}{9} \pi ^4 c z^2+\frac{1}{810} \pi ^8
   c (5 c+11) z^4+\frac{32 \pi ^{12} c \left(175 c^2+1155 c+1322\right) z^6}{382725}+\cdots,
\end{equation}
 where we've multiplied products of cylinder one-point functions with inverse two-point coefficients. 
 Note that $h(z)$ can indeed be written in the form $h(z)=e^{cf(z)}$
 with 
 \be
 f(z) &=\frac{ \pi ^4 z^2}{9}+\frac{11 \pi ^8 z^4}{810}+\frac{661 \pi ^{12} z^6}{382725}+\dots.
 \ee
 In 2 dimensions, the genus-2 crossing equation also includes a nontrivial contribution from the Weyl anomaly which itself has the form $e^{S_\mathrm{Weyl}}=e^{c g(z)}$ for some $g(z)$. So $f(z)$ is only a piece of the structure of the thermal flat limit in 2d.
 
 \begin{table}
 \centering
 \begin{tabular}{c|c|c|c}
 $\cO$ & $\De$ & spin & $b_\cO$\\
 \hline
 $\e$ & 1.41262528(29) \cite{Chang_2025} & 0 & $0.667(3)$ \cite{HasenbuschPrivate} \\
 $T$ & 3 & 2 & $-0.459(3)$ \cite{PhysRevE.79.041142,PhysRevE.53.4414,PhysRevE.56.1642} \\
 $\e'$ & 3.82951(61) \cite{Reehorst:2021hmp} & 0 & $3.6\pm 1.1$ \cite{Barrat:2025wbi}; $3.2\pm 0.4$ \cite{PhysRevE.79.041142,PhysRevE.53.4414,PhysRevE.56.1642,Iliesiu:2018zlz}
 \\
 $C_4$ & $5.022665(28)$ \cite{Simmons-Duffin:2016wlq} & 4 & 31(?) \cite{Iliesiu:2018zlz}\\
 $T'$ & $5.499(17)$ \cite{Reehorst:2021hmp} & 2 & ? \\
 $C_4'$ & $6.42065(65)$ \cite{Simmons-Duffin:2016wlq} & 4 & ? \\
 $\e''$ & $6.8956(43)$ \cite{Simmons-Duffin:2016wlq} & 0 & ? \\
$C_6$ & $7.028488(16)$ \cite{Simmons-Duffin:2016wlq} & 6 & 234(?) \cite{Iliesiu:2018zlz} \\
$C_8$ & $9.031023(30)$ \cite{Simmons-Duffin:2016wlq} & 8 & 1770(?) \cite{Iliesiu:2018zlz} \\
$C_{10}$ & $11.0324141(99)$ \cite{Simmons-Duffin:2016wlq} & 10 & 13500(?) \cite{Iliesiu:2018zlz} \\
 \end{tabular}
 \caption{\label{tab:isingdata}Low-dimension operators in the 3d Ising CFT and their thermal one-point functions. For $b_\e,b_T$ we have included the current most precise determinations, which come from Monte-Carlo methods. Error bars have different meanings for different quantities. The errors from \cite{Reehorst:2021hmp} are rigorous, while errors from \cite{Simmons-Duffin:2016wlq} are non-rigorous estimates from the extremal functional method. Errors from \cite{HasenbuschPrivate,PhysRevE.79.041142,PhysRevE.53.4414,PhysRevE.56.1642} are statistical. Errors from \cite{Iliesiu:2018zlz} and \cite{Barrat:2025wbi} come from approximation methods used in solving the thermal bootstrap crossing equations. When no good error estimates are available, we indicate that with a question mark. The first few operators where estimates of thermal one-point functions are not available are $T',C_4',\e'',...$. However, estimates for some higher dimension operators --- specifically the approximately conserved currents $C_6,C_8,\dots$ are available because they play an important role in the thermal bootstrap for the two-point function of $\sigma$. Note that all operators in this table are $\Z_2$ and parity even.
 }
 \end{table}
 
\begin{figure}
\centering
\includegraphics[width=0.8\textwidth]{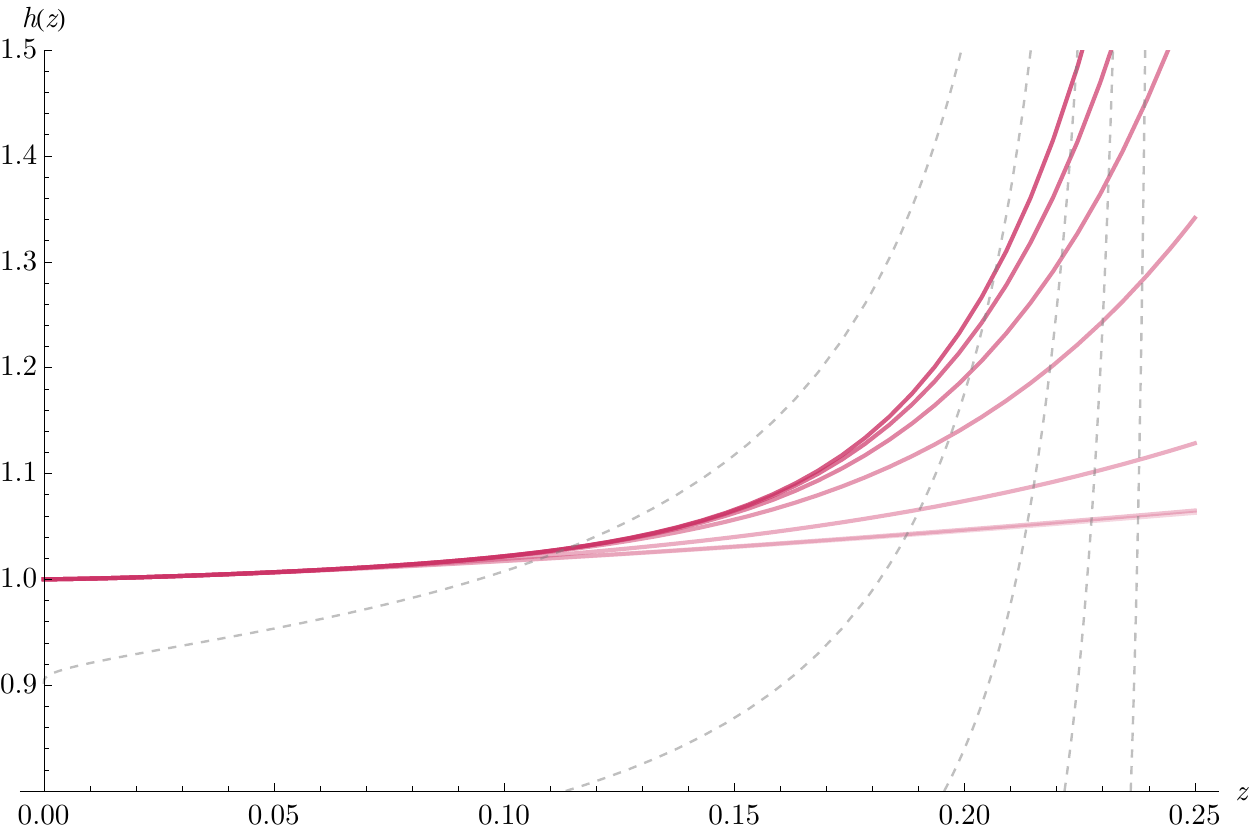}
\includegraphics[width=0.8\textwidth]{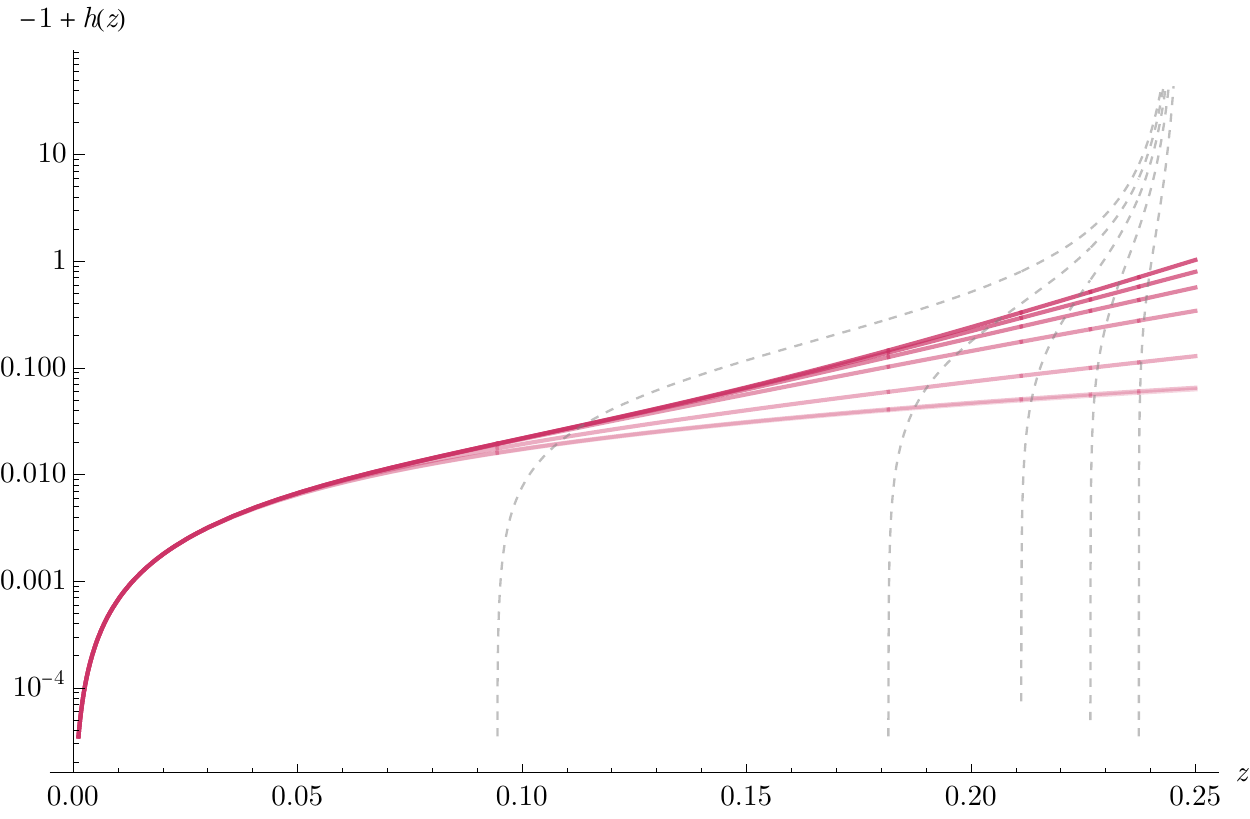}
\caption{\label{fig:hising}An estimate of $h(z)$ for the 3d Ising CFT, using data in table~\ref{tab:isingdata}. On top, we plot $h(z)$, while below we show $h(z)-1$ on a log scale. The red curves are the result of summing contributions from known thermal one-point functions, including more and more terms, with the highest curve including the contributions of $1,\e,T,\e',C_4,C_6,C_8,C_{10}$. Unknown contributions from $T',C_4',\e'',\dots$, are not included, so this is an estimated lower bound on $h(z)$. The gray dashed curves show the expected singular behavior (\ref{eq:hotspoth}) near $z=1/4$, with different curves corresponding to different overall multiplicative factors.}
\end{figure}

 It is also interesting to estimate $h(z)$ in a 3d CFT, such as the 3d Ising model. In this case, only a few thermal one-point coefficients are known, see table~\ref{tab:isingdata}. We can use this data to estimate $h(z)$ up to order $\sim z^{5}$. However, some thermal one-point functions of higher dimension operators have been estimated as well, and it is reasonable to include these in our computation of $h(z)$, since all contributions are sign-definite. We plot the resulting estimates in figure~\ref{fig:hising}.
 
While summing over operators gives an estimate for $h(z)$ at small $z$, the behavior of $h(z)$ near $z=1/4$ (the maximum value) can be computed using the hot-spot conjecture and thermal effective action \cite{Benjamin:2023qsc}. It is given by
\be
\label{eq:hotspoth}
h(z) &\sim \exp\p{\frac{f\vol\, S^2}{\b_\mathrm{rel}^2} + \textrm{nonsingular}} \qquad (\textrm{near $z=1/4$}),\\
\b_\mathrm{rel} &= \cosh^{-1}\p{1-\frac{2}{z}+\frac{1}{2z^2}}.
\ee
In the 3d Ising CFT, we have $f\approx 0.153$.
The ``nonsingular" contributions come from higher-order terms in the thermal effective action, such as $c^{(1)}$, but also parts of the geometry away from the hotspot whose contributions we don't know how to estimate. In figure~\ref{fig:hising}, we show the behavior (\ref{eq:hotspoth}) with different multiplicative contributions representing the unknown effects of the ``nonsingular" terms.

\subsection{Saddles from moments }\label{subsec:saddles from moments}
From a detailed analysis in the dumbbell channel we've determined the large-$r$ asymptotics of $Z(M_2)$.
We now turn to the sunrise channel decomposition and ask
what kind of distribution over the triplet $((\Delta_1,\lambda_1),(\Delta_2,\lambda_2),(\Delta_3,\lambda_3))$ can reproduce such asymptotics. We will
assume that in the thermal flat limit, the integral over the scaling dimensions $\Delta_i$ as well as the sum over $SO(d)$ representations
 $\lambda_i$ are dominated by a single saddle point 
$((\Delta_1^*,\lambda_1^*),(\Delta_2^*,\lambda_2^*),(\Delta_3^*,\lambda_3^*))$.
We will identify the position of this saddle via 
expectation values of Casimirs.

 Furthermore, the genus-2 block $B^{s,s'}_{123}$ can be viewed as a sum over states in the conformal multiplet 
$\mathcal{R}_{(\Delta_1,\lambda_1)}\otimes \mathcal{R}_{(\Delta_2,\lambda_2)}\otimes \mathcal{R}_{(\Delta_3,\lambda_3)}$. Within the 
dominating multiplet $\mathcal{R}_{(\Delta_1^*,\lambda_1^*)}\otimes \mathcal{R}_{(\Delta_2^*,\lambda_2^*)}\otimes \mathcal{R}_{(\Delta_3^*,\lambda_3^*)}$, 
the sum also localizes to a coherent state labeled by a triplet of coordinates $(x_1^*,x_2^*,x_3^*)$. 
Such a coherent state can be identified from moments of the conformal generators. To keep the presentation simple, we will illustrate this technique mostly
in 1d and 2d throughout the remainder of the section, while noting that the method generalizes straightforwardly to higher dimensions.
\subsubsection{Saddles in $\Delta_i$}\label{subsubsec:saddles in Delta}
The genus-2 blocks are simultaneous eigenfunctions of the three Casimirs, namely:
\begin{equation}
  \mathcal{C}_2^{(i)} B^{s,s'}_{123}=(\Delta_i(\Delta_i-d) +C_2(\lambda_i) )B^{s,s'}_{123},
\end{equation}
where $C_2(\lambda_i)$ denotes the quardratic Casimir eigenvalue of the $SO(d)$ irrep $\lambda_i$. (For 
symmetric traceless tensors, it is given by $J(J+d-2)$.) Acting with $\mathcal{C}_2^{(i)}$ on both sides of the sunrise channel
decomposition, we find:
\begin{equation}
    \begin{split}
  \langle\mathcal{C}_2^{(i)} \rangle \equiv
   \frac{ \mathcal{C}_2^{(i)}Z([g_1,g_2,g_3])}{Z([g_1,g_2,g_3])}&= Z^{-1}\sum_{\lambda_1,\lambda_2,\lambda_3}\int d\Delta_1 d\Delta_2 d\Delta_3 C_2(\Delta_i,\lambda_i) P^{s,s'}_{123}B^{s',s}_{123}([g_1,g_2,g_3])\\
   &\sim C_2(\Delta_i^*,\lambda_i^*),
    \end{split}
\end{equation}
where we've assumed that the action of the Casimir operators do not shift the position of the saddle. 

Now let us derive an expression for the Casimir operator $\mathcal{C}_2^{(i)}$. We will first explain how this works in 1d. 
We write the partition function and the block as functions of the ``loop coordinates'' $\chi_\rho(a),\chi_\rho(b),\chi_\rho(ab)$ introduced 
earlier. To simplify notation, we call them $\tau_{13},\tau_{32},\tau_{12}$, respectively, from now on. The conformal generators at each site are given by the following differential operators:
\begin{equation}
        \mathcal{L}^{(i)}_{AB}=\sum_{jk}(\mathcal{L}^{(i)}_{AB}\tau_{jk})\frac{\partial}{\partial \tau_{jk}},
        \quad
        \mathcal{L}_{AB}^{(i)}\tau_{jk}
        =
        \left\{
\begin{aligned}
 -\text{tr}(g_j^{-1} L_{AB}g_k), \quad  &i=j \\
 \text{tr}(g_j^{-1} L_{AB}g_k), \quad &i=k \\
 0, \quad &\text{otherwise}.
\end{aligned}
\right .
\end{equation}
Combining the generators into quadratic Casimirs, we find:
\begin{equation}\label{eqn:Casimir}
    \begin{split}
    \mathcal{C}_2^{(i)}=-\frac{1}{2}\mathcal{L}_{AB}^{(i)}\mathcal{L}^{AB,(i)}
    &=-\frac{1}{2}(\mathcal{L}_{AB}^{(i)}\tau_{jk})(\mathcal{L}^{AB,(i)}\tau_{\ell m})\frac{\partial^2}{\partial \tau_{jk}\partial \tau_{\ell m}}
    -\frac{1}{2}(\mathcal{L}_{AB}^{(i)}\mathcal{L}^{AB,(i)}\tau_{jk})\frac{\partial}{\partial \tau_{\ell m}},\\
    (\mathcal{L}_{AB}^{(i)} \mathcal{L}^{AB,(i)}\tau_{jk})&=
    \left\{
\begin{aligned}
 \text{tr}(g_j^{-1} L_{AB} L^{AB}g_k)=-2 C_2(\rho)\tau_{jk}, \quad  &i\in\{j,k\} \\
 0, \quad &\text{otherwise}.
\end{aligned}
\right.
    \end{split}
\end{equation}
The term $(\mathcal{L}_{AB}^{(i)}\tau_{jk})(\mathcal{L}^{AB,(i)}\tau_{\ell m})$ can be further simplified into polynomials of 
$\tau_{ij}$. We present the detailed forms of various Casimir operators in Appendix \ref{appendix:Casimir}.

For now,
let us make a simplifying observation: when acting on a function of the form $\exp(f(\tau_{ij}))$ with $f(\tau_{ij})\rightarrow \infty$
in the relevant kinematic limit, the first term on the right hand side of  (\ref{eqn:Casimir}) dominates over the second. 
Furthermore:
\begin{equation}
    \frac{\partial^2 \exp(f)}{\partial \tau_{jk}\partial \tau_{\ell m}}  \sim \exp(f) \frac{\partial f}{\partial \tau_{ij}} \frac{\partial f}{\partial \tau_{\ell k}}.
\end{equation}
This suggests that at leading order, we can write the expectation value of Casimirs as polynomials in expectation values of the conformal generators $\langle \mathcal{L}_{AB}^{(i)}\rangle$:
\begin{equation}
    \langle \mathcal{C}_2^{(i)}\rangle \sim -\frac{1}{2} \langle \mathcal{L}_{AB}^{(i)}\rangle \langle \mathcal{L}^{AB,(i)}\rangle.
\end{equation}
In other words, the conformal generators can be treated classically and replaced with their moments.
The same argument also applies to higher Casimirs. 

\paragraph{An example in 3d:}
Making the choice $\rho=\square$ and turning off angular fugacities, the moments of the generators $\vec{P},\vec{K},D,J_{12}$ are given as follows:
\be
\label{eqn:moment_transvers}
    \langle\mathcal{P}^{(i)}_{2}\rangle&=\langle \mathcal{K}^{(i)}_{2}\rangle=\langle\mathcal{P}^{(i)}_{3}\rangle=\langle \mathcal{K}^{(i)}_{3}\rangle
    =\langle \mathcal{J}^{(i)}_{12}\rangle=\langle \mathcal{J}^{(i)}_{13}\rangle=\langle \mathcal{J}^{(i)}_{23}\rangle
    =0,
\\
\label{eqn:moment_nonzero}
    % P1
   \langle \mathcal{P}^{(1)}_{1} \rangle
        &= -\frac{\tilde{f}_{1 3}}{\beta _{13}^3 \beta _{23} \sqrt{z}}+\frac{\left(\beta _{13}^2+24 \beta _{23} \beta _{13} \left(z+\sqrt{z}\right)+\beta _{23}^2 (24 z-1)\right) \tilde{f}_{1 3}}{96 \beta _{13}^3
   \beta _{23} \sqrt{z}}
           +O(r), \\
    % K1
    \quad
    \langle \mathcal{K}^{(1)}_{1} \rangle
        &= -\frac{4 \tilde{f}_{1 3}}{\beta _{13}^3 \beta _{23} \sqrt{z}}+\frac{\left(\beta _{13}^2+24 \beta _{23} \beta _{13} \left(z-\sqrt{z}\right)+\beta _{23}^2 (24 z-1)\right) \tilde{f}_{1 3}}{24 \beta _{13}^3
   \beta _{23} \sqrt{z}}
       +O(r), \\
   % D
   \quad
   \langle \mathcal{D}^{(1)} \rangle
        &=\frac{2 \tilde{f}_{13}}{\beta _{13}^3 \beta _{23} \sqrt{z}}-\frac{\left(\beta _{13}-\beta _{23}\right) \left(\beta _{13}+\beta _{23}+24 \beta
   _{23} z\right) \tilde{f}_{1 3}}{48 \beta _{13}^3 \beta _{23} \sqrt{z}}
        +O(r),
    \\
    % P1 (2)
    \langle \mathcal{P}^{(2)}_{1} \rangle
        &=\frac{\tilde{f}_{2 3}}{\beta _{13} \beta _{23}^3 \sqrt{z}}-\frac{\left(\beta _{23}^2+\beta _{13}^2 (24 z-1)+24 \beta _{23} \beta _{13}
   \left(z+\sqrt{z}\right)\right) \tilde{f}_{2 3}}{96 \beta _{13} \beta _{23}^3 \sqrt{z}}
    +O(r), 
    \\
    \quad
    \langle \mathcal{K}^{(2)}_{1} \rangle
        &= \frac{4 \tilde{f}_{2 3}}{\beta _{13} \beta _{23}^3 \sqrt{z}}-\frac{\left(\beta _{23}^2+\beta _{13}^2 (24 z-1)+24 \beta _{23} \beta _{13}
   \left(z-\sqrt{z}\right)\right) \tilde{f}_{2 3}}{24 \beta _{13} \beta _{23}^3 \sqrt{z}}
       +O(r), \\
    \quad
    \langle \mathcal{D}^{(2)} \rangle
        &= \frac{2 \tilde{f}_{2 3}}{\beta _{13} \beta _{23}^3 \sqrt{z}}+\frac{\left(\beta _{13}-\beta _{23}\right) \left(\beta _{13}+\beta _{23}+24 \beta _{13} z\right) \tilde{f}_{2 3}}{48 \beta _{13} \beta _{23}^3
   \sqrt{z}}
       +O(r),
\ee
along with
\small
\be
\label{eqn:nonzero_moments_3}
    % P1
   \langle \mathcal{P}^{(3)}_{1} \rangle
    &= \frac{\beta _{23}^2 \tilde{f}_{1 3}-\beta _{13}^2 \tilde{f}_{2 3}}{\beta _{13}^3 \beta _{23}^3 \sqrt{z}}
    \nn \\ 
    &+\frac{\beta _{13}^2 \left(\beta _{23}^2+\beta_{13}^2 (24 z-1)+24 \beta _{23} \beta _{13} \left(z+\sqrt{z}\right)\right) \tilde{f}_{2 3}-\beta _{23}^2 \left(\beta _{13}^2+24 \beta _{23}
   \beta _{13} \left(z+\sqrt{z}\right)+\beta _{23}^2 (24 z-1)\right) \tilde{f}_{1 3}}{96 \beta _{13}^3 \beta _{23}^3 \sqrt{z}}\\
   &+O(r), 
   \nn \\
    \quad
    % K1
    \langle \mathcal{K}^{(3)}_{1} \rangle
    &=\frac{4 \beta _{23}^2 \tilde{f}_{1 3}-4 \beta _{13}^2 \tilde{f}_{2 3}}{\beta _{13}^3 \beta _{23}^3 \sqrt{z}}
    \nn \\
    &+\frac{\beta _{23}^2 \left(-\beta _{13}^2-24
   \beta _{23} \beta _{13} \left(z-\sqrt{z}\right)+\beta _{23}^2 (1-24 z)\right) \tilde{f}_{1 3}+\beta _{13}^2 \left(\beta _{23}^2+\beta _{13}^2
   (24 z-1)+24 \beta _{23} \beta _{13} \left(z-\sqrt{z}\right)\right) \tilde{f}_{2 3}}{24 \beta _{13}^3 \beta _{23}^3 \sqrt{z}}
   \nn \\
    &+O(r), \\
   % D
   \quad
   \langle \mathcal{D}^{(3)} \rangle
    &=-\frac{2\left(\beta _{23}^2 \tilde{f}_{1 3}+\beta _{13}^2 \tilde{f}_{2 3}\right)}{\beta _{13}^3 \beta _{23}^3 \sqrt{z}}+\frac{\left(\beta _{13}-\beta _{23}\right) \left(\beta _{23}^2 \left(\beta _{13}+\beta _{23}+24 \beta _{23} z\right) \tilde{f}_{1 3}-\beta
   _{13}^2 \left(\beta _{13}+\beta _{23}+24 \beta _{13} z\right) \tilde{f}_{2 3}\right)}{48 \beta _{13}^3 \beta _{23}^3 \sqrt{z}}
   \nn \\ 
   &+O(r).
\ee
\normalsize
Notice that 
\be
\sum_{i=1}^3 \langle\cP_1^{(i)} \rangle=
\sum_{i=1}^3 \langle\cK_1^{(i)} \rangle=
\sum_{i=1}^3 \langle\cD^{(i)} \rangle=0,
\ee
which is expected since the partition function is invariant under simultaneous left action of any conformal group element. 

In computing the moments listed above, we used the following asymptotic formula for $Z(M_2)$:
\begin{equation}
    Z(M_2)=h(z) \exp\left(\frac{\tilde{f}_{13}}{\beta_{13}^2}+ \frac{\tilde{f}_{32}}{\beta_{32}^2}\right),\quad \tilde{f}_{ij}\equiv f_{ij} \vol S^1,
\end{equation}
where for generality, we made the free energy densities $f_{ij}$ at the two hot spots different. (This could happen if 
topological defects are inserted onto $M_2$.)
As long as $z$ is away from $1/4$, the function $h(z)=\sum_{\mathcal{O}}b^2_{\mathcal{O}} q_{J_\cO} z^{\Delta_{\mathcal{O}}}$  remains finite as $r\to \oo$. As a consequence, the dependence on $h(z)$
 only appears at subleading order.
 
 Combining  generators into various Casimir invariants, we find:
 \begin{equation}
    \langle \mathcal{C}_2^{(1)} \rangle \sim \frac{4 \tilde{f}_{1 3}^2}{\beta _{13}^6}, \quad
    \langle \mathcal{C}_2^{(2)} \rangle \sim \frac{4\tilde{f}_{32}^2}{\beta_{32}^6},\quad
    \langle \mathcal{C}_2^{(3)} \rangle \sim \frac{16 \tilde{f}_{13} \tilde{f}_{32}}{\beta _{13}^4 \beta _{23}^4 z},
 \end{equation}
  \begin{equation}
    \langle \mathcal{C}_4^{(1)} \rangle \sim \frac{16\tilde{f}_{13}^4}{\beta_{13}^{12}}, \quad
    \langle \mathcal{C}_4^{(2)} \rangle \sim \frac{16\tilde{f}_{32}^4}{\beta_{32}^{12}},\quad
    \langle \mathcal{C}_4^{(3)} \rangle \sim \frac{256 \tilde{f}_{13}^2 \tilde{f}_{32}^2}{\beta _{13}^8 \beta _{23}^8 z^2}.
 \end{equation}
From these values, we can determine the saddles in the quantum numbers:
 \begin{equation}\label{Delta_saddle_2d}
   \Delta_1^*\sim \frac{2 \tilde{f}_{13}}{\beta_{13}^3},
   \quad
   \Delta_2^*\sim \frac{2\tilde{f}_{32}}{\beta_{32}^3},
   \quad
   \Delta_3^*\sim \frac{4\tilde{f}_{32}^{1/2} \tilde{f}_{13}^{1/2}}{\beta_{32}^{2} \beta_{13}^{2}\sqrt{z}},
   \quad
   J_i^* = 0.
\end{equation}
Note that the saddle values of the spins $J_i$ vanish because the moments of the quartic Casimirs are squares of the moments of the quadratic Casimirs.

 This calculation easily generalizes to other dimensions $d$. As long 
as the angular fugacities are turned off, only the generators $\mathcal{P}^{(i)}_{1},
\mathcal{K}^{(i)}_{1},\mathcal{D}^{(i)}$ can have non-zero moments. This is because when all angular twists
are absent, the geometry is invariant under the action of the $x^1$-preserving $\SO(d-1)$ subgroup. However,
 under its adjoint action, 
the generators $K_a$,$P_a$,$J_{1a}$ with  $a,b>1$ transform according to the vector representation
and the generators $J_{ab}$ transform according to the adjoint representation. Their expectation values are thus forced to vanish.
More explicitly, let $g$ be any element in the $SO(1,2)$ subgroup generated by $P_1,K_1,D$. We have:
\begin{equation}
    \text{tr}(P_a g)=\text{tr}(h P_a gh^{-1})=\pi(h)_a^b\text{tr}(P_b h gh^{-1})=\pi(h)_a^b\text{tr}(P_b g),\quad \forall h\in \SO(d-1),
\end{equation}
which immediately implies $\text{tr}(P_a g)=0$. Similarly, $\text{tr}(K_a g)=\text{tr}(J_{1a} g)=\text{tr}(M_{ab} g)=0$.
The calculation 
of $\langle \mathcal{C}_2^{(i)}\rangle$ then reduces to an effective 1d computation. Ultimately, for generic $d$, we find:
\begin{equation}\label{eq:Delta_i_saddle}
\Delta_1^*\sim \frac{(d-1)\tilde{f}_{13}}{\beta_{13}^d},\quad 
\Delta_2^*\sim \frac{(d-1)\tilde{f}_{32}}{\beta_{32}^d},\quad
\Delta_3^*\sim \frac{2(d-1)}{\sqrt{z}}\left( \frac{\tilde{f}_{13}\tilde{f}_{32}}{\beta_{13}^{d+1}\beta_{32}^{d+1}}\right)^{1/2},\quad
C_2(\lambda_i^*)=0.
\end{equation}

Recall from (\ref{eq:thermaflatbetas}) that $\b_{13}$ and $\b_{23}$ become small in the thermal flat limit. Thus, we see that in terms of quantum numbers in the sunrise channel, this corresponds to $\De_1,\De_2,\De_3$ becoming large, with the ratio $\De_1 \De_2/\De_3^{\frac{2d}{d+1}}$ held fixed. We refer to this as the ``heavy-heavy-heavier" regime.

\subsubsection{Saddles in the shadow integral}\label{subsubsec:saddles in the shadow integral}

The expectation values (\ref{eqn:moment_transvers})-(\ref{eqn:nonzero_moments_3})
not only fix the dominant quantum numbers in the sunrise channel; they also locate the saddle of the shadow integral for the sunrise block. Recall that when the $\Delta_i$'s are on the principal series $\frac{d}{2}+\mathrm{i} \mathbb{R}$, the 
shadow integral evaluates the genus-2 partial wave, which can be 
expressed as a linear combination of the genus-2 block and seven shadow blocks.
We refer to the coefficients multiplying these blocks as \textit{triple shadow coefficients}. When the scaling dimensions are parametrically large, 
the integrand of the shadow integral becomes rapidly oscillatory, and the 
genus-2 partial wave is well-approximated by a saddle-point approximation. The shadow integrand typically has multiple 
saddle points. To isolate the contribution of the 
genus-2 block from the shadow blocks, we need to identify the appropriate saddle point. 

The physical genus-2 block is then obtained by analytically continuing all $\Delta_i$ from $\frac{d}{2}+\mathrm{i}\mathbb{R}$
to $\mathbb{R}$.
Further details are deferred to section \ref{subsection:saddle_point_approximation_of_block}. 
For now, we write schematically:
\be
 &B_{\Delta_1^*,\Delta_2^*,\Delta_3^*}([g_1,g_2,g_3])
 \nn\\
 &\sim
         (S^3_{\Delta_1^*,\Delta_2^*,\Delta_3^*})^{-1} V_{d-\Delta_1^*,d-\Delta_2^*,d-\Delta_3^*}(x_1,x_2,x_3) 
         g_1g_2 g_3 \cdot 
        V_{\Delta_1^*,\Delta_2^*,\Delta_3^*}(x_1,x_2,x_3)\Bigg|_{(x_1^*,x_2^*,x_3^*)},
\ee
here $B_{\Delta_1^*,\Delta_2^*,\Delta_3^*}$ denotes a scalar genus-2 block, $S^3_{\Delta_1^*,\Delta_2^*,\Delta_3^*}$
is the corresponding triple shadow coefficient and $(x_1^*,x_2^*,x_3^*)$ denotes the saddle point 
corresponding to the genus-2 block. 
At leading order, the expectation values of the conformal generators are given by:
\be
    \langle \mathcal{L}_{AB}^{(i)}\rangle &= \frac{\mathcal{L}_{AB}^{(i)}Z([g_1,g_2,g_3])}{Z([g_1,g_2,g_3])}
    \nn\\
    &\sim \frac{\mathcal{L}_{AB}^{(i)} B_{\Delta_1^*,\Delta_2^*,\Delta_3^*}([g_1,g_2,g_3])}{B_{\Delta_1^*,\Delta_2^*,\Delta_3^*}([g_1,g_2,g_3])}
    \sim  \frac{\mathcal{L}^{(i)}_{AB}\left( g_1 g_2 g_3 \cdot V_{\Delta_1^* \Delta_2^* \Delta_3^*}(x_1,x_2,x_3) \right)}
    { g_1 g_2 g_3 \cdot V_{\Delta_1^*\Delta_2^*\Delta_3^*}(x_1,x_2,x_3) }\Big|_{(x_1^*,x_2^*,x_3^*)}.
\ee
To evaluate the rightmost expression,
we no longer treat $\mathcal{L}_{AB}^{(i)}$ as a differential operator. Instead, we multiply 
the corresponding $g_i$ on the left by $L_{AB}^{(i)}$, yielding a rational function of 
 $(x_1^*,x_2^*,x_3^*)$. Matching them with the previously computed moments $\langle \mathcal{L}_{AB}^{(i)}\rangle$
 determines the saddle values. 

Let us illustrate this procedure for $d=2$. 
By $SO(d-1)$ invariance, the saddle points lie on the $x^1$ axis, which greatly simplifies the rational functions.
We show the action of $\mathcal{P}^{(1)}_1$ for the purpose of illustration. The expression for
the rest of the conformal generators are similar in flavor:
\be
    &\frac{\mathcal{P}^{(1)}_{1}\left( g_1 g_2 g_3 \cdot V_{\Delta_1^* \Delta_2^* \Delta_3^*}(x_1,x_2,x_3) \right)}
    { g_1 g_2 g_3 \cdot V_{\Delta_1^*\Delta_2^*\Delta_3^*}(x_1,x_2,x_3) }\Bigg|_{x_1^2=x_2^2=x_3^2=0}= \frac{N}{D}+ O(r^2),
\ee
where
\be
 N &= \sqrt{\tilde{f}_{13}\tilde{f}_{23}} \left(\beta _{13} \left(x_1^1 +2\right)+2 \beta _{23} x_1^1  \sqrt{z}\right){}^2 \left(\beta _{32} \left(x_2^1-2\right)
   \left(x_3^1 -2\right)+4 \beta _{13} \left(x_3^1 -x_2^1 \right) \sqrt{z}\right),
\\
D &=\beta _{13}^2 \beta _{23}^2 \sqrt{z} 
    \left(\beta _{13} \left(x^1_1+2\right)
   \left(x^1_3+2\right)+4 \beta _{23} \sqrt{z} \left(x^1_1-x^1_3\right)\right)
  \nn \\ &\quad\times \Big(\beta _{13} \beta _{23} \left(x^1_1+2\right) \left(x^1_2-2\right)+4 \beta
   _{13} \beta _{23} z \left(x^1_1-x^1_2\right)
   \nn\\
   &\quad\quad\quad+\sqrt{z} \left(\beta _{13}^2
   \left(x^1_1+2\right) \left(x^1_2+2\right)+\beta _{23}^2 \left(x^1_1-2\right)
   \left(x^1_2-2\right)\right)\Big).
\ee

Equating $\mathcal{L}^{(i)}_{AB}(g_1 g_2 g_3\cdot V_{\Delta_1^*,\Delta_2^*,\Delta_3^*}(x_1,x_2,x_3))/(g_1 g_2 g_3\cdot V_{\Delta_1^*,\Delta_2^*,\Delta_3^*}(x_1,x_2,x_3))|_{(x_1^*,x_2^*,x_3^*)}$ to the corresponding expectation values (\ref{eqn:moment_nonzero})-(\ref{eqn:nonzero_moments_3}), we find the following solution for 
$x_1^{1*},x_2^{1*},x_3^{1*}$:
\begin{equation}\label{classical_saddle}
(x_1^{1*},x_2^{1*},x_3^{1*})=\left(-2,2,-2\frac{(\beta _{23}^2\tilde{f}_{13})^{1/2} -(\beta_{13}^2\tilde{f}_{32})^{1/2} }{(\beta _{13}^2\tilde{f}_{32})^{1/2}+(\beta _{23}^2\tilde{f}_{13})^{1/2}}
\right).
\end{equation}
As we will show later, the saddle-point equations of the shadow integral  can be solved order by order in the thermal 
flat limit, with $\Delta_1\sim r^d,\Delta_2\sim r^d,\Delta_3\sim r^{d+1}$. To leading order,
the saddle point associated with the genus-2 block is:
\begin{equation}
    \left(-2,2,2\frac{\gamma_2-\gamma_1}{\gamma_1+\gamma_2}\right).
\end{equation}
Combining  (\ref{Delta_saddle_2d}) and  (\ref{gamma2Delta}), one verifies that 
 (\ref{classical_saddle}) indeed yields the correct saddle value.
\section{OPE asymptotics }\label{sec:OPE asymptotic}

In this section, we invert the genus-2 sunrise block expansion of the partition function and 
relate the heavy-heavy-heavier OPE coefficients to thermal one-point coefficients.
The relevant background is explained in section 7 of \cite{Benjamin:2023qsc}.
For the purpose of self-consistency, however, we start this section with a lightning review.
\subsection{A lighting review of the genus-2 sunrise block}
Given a triplet of conformal multiplets $\pi_i =(\Delta_i,\lambda_i),i=1,2,3$ and a pair of three-point structures $s,s'$,
the genus-2 block is a simultaneous eigenfunction of the three Casimirs:
\be 
\cC_2^{(i)} B_{123}^{s,s'}(g_1,g_2,g_3) = C_2(\pi_i) B_{123}^{s,s'}(g_1,g_2,g_3), \quad i=1,2,3.
\ee 
It satisfies a boundary condition consistent with its definition as a sum over states (see the discussion around (\ref{eq:sunrisesumoverstates})).
To compute $B_{123}^{s,s'}$ in the large quantum-number limit, we can focus on a closely related 
object --- the genus-2 partial wave:
\be\label{eq:genus2partialwave}
\Psi^{s,s'}_{123}(g_1,g_2,g_3) = 
\int d^d x_1 d^d x_2 d^d x_3
        V^{s,*}_{abc}(x_1,x_2,x_3)g_1 g_2 g_3 \cdot V^{s,abc} (x_1,x_2,x_3),
\ee
where $a,b,c$ are spin indices and
\be 
g_1g_2g_3 \cdot V^{s,a_1 a_2 a_3}(x_1,x_2,x_3) 
    = 
\left[ 
    \prod_{i=1}^3 
    \Omega(x_i')^{\Delta_i}
    (\lambda_i)^{a_i}_{\ \  b_i}(h_i^{-1}( x_i'))
\right ] V^{s,b_1 b_2 b_3}(x_1',x_2',x_3'),
\ee
with $x_i' = g_i\cdot x_i$. By construction, $ \Psi_{123}^{s,s'}$ solves the same set of 
 Casimir equations as the genus-2 block. However, it is different from $B_{123}^{s,s'}$ in that
\begin{itemize}
\item The integral for $\Psi_{123}^{s,s'}$ is defined for principal series representations where $\Delta_i = \frac{d}{2}+i\mathbb{R}$, while we want to compute $B_{123}^{s,s'}$ for physical representations (real $\De$ above the unitarity bound).

\item  $\Psi_{123}^{s,s'}$  satisfies different boundary conditions from $B_{123}^{s,s'}$. 
\end{itemize}
Analogous to how four-point conformal blocks are related to four-point partial waves, the 
following identity is expected to hold:
\be 
\label{eq:partialwaveexpression}
 \Psi_{123}^{s',s} = (I^{-3}S^3_{\tilde{1}^\dagger \tilde{2}^\dagger \tilde{3}^\dagger})^{s'}_{\ \ t'}B_{123}^{t's}+\text{7 shadow blocks}.
\ee
where:
\begin{itemize}
\item For a principal series representation $\pi_i=(\frac{d}{2}+i\nu_i,\lambda_i)$, its shadow representation 
$\tilde{\pi}_i$ is defined as $(\frac{d}{2}-i \nu_i,\lambda_i^R)$ where $\lambda_i^R$ denotes the reflected representation of $\lambda_i$.
The shadow blocks are obtained from $B_{123}^{s,s'}$ by making the replacement $\pi_i\rightarrow \tilde{\pi}_i$. 
\item 
Let us write $[\pi_1\otimes \pi_2\otimes \pi_3]$ for the space of solutions to the conformal Ward identities for three-point functions of those representations.
An element in $[\pi_1\otimes \pi_2\otimes \pi_3]$ is $V^{s,abc}(x_1,x_2,x_3)\equiv \langle \cO^a(x_1)\cO^b(x_2)\cO^c(x_3)\rangle^{(s)}$
where $\cO^a(x_1),\cO^b(x_2),\cO^c(x_3)$ transform according to $\pi_1,\pi_2,\pi_3$ respectively
and $s$ labels 
an element in $(\lambda_1\otimes \lambda_2\otimes \lambda_3)^{\SO(d-1)}$.\footnote{We use the following notation from \cite{Karateev:2018oml}: if an operator $\cO^a(x)$ transforms in the representation 
$\pi = (\Delta,\lambda)$, we denote its shadow in the representation $\tl\pi=(\tl \De,\l^R)$ by $\widetilde{\cO}^{\bar{a}}$, where the barred index indicates the reflected $\SO(d)$ representation.
We denote an operator that transforms according to 
$\widetilde{\pi}^\dagger \equiv (\widetilde{\Delta},\lambda^*)$ by $\widetilde{\cO}^\dagger_a(x)$ and an operator that transforms 
according to $\pi^\dagger \equiv (\Delta,(\lambda^*)^R)$ by $\cO^\dagger_{\bar{a}}$.}

In (\ref{eq:partialwaveexpression}), the coefficients $(I^{-3} S^3_{\tilde{1}^\dagger \tilde{2}^\dagger \tilde{3}^\dagger})^{s'}_{t'}$ are components of the following map:
\be 
(I^{-3} S^3_{\tilde{1}^\dagger \tilde{2}^\dagger \tilde{3}^\dagger}) :
     [\pi_1^* \otimes \pi_2^* \otimes \pi_3^* ]
            \rightarrow
     [\tilde{\pi}_1^\dagger \otimes \tilde{\pi}_2^\dagger\otimes \tilde{\pi}_3^\dagger],
\ee
where $\pi_i^*\equiv (\Delta_i,\lambda_i^*)$, $\tilde{\pi}_i^\dagger \equiv (d-\Delta_i,\lambda_i^*)$
with $\lambda_i^*$ being the dual representation of $\lambda_i$.
$S^3_{\tilde{1}^\dagger \tilde{2}^\dagger \tilde{3}^\dagger}: 
[\otimes_{i=1}^3 \pi_i^\dagger] 
\rightarrow 
[\otimes_{i=1}^3 \tilde{\pi}_i^\dagger]$ is the composition 
of three ``shadow maps'':
\be 
(S^3_{\tilde{1}^\dagger \tilde{2}^\dagger \tilde{3}^\dagger})=
S([\tilde{\pi}_1^\dagger] \tilde{\pi}_2^\dagger  \tilde{\pi}_3^\dagger)
\circ S(\pi_1^\dagger [\tilde{\pi}_2^\dagger]\tilde{\pi}_3^\dagger)
\circ S(\pi_1^\dagger\pi_2^\dagger [\tilde{\pi}_3]^\dagger),
\ee
where $S(\pi_1\pi_2[\pi_3]):[\pi_1\otimes \pi_2\otimes \tilde{\pi}_3]\rightarrow [\pi_1\otimes \pi_2\otimes \pi_3]$
is defined using a shadow transform:
\be 
\int d^dy 
\langle \widetilde{\cO}^{\bar{c}}(x_3)\widetilde{\cO}^\dagger_c(y) \rangle 
\langle \cO^a(x_1)\cO^b(x_2)\cO^c(y) \rangle^{(s)}
\equiv S(\pi_1 \pi_2 [\pi_3])^s_t \langle \cO^a(x_1) \cO^b(x_2) \widetilde{\cO}^{\bar{c}}(x_3)\rangle^{(t)}.
\ee
Given an element in $[\otimes_{i=1}^3 \pi_i^*]$,
contracting the three spin indices simultaneously with the inverse inversion tensor $I^{-1}$ will reflect the $\SO(d)$
representations $\lambda_i^*$, yielding an element in the desired space $[\otimes_{i=1}^3 \pi_i^\dagger]$ . 
\end{itemize}

When the $\Delta_i$'s are large, the integrand in (\ref{eq:genus2partialwave}) develops 
multiple saddle points. Expanding around different saddle points and 
contracting with the (inverse of) the triple shadow coefficients, we get leading asymptotics
of different blocks. Even though $\Psi_{123}^{s',s}$ is initially defined for principal series 
representations, after multiplying by the inverse triple shadow coefficient 
$((I^{-3} S^3_{\tilde{1}^\dagger\tilde{2}^\dagger\tilde{3}^\dagger})^{-1})^{s'}_{\ \ t'}$ we should be able to analytically continue the resulting expression onto the real $\Delta$-axis.

An important feature of the genus-2 sunrise block is the following orthogonality relation:
\begin{equation}\label{block_orthogonality}
    \oint d\mu 
            B_{\pi_1\pi_2\pi_3}^{s's}
            B_{ \tilde{\pi}_1'^{\dagger} 
                \tilde{\pi}_2'^{\dagger}
                \tilde{\pi}_3'^{\dagger} 
                }^{t'^{*}t^*}
    =
        T^{ts} T^{s't'}
        \prod_{i=1}^3 2 \pi \delta(\Delta_i-\Delta_i')
        \frac{2^d \vol\ \SO(d)}{\dim \lambda_i}\delta_{\lambda_i,\lambda_i'},
\end{equation}
 The three-point pairing matrix $T^{{ts}}$ is defined as:
\begin{equation}
    T^{ts}
    =
    \frac{
        V^t(0,e,\infty)^* V^s(0,e,\infty)
        }{2^d \vol\ \SO(d-1)},
\end{equation}
where $V^s(0,e,\infty)$ is a tensor transforming under $\lambda_1\otimes \lambda_2\otimes \lambda_3$ and $V^{t*}(0,e,\infty)$ transforms under
$\lambda_1^*\otimes \lambda_2^*\otimes \lambda_3^*$. The spin indices of $V^{t*}$ and $V^s$ are implicitly contracted.
This orthogonality relation is derived by deforming the integration contour into the ``low temperature region'' where all three cylinders $C_i$ become long, and using the low temperature limit of the block.

The integration measure $d\mu$ in  (\ref{block_orthogonality}) is the natural quotient measure on the moduli space $\mathcal{M}=G\backslash (G^{-})^3/ G$. 
In practice, however, we want to work with a partially gauge-fixed measure $d a_1 d a_2 d \beta_3 d\vec{\alpha} d\vec{\Phi}$ which is better suited  for our parameterization 
of the genus-2 sunrise geometry. To relate the measures, we go through the Faddeev-Popov procedure and find:
\be\label{FP}
    d\mu &= Q(a_1,a_2,\beta_3,\vec{\alpha}_i,\vec{\Phi}_i)da_1 da_2 d\beta_3 d \vec{\alpha}_i d\vec{\Phi}_i,
\ee
where
\be
 Q(a_1,a_2,\beta_3,\vec{\alpha}_i,\vec{\Phi}_i)
 & = \frac{2^{3d+5}(1+a_1)^{2d-1}(1+a_2)^{2d-1}(a_1+a_2-2)^{d-1}\beta_3}{(\vol\ \SO(d-1))^2(1-a_1)^{2d+1} (1-a_2)^{2d+1}} + O(\alpha,\Phi,\beta_3^2).
\ee
For more details, we  refer the reader to \cite{Benjamin:2023qsc}.

Equipped with the orthogonality relation (\ref{block_orthogonality}) and FP determinant (\ref{FP}),  we can invert the genus-2 block expansion of the partition function:
\begin{equation}
    Z= \sum_{\lambda_1,\lambda_2,\lambda_3} \ \int \ d\Delta_1 d\Delta_2 d\Delta_3 P^{s s'}_{123} B^{s's}_{123},
\end{equation}
to  get the genus-2 block coefficient $P_{123}^{s s'}$:
\be
P_{123}^{ss'}=(T^{-1})^{st}(T^{-1})^{t' s'} \frac{1}{(2\pi)^3} \prod_{i=1}^3 
\left( \frac{\dim \lambda_i}{2^d \vol \ \SO(d)} \right) \mathcal{I}_{123}^{t t'},\\
\mathcal{I}_{123}^{t t'}=
\int da_1 da_2 d\beta_3 \prod_{i=1}^3 d\vec{\alpha}_id\vec{\Phi}_i Q  Z  
B_{ \tilde{1}'^{\dagger} 
\tilde{2}'^{\dagger}
\tilde{3}'^{\dagger} 
}^{t't}.
\ee
The majority of this section will be devoted to the computation of $ \mathcal{I}_{123}^{tt'}$.

Here is an outline of the computation:
\begin{enumerate}
    \item 
    Compute the asymptotics of $B^{s,s'}_{123}$ and $B^{s,s'}_{\tilde{1}\tilde{2}\tilde{3}}$ in the heavy-heavy-heavier 
    regime. As explained before, we do this via saddle point approximation of the partial wave integral, namely:
    \be
    B^{s,s'}_{123} = 
    ((I^{-3} S^3_{\tilde{1}^\dagger \tilde{2}^\dagger \tilde{3}^\dagger})^{-1})^{s}_{\ \ t}\Psi^{t,s'}_{123}\Big|_{\text{correct saddle}}. 
    \ee  
We divide this computation into three steps:
    \begin{itemize}
        \item 
        Set $\Delta_i = \frac{d}{2}+ i\nu_i$ and locate saddle points of the partial-wave integrand in 
        the limit where $\nu_1\sim r^d,\nu_2\sim r^d,\nu_3\sim r^{d+1}$, with $\nu_1 \nu_2 /\nu_3^{2 d/(d+1)}$ held fixed. Note that the positions of the saddle points, to leading order at large $\nu_i$, depend only on the part of the integrand that is exponential in the $\nu_i$ (the ``rapidly-varying part"). Thus, for example, we can replace $d-\De_i\to -\De_i$ in the saddle point equations. This ensures the equations are linear in the $\De_i$, and thus their solutions only depend on ratios of the $\De_i$.
        \item 
        Identify the correct saddle point by matching $\vec{q}\left( \Delta_i^*\right)$ with (\ref{classical_saddle}). The ``dominant quantum numbers" $\Delta_i^*$ are given
        in (\ref{eq:Delta_i_saddle}), not to be confused with the complex conjugates of $\Delta_i$. 
        \item 
        Compute leading asymptotics of the triple shadow coefficients $(I^{-3} S^3_{\tilde{1}^\dagger \tilde{2}^\dagger \tilde{3}^\dagger})^{s}_{\ \ t}$
        in the appropriate large $\nu_i$ limit.
        Analytically continue $((I^{-3} S^3_{\tilde{1}^\dagger \tilde{2}^\dagger \tilde{3}^\dagger})^{-1})^{s}_{\ \ t} \Psi^{t,s'}_{123}\Big |_{\text{correct saddle}}$
        from $\Delta_i\in \frac{d}{2}+i\mathbb{R}$ to $\Delta_i\in \mathbb{R}$ by simultaneously rotating the $\Delta_i$'s to the real axis.
    \end{itemize}
We can then obtain the shadow blocks by replacing $\pi_i\to \tl \pi_i$, or alternatively by repeating the above computation with a different saddle.
\item Compute the large-$r$ asymptotics  of $Z(M_2)$ using the thermal effective action and re-write the asymptotic
expansion in terms of the sunrise channel coordinates $a_1,a_2,\beta_3,\vec{\alpha},\vec{\Phi}$. 
\item Perform the integral $\cI^{t t'}_{123}$ using results from the previous steps. This integral is again 
well-approximated via a saddle point. 
\end{enumerate}

\subsection{Saddle point approximation of the block}\label{subsection:saddle_point_approximation_of_block}
To find the saddle points of the partial wave integral, we start with the limit $\beta_3=0$ and $\vec{\alpha}_i=0,\vec{\Phi}_i=0$, then
solve perturbatively in small $\beta_3$ and small $\alpha,\Phi$. When all the angles are set to zero, $\SO(d-1)$ invariance of the partial wave integrand 
forces the saddle points to lie on the $x^1$ axis. Collecting the unknowns together into a triplet $\vec q=(x_1^1,x_2^1,x_3^1)$, we find the following four
solutions:
\be\label{saddles}
\vec{q}_1&=\left(-2,2, \frac{2(\gamma_2-\gamma_1)}{\gamma_1+\gamma_2}\right)+O(1/r),
\\
\vec{q}_2&=\left(-2,2, \frac{2(\gamma_2+\gamma_1)}{\gamma_1-\gamma_2}\right)+O(1/r),
\\
\vec{q}_3&=\left(-2a_1,2a_2,\frac{2\left(a_1 a_2-1-\sqrt{(1-a_1^2)(1-a_2^2)}\right)}{a_1-a_2} \right)+O(1/r),
\\
\vec{q}_4&=\left(-2 a_1,2a_2, \frac{2\left(a_1 a_2-1+\sqrt{(1-a_1^2)(1-a_2^2)}\right)}{a_1-a_2} \right)+O(1/r),
\ee
where for brevity and future convenience, we define
\begin{equation}\label{gamma2Delta}
\begin{split}
 \gamma_1&\equiv\sqrt{\left(1-a_2\right) \left(4 \left(a_1+1\right) \Delta _1^2+\left(1-a_1\right) \beta _3 \Delta _3^2\right)},\\
 \gamma_2&\equiv\sqrt{\left(1-a_1\right) \left(4 \left(a_2+1\right) \Delta _2^2+\left(1-a_2\right) \beta _3 \Delta _3^2\right)},\\
 \gamma_3&\equiv\left(1-a_1\right) \left(1-a_2\right) \beta _3 \Delta _3^2.
 \end{split}
\end{equation}

Due to shadow symmetry of the partial wave integral, the saddle points transform among themselves under the action of a $\mathbb{Z}_2^3$ group, described in \cite{Benjamin:2023qsc}. 
Let's denote the generators by $\sigma_1,\sigma_2,\sigma_3$, with $\sigma_i$ related to a shadow transform at site $i$. It can be checked explicitly
that the four saddle points above are each fixed by $\sigma_1$ and $\sigma_2$, while $\sigma_3$ exchanges $\vec{q}_1$ with $\vec{q}_2$ and $\vec{q}_3$ with $\vec{q}_4$.

There are four additional saddle points located at:
\begin{equation}
\begin{split}
\vec q_5&=\left(-2, \frac{-2 \left(a_2+1\right){}^2 \gamma _3+2 \left(a_2-1\right){}^2 \gamma _2^2-4
   \left(a_2^2-1\right) \gamma _2 \sqrt{\gamma _2^2-\gamma _3}}{\left(a_2-1\right)
   \left(a_2+3\right) \gamma _2^2-\left(a_2+1\right){}^2 \gamma _3},-2\right) + \dots \\ 
\vec q_6&= \left(-2,\frac{-2 \left(a_2+1\right){}^2 \gamma _3+2 \left(a_2-1\right){}^2 \gamma _2^2+4
   \left(a_2^2-1\right) \gamma _2 \sqrt{\gamma _2^2-\gamma _3}}{\left(a_2-1\right)
   \left(a_2+3\right) \gamma _2^2-\left(a_2+1\right){}^2 \gamma _3},-2\right)  + \dots \\
\vec q_7&= \left(\frac{2 \left(a_1+1\right){}^2 \gamma _3-2 \left(a_1-1\right){}^2 \gamma _1^2-4
   \left(a_1^2-1\right) \gamma _1 \sqrt{\gamma _1^2-\gamma _3}}{\left(a_1-1\right)
   \left(a_1+3\right) \gamma _1^2-\left(a_1+1\right){}^2 \gamma _3},2,2 \right) + \dots \\ 
\vec q_8&= \left(\frac{2 \left(a_1+1\right){}^2 \gamma _3-2 \left(a_1-1\right){}^2 \gamma _1^2+4
   \left(a_1^2-1\right) \gamma _1 \sqrt{\gamma _1^2-\gamma _3}}{\left(a_1-1\right)
   \left(a_1+3\right) \gamma _1^2-\left(a_1+1\right){}^2 \gamma _3},2,2 \right)  + \dots.
\end{split}
\end{equation}
These saddle points scale towards the singularities of the saddle point equations and cannot be found via naive expansion at generic values of $\vec q$. 
The saddle points $\vec q_5$ and $\vec q_6$ are exchanged by $\sigma_2$, while they are invariant under $\sigma_1$ and $\sigma_3$. Similarly, $\vec q_7$
and $\vec q_8$ are exchanged by $\sigma_1$ and invariant under $\sigma_2,\sigma_3$.  

We will make the assumption that the large-quantum-number block corresponds to a single saddle point in the partial wave integral. This is verified analytically at low temperatures, and supported numerically at high temperatures in \cite{Benjamin:2023qsc}, and we assume that it continues to be true in the regime of interest here.
To determine which saddle point corresponds to the block, we can take the infinite temperature limit within our thermal flat limit and match it to the infinite 
temperature limit within the high-temperature limit of \cite{Benjamin:2023qsc}. Recall that in \cite{Benjamin:2023qsc}, the correct high-temperature saddle $\vec{q}_0=\left(\frac{2\Delta_3}{2\Delta_2-\Delta_3},-\frac{2\Delta_3}{2\Delta_1-\Delta_3},\frac{2(\Delta_2-\Delta_1)}{\Delta_1+\Delta_2} \right)$
was found by numerically tracking the correct low temperature saddle to high temperature. In the regime where $\Delta_3$ is parametrically larger than $\Delta_1,\Delta_2$, we can re-expand $\vec{q}_0$ to be:
\begin{equation}
    \left(-2,2,\frac{2(\Delta_2-\Delta_1)}{\Delta_1+\Delta_2}\right)+\dots,
\end{equation}
which matches with $\vec{q}_1$ when $a_1=a_2=\beta_3=0$. 
As a separate check, one can verify that $\vec{q}_1$ indeed 
agrees with  (\ref{classical_saddle}) when the $\Delta_i$'s are taken
to be their saddle values.  Thus, we identify $\vec{q}_1$ as the saddle point corresponding to the block in the thermal flat limit, and we focus on it henceforth. 

Starting from the leading order expression for $\vec{q}_1$ in  (\ref{saddles}), it is straightforward to
compute corrections in $\beta_3$ order by order. To evaluate the partial wave integral, we divide the integrand into 
a rapidly-varying piece and a slowly-varying piece. The slowly-varying piece is evaluated directly at the saddle point while
the exponent of the rapidly-varying piece is expanded to second order around the saddle point, giving rise to a gaussian integral with an associated one-loop determinant. Combining the partial wave integral with the inverse triple shadow coefficient
and continuing the $\Delta_i$ from principal axis to the real axis, we find the following asymptotic expression for the genus-2 sunrise block in the scaling regime appropriate to the thermal flat limit:
\be
&B_{123}^{s's}\Big|_{h_i=1}\nn\\
&=\pi ^{-d} 2^{2 \left(\Delta _1+\Delta _2+\Delta _3\right)-\frac{7 d}{2}}
\left(\frac{\left(1-a_1\right) \left(1-a_2\right) \sqrt{\gamma _1 \gamma
   _2}}{\sqrt{\left(a_1+1\right) \left(a_2+1\right) \left(\gamma _1^2 \left(2 \gamma
   _2^2-\gamma _3\right)-\gamma _2^2 \gamma _3\right)}}
\right)^d \nn\\
&\quad\times
\left(-\Delta _1-\Delta _2+\Delta _3\right)^{-\Delta _1-\Delta _2+\Delta _3}
\left(\Delta _1-\Delta _2+\Delta _3\right)^{-\Delta _1+\Delta _2-\Delta _3}
\left(-\Delta _1+\Delta _2+\Delta _3\right)^{\Delta _1-\Delta _2-\Delta _3} \nn\\
&\quad\times
\left(\Delta _1+\Delta _2+\Delta _3\right)^{-\Delta _1-\Delta _2-\Delta _3} \Delta
_1^{2 \Delta _1-\frac{d}{2}} \Delta _2^{2 \Delta _2-\frac{d}{2}} \Delta _3^{2
\left(d+\Delta _3\right)}
\times 
\exp
\left(
-\sqrt{\frac{\beta_3}{\gamma_3}}\frac{\gamma _1 \gamma _2}{\sqrt{\left(a_1-1\right) \left(a_2-1\right)} }
\right)
\nn\\
&\quad\times
V^{s'}(0,e,\infty)^* \cdot  V^s(0,e,\infty).
\ee
We stress that the factors 
$(-\Delta_1-\Delta_2+\Delta_3)^{-\Delta_1-\Delta_2+\Delta_3}$,
$(\Delta_1-\Delta_2+\Delta_3)^{-\Delta_1+\Delta_2-\Delta_3}$, and
$(-\Delta_1+\Delta_2+\Delta_3)^{\Delta_1-\Delta_2-\Delta_3}$
are all positive, since we are working in a scaling limit where $\Delta_3$ is parametrically larger 
than $\Delta_1,\Delta_2$.

The expression above is valid when all angular fugacities are turned off. However, in order to compute 
the inversion integral, we also need the angular dependence of the ``classical piece"  $\exp(\cdots)$.
Once the angular fugacities are turned on, the saddle points move away from the $x^1$ axis. To leading order in the angles, we find:
\begin{equation}
    \vec{x}_1^\perp= \frac{\Delta _1 \Delta _3\left(1-a_1\right) \left(1-a_2\right)  \left(\left(1-a_1\right) 
   \gamma _1^2 \vec{\alpha}_1-\left(1-a_2\right)  \gamma _2^2 \vec{\alpha}_2-2\left(\gamma _1^2-\gamma _2^2\right)\vec{\alpha}_3 \right)}{\gamma _1^3 \gamma _2}
   +O(\alpha^2,\Phi^2)
\end{equation}
\begin{equation}
    \vec{x}_2^\perp=\frac{\Delta _2 \Delta _3 \left(1-a_1\right) \left(1-a_2\right)  \left(\left(1-a_1\right) 
   \gamma _1^2 \vec{\alpha}_1-\left(1-a_2\right)  \gamma _2^2 \vec{\alpha}_2-2\left(\gamma _1^2-\gamma _2^2\right)\vec{\alpha}_3 \right)}{\gamma _1 \gamma _2^3}
   +O(\alpha^2,\Phi^2)
\end{equation}
\begin{equation}
    \vec{x}_3^\perp=
    -\frac
    {   \Delta _3^2 \left(1-a_1\right) \left(1-a_2\right)\left(\left(1-a_1\right) 
        \gamma _1^2 \vec{\alpha}_1-\left(1-a_2\right)  \gamma _2^2 \vec{\alpha}_2-2\left(\gamma _1^2-\gamma _2^2\right)\vec{\alpha}_3 \right)
    }
   {   
    \gamma _1 \gamma _2 \left(\gamma _1+\gamma_2\right){}^2
   }+O(\alpha^2,\Phi^2).
\end{equation}
where $\vec{x}_i^{\perp}$ and $\vec{\alpha}_i$ are $(d-1)- $dimensional vectors and $\vec{\alpha}_i$ 
are defined in  (\ref{eqn:sunrise_channel_angular_twist}).
Plugging in the expansion of $\vec{q}_1$ up to second order in $\vec{\alpha}$ and $\vec{\Phi}$,
we find the following correction term in the classical piece of the block:
\begin{equation}
\label{eq:equationwithangularcoefficients}
    B_{123}^{s's}\Big|_{\exp(\cdots)}= \exp
\left(
-\sqrt{\frac{\beta_3}{\gamma_3}}\frac{\gamma _1 \gamma _2}{\sqrt{\left(a_1-1\right) \left(a_2-1\right)} }
+\sum_{\substack{i,j=1 \\  i\leq j}}^3 C_{i,j}\vec{\alpha}_i\cdot \vec{\alpha}_j 
\right),
\end{equation}
where $C_{i,j}$ are $\De$-dependent coefficients that we write explicitly in Appendix \ref{app:coefficientsCij}.

\subsection{The inversion integral}
In the thermal flat limit, the partition function takes the form:
\begin{equation}\label{partition_function}
    \begin{split}
    Z = h(z) \exp\p{-S_{13}-S_{23}},
    \end{split}
\end{equation} 
where $h(z)=\sum_\mathcal{O} b_\mathcal{O}c_\cO^{-1} q_{J_\cO} z^{\Delta_\cO}$ 
and $S_{13}$, $S_{23}$ are the thermal effective actions of the two dumbbells. We can write them 
in terms of the relative coordinates:
\begin{equation}
S_{ij}=-\frac{\vol\ S^{d-1}}{\prod_{a=1}^n(1+\Omega^a_{ij})}\left[\frac{f_{ij}}{\beta_{ij}^{d-1}}-\frac{d-2}{\beta_{ij}^{d-3}}\left( (d-1) c^{(1)}_{ij}+(2c^{(1)}_{ij} +\frac{8}{d} c^{(2)}_{ij})\sum_{a=1}^n (\Omega^a_{ij})^2 \right) +\cdots \right],
\end{equation}
where $n=\lfloor d/2 \rfloor$ is the rank of the orthogonal group $SO(d)$. Here, $f_{ij}$, $c^{(1)}_{ij}$, and $c^{(2)}_{ij}$
are Wilson coefficients associated to the cosmological constant term, the Einstein term $\hat{R}$, and Maxwell term $F^2$ in the thermal effective action. For generality, we assume that the Wilson coefficients can be different in the two dumbbells, as would be the case if different topological defects were inserted on the dumbbells.

It is useful to re-write the thermal effective actions in terms of the coordinates $a_1,a_2,\beta_3$,$\vec{\alpha}$ and $\Phi$.
For example, the cosmological constant term reads
\be
\label{partition_function}
  S_{13}\Big|_{\text{leading}}&=-f_{13}\vol S^{d-1} 
        \left(
            \frac{1-a_1}{4 \beta_3 (1+a_1)}\right)^{\frac{d-1}{2}}
         \nn\\
         &\quad \x
            \left( 1-\frac{(1-a_1)(\vec{\Phi}_1-\vec{\Phi}_3)^2}{4(1+a_1)\beta_3}
        -\Gamma^{1,1}_{\Lambda} |\vec{\alpha}_1|^2 -\Gamma^{3,3}_{\Lambda}|\vec{\alpha}_3|^2 -\Gamma^{1,3}_{\Lambda} (\vec{\alpha}_1\cdot \vec{\alpha}_3)
        \right),
\ee
where the coefficients $\Gamma_{\Lambda}^{1,1}, \Gamma_{\Lambda}^{3,3}, \Gamma_{\Lambda}^{1,3}$ are given by
\begin{equation}
\begin{split}
    \Gamma_{\Lambda}^{1,1}&=
                    (d+1)\frac{(1-a_1)^2}{32 \beta_3^2}
                    -\frac{(1-a_1)}{384 (1+a_1)\beta_3}(a_1^2 \left(d^2-8 d-1\right)-4 a_1 \left(d^2+4 d-1\right)+d^2+16 d-73)+O(1),\\
    \Gamma_{\Lambda}^{3,3}&=
                    \frac{d+1}{8 \beta_3^2}
                    -\frac{(a_1^2-4 a_1+1)(d^2+4 d-13)}{96 (1-a_1^2)\beta_3}+O(1),\\
    \Gamma_{\Lambda}^{1,3}&=
                    -(d+1)\frac{(1-a_1)}{8 \beta_3^2}
                    +\frac{a_1^2 \left(d^2+4 d-13\right)-4 a_1 \left(d^2+d-4\right)+d^2+16 d-49}{96 \left(a_1+1\right) \beta _3}+O(1).
\end{split}
\end{equation}
The cosmological constant term in  $S_{23}$ has a similar expansion, with $f_{13},a_1,\vec{\alpha}_1,\vec{\Phi}_1$ replaced by $f_{23},a_2,\vec{\alpha}_2,\vec{\Phi}_2$.
Recall that when $a_1=a_2=0$,\footnote{This is equivalent to setting $\beta_{13,0}=\beta_{23,0}$, $z=1/4$.} the geometry of the genus-2 manifold reduces to the one discussed in \cite{Benjamin:2023qsc}, where all three balls
are tangent to each other. In this limit, $\beta_3=\beta_{13,0}^2/4=\beta_{23,0}^2/4$, and $S_{13}\Big |_{\text{leading}}$ becomes:
\begin{equation}
    -\frac{f_{13}\vol S^{d-1}}{\beta_{13,0}^{d-1}}\left(1-\frac{(\vec{\Phi}_1-\vec{\Phi}_3)^2}{\beta_{13,0}^2}-8(d+1) \frac{(\frac{1}{4}\vec{\alpha}_1-\frac{1}{2}\vec{\alpha}_3)^2}{\beta^4_{13,0}} +\cdots \right),
\end{equation}
this agrees with the hotspot formula in the high-temperature limit. 

The Einstein and Maxwell terms are similarly given by:
\be
    S_{13}\Big|_{R}&= c^{(1)}_{13}(d-1)(d-2)  \vol \ S^{d-1} \left(\frac{1-a_1}{4\beta _3(a_1+1)}\right)^{\frac{d-3}{2}}
    \nn\\
    &\quad \x\Big(1-\frac{(1-a_1)(d-3)(\vec{\Phi}_1-\vec{\Phi}_3)^2}{4(1+a_1)(d-1)\beta_3}
    -\Gamma^{1,1}_{R} |\vec{\alpha}_1|^2
    -\Gamma^{3,3}_{R}|\vec{\alpha}_3|^2 -\Gamma_{R}^{1,3} (\vec{\alpha}_1\cdot \vec{\alpha}_3)\Big),
\\
    S_{13}\Big|_{F^2}&= \frac{8(d-2)c^{(2)}_{13}}{d} \vol \ S^{d-1} \left(\frac{1-a_1}{4\beta _3(a_1+1)}\right)^{\frac{d-1}{2}}
    \nn\\
    &\quad \x \Big(
    (\vec{\Phi}_1-\vec{\Phi}_3)^2
     -\Gamma^{1,1}_{F^2} |\vec{\alpha}_1|^2-\Gamma^{3,3}_{F^2}|\vec{\alpha}_3|^2 -\Gamma_{F^2}^{1,3} (\vec{\alpha}_1\cdot \vec{\alpha}_3) 
    \Big),
\ee
with
\be
    \Gamma_{R}^{1,1}&=\frac{\left(a_1-1\right){}^2  (d-3)(d+1)}{32 (d-1) \beta _3^2 }
    \nn\\
    &\quad +\frac{\left(a_1-1\right) (d-3) \left(a_1^2 \left(d^2-10 d-3\right)-4
   a_1 \left(d^2+2 d-3\right)+d^2+14 d-75\right)}{384(d-1) \left(a_1+1\right) \beta _3 }+O(1),
   \nn\\
   \Gamma_{R}^{3,3}&=-\frac{(d-3)(d+1)}{8(1-d)\beta _3^2}+\frac{\left(a_1^2-4 a_1+1\right) (d+5) (d-3)^2}{96 \left(a_1-1\right) \left(a_1+1\right)(d-1) \beta _3}+O(1),
   \nn\\
   \Gamma_{R}^{1,3}&=\frac{\left(a_1-1\right) \left(d-3\right)\left(d+1\right)}{8 (d-1)\beta _3^2}+\frac{(d-3)^2 \left(a_1^2 (d+5)-4 a_1 (d+2)+d+17\right)}{96
   \left(a_1+1\right)(d-1)\beta_3}+O(1),
\ee
and
\begin{equation}
\begin{split}
    \Gamma_{F^2}^{1,1}=\frac{a_1^2-1}{4 \beta _3}+O(1), \quad
   \Gamma_{F^2}^{3,3}=\frac{a_1+1}{\left(a_1-1\right) \beta_3}+O(1),\quad
   \Gamma_{F^2}^{1,3}=\frac{a_1+1}{\beta _3}+O(1).
\end{split}
\end{equation}

At leading order in the $1/r$ expansion, the saddle points of the inversion integral and the associated 
one-loop factors are determined by the cosmological constant terms. However, the contributions of the subleading terms are 
comparable to the contribution of $h(z)$, so we will keep them in the final formula.  

Finally, we have all the ingredients needed to evaluate the inversion integral $\mathcal{I}_{123}^{s,s'}$. We specialize to $d=3$ in what follows, though the computation generalizes straightforwardly to other dimensions. 
We will also now assume $f_{13}=f_{23}=f$.

We first perform the integral over the angles $\vec{\alpha}$ and $\vec{\Phi}_i$ via saddle point approximation. Using the small-angle expansion 
of $Z$ and $B^{s,s'}_{\tilde{1}\tilde{2}\tilde{3}}$, we find the one-loop factors to be:
\begin{equation}
\text{one-loop factor in $\Phi_i$}=\vol\ \SO(2) \times \frac{4 \pi ^{4/3} f^{1/3}}{\Delta _1^{2/3} \Delta _2^{2/3}}.
\end{equation}
The $\Phi$ part of the integrand only depends on the differences $\vec{\Phi}_1-\vec{\Phi}_3$ and $\vec{\Phi}_2-\vec{\Phi}_3$. This is a consequence 
of the $SO(d-1)$ gauge symmetry. To evaluate its one-loop factor, we first gauge fix $ \vec{\Phi}_3$ to be zero and 
evaluate the hessian with respect to $\vec{\Phi}_1$,$\vec{\Phi}_2$, then multiply the final result by $\vol\ SO(d-1)$.

For the $\alpha$ integral, we find:
\be
&\text{one-loop factor in $\alpha_i$}\nn\\
&=
\frac{64 f^{1/3} \pi ^{10/3} \left(\Delta _1^{2/3}+\Delta _2^{2/3}\right)  \left(2 \Delta_2^{1/3}
   \Delta _1+(\pi f)^{1/3} \Delta _3 \right)^2 \left(2 \Delta_1^{1/3} \Delta
   _2+(\pi f)^{1/3} \Delta _3 \right)^2}{\Delta _1^{7/3}\Delta _2^{7/3} \Delta _3^4 \left(\Delta_1^{1/3}-\Delta_2^{1/3}\right)^2 }.
\ee
%where
% \be
% \label{eqn:p}
%     p(\Delta_1,\Delta_2,\Delta_3)
%     &=18432\ 2^{2/3} \Delta _1^4 \Delta _2^4 \left(\Delta _1^{4/3}+16 \Delta _2^{2/3} \Delta _1^{2/3}+\Delta
%    _2^{4/3}\right)
%    \nn\\
%    &\quad +896\ 2^{2/3} \pi ^{5/3} \Delta _1 \left(\Delta _1^{2/3}+\Delta _2^{2/3}\right) \Delta _2
%    \Delta _3^5 f^{5/3}
%    \nn\\
%    &\quad +256\ 2^{2/3} \pi ^{4/3} \Delta _1^{4/3} \Delta _2^{4/3} \left(8 \Delta _1^{4/3}+63
%    \Delta _2^{2/3} \Delta _1^{2/3}+8 \Delta _2^{4/3}\right) \Delta _3^4 f^{4/3}
%    \nn\\
%    &\quad +768 (2 \pi )^{2/3} \Delta
%    _1^{8/3} \Delta _2^{8/3} \left(79 \Delta _1^{4/3}+310 \Delta _2^{2/3} \Delta _1^{2/3}+79 \Delta
%    _2^{4/3}\right) \Delta _3^2 f^{2/3}
%    \nn\\
%    &\quad +16\ 2^{2/3} \pi ^2 \left(\Delta _1^{2/3}+\Delta _2^{2/3}\right){}^2
%    \Delta _3^6 f^2
%    \nn\\
%    &\quad +58368\ 2^{2/3} \pi  \Delta _1^{7/3} \left(\Delta _1^{2/3}+\Delta _2^{2/3}\right) \Delta
%    _2^{7/3} \Delta _3^3 f
%    \nn\\
%    &\quad +276480\ 2^{2/3} \pi^{1/3} f^{1/3}\Delta _1^{11/3} \left(\Delta _1^{2/3}+\Delta
%    _2^{2/3}\right) \Delta _2^{11/3} \Delta _3.
% \ee
Note that at leading order, the hessian in the $\alpha$ integral has vanishing determinant if $\Delta_1=\Delta_2$. Something must regulate this singularity when $\De_1=\De_2$. We leave this question for the future, and for now we assume $\Delta_1\neq \Delta_2$. 

For the integral over $a_1,a_2,\beta_3$, the saddle point is located at
\begin{equation}
\begin{split}
a_1^*&=1-\frac{4 \Delta _1 \Delta_2^{1/3}}{2 \Delta_2^{1/3} \Delta _1+(\pi f)^{1/3} \Delta _3 }
+O(r^{-2}),\\
a_2^*&=1-\frac{4 \Delta_1^{1/3}\Delta _2}{2 \Delta_1^{1/3} \Delta _2+(\pi f)^{1/3} \Delta _3 }
+O(r^{-2}),\\
\beta_3^*&=
\frac{2 (\pi f \Delta_1 \Delta_2)^{1/3}}{\Delta _3}+O(r^{-4}),
\end{split}
\end{equation}
and the corresponding one-loop factor is:
\begin{equation}
\begin{split}
    \text{one loop factor in $a_1,a_2,\beta_3$ }=
    \frac{64 \sqrt{f} \pi ^2 \Delta _1 \Delta _2 \Delta _3  \sqrt{\Delta _1^{2/3}+\Delta _2^{2/3}} }{3 (2
   \Delta_2^{1/3}\Delta _1+(\pi f)^{1/3} \Delta _3 )^2 (2 \Delta_1^{1/3}
   \Delta _2+(\pi f)^{1/3} \Delta _3 )^2}.
\end{split}
\end{equation}
To ensure $a_1^*>0$ and $a_2^*>0$, the following inequality should be satisfied:
\begin{equation}
\frac{\Delta_1^2\Delta_2^2}{\Delta_3^3} \max\left(\frac{\Delta_2}{\Delta_1},\frac{\Delta_1}{\Delta_2}\right)<\frac{\pi f}{8}.
\end{equation}

Combining everything together, we find the following asymptotic formula for the genus-2 sunrise block coefficient
in 3d, assuming $\Delta_1\neq \Delta_2$:
\begin{equation}\label{eqn:main_formula}
    \begin{split}
    &P^{s,s'}_{123}\\
    &\sim 
    (T^{-1})^{s,s'}\prod_{i=1}^3 \left( \frac{\dim \lambda_i}{\vol \SO(3)}\right)
    \left[\sum_{\mathcal{O}}b_\mathcal{O}^2 q_{J_\cO}
        \left(\frac{\Delta _1^{2} \Delta _2^{2}}{\pi f  \Delta _3^3}\right)^{\frac{2\Delta_\mathcal{O}}{3}}
    \right]
    \exp \left( (f\pi)^{1/3}(\Delta_1^{2/3}+\Delta_2^{2/3})-16 \pi c^{(1)}\right)\\
    &\quad\times
    \pi ^{17/6} f^{10/3} 
    2^{-2 \Delta _1-2 \Delta _2-2 \Delta _3+15} 
    \Delta _1^{-2 \Delta _1-\frac{23}{6}} 
    \Delta _2^{-2 \Delta_2-\frac{23}{6}} 
    \Delta _3^{\frac{27}{2}-2 \Delta _3}
    \left(-\Delta _1-\Delta _2+\Delta _3\right)^{\Delta _1+\Delta _2-\Delta _3-3}\\
    &\quad\times \left(\Delta _1-\Delta _2+\Delta _3\right)^{\Delta _1-\Delta _2+\Delta _3-3} 
    \left(-\Delta _1+\Delta _2+\Delta _3\right)^{-\Delta_1+\Delta _2+\Delta _3-3} 
    \left(\Delta _1+\Delta _2+\Delta _3\right)^{\Delta _1+\Delta _2+\Delta _3-9}  \\
   &\quad\times \frac{
                \left(4 \Delta _1 \Delta _2+(\pi f)^{1/3}\Delta _3 (\Delta _2^{2/3}+\Delta_1^{2/3})\right)^2}
                {
                3 \left(\Delta _1^{1/3}-\Delta _2^{1/3}\right)^2
                }.
    \end{split}
\end{equation}
Again, this formula is valid in the thermal flat limit $\De_1,\De_2,\De_3$ large with fixed ratio $\De_1^2 \De_2^2/\De_3^3$, while the spins $\l_i$ are kept finite.

\section{Discussion}

In this work, we initiated the study of a genus-2 crossing equation in $d\geq 2$. This equation can be understood as arising from the usual four-point crossing equation after contracting pairs of external states. We described natural coordinates for both channels (dumbbell and sunrise) and the relationship between them, described the conformal blocks that appear on both sides, and discussed mapping class group invariance of the genus-2 partition function in 3d.
In the thermal flat limit, the function $h(z)$ encoding squares of thermal one-point coefficients arises naturally 
on the dumbbell side of the crossing equation. This leads to a relation between asymptotic heavy-heavy-heavier  OPE coefficients and squared thermal one-point functions in 3d CFTs.

Our formula for OPE asymptotics joins others in the literature for different regimes of quantum numbers in higher dimensional CFTs. 
For example, \cite{Benjamin:2023qsc} studies a heavy-heavy-heavy limit where all operator dimensions become large at the same rate (with fixed spins). 
It is natural to 
expect a direct interpolation between the heavy-heavy-heavy limit (more concretely, see equation (7.51) in \cite{Benjamin:2023qsc})
and the heavy-heavy-heavier limit studied in this work. 
The formula for leading OPE asymptotics don't seem to be connected in a simple way. However,
as we discussed in section \ref{sec:OPE asymptotic}, saddle point positions in the partial wave integral 
as well as the expansion of the genus-2 partition function do interpolate nicely. It would be interesting to systematically compute subleading corrections 
and understand how OPE asymptotics in the two scaling limits are connected. Relatedly, it would be nice to develop more systematic methods for solving the saddle point equations.

Meanwhile, \cite{Delacretaz:2020nit} uses hydrodynamics to make predictions for OPE coefficients $c_{LHH'}$ where $H$ and $H'$ are potentially different heavy operators with similar energies.
Our derivation of OPE asymptotics was based on applying a kind of inversion formula to the genus-2 partition function, and looking for saddle points in the inversion integral. In general, such methods lead to rough asymptotics but not precise statements. To be more precise, one could try to characterize the behavior of averages over ``windows" in the space of states, analogous to \cite{Pal:2019zzr,Mukhametzhanov:2020swe,Pal:2025yvz} for the modular bootstrap in 2d, or \cite{Pal:2022vqc} for the lightcone bootstrap. We expect that gaining control over small windows in the space of operator energies might be important for making contact with hydrodynamics \cite{Delacretaz:2020nit}. Reference \cite{Anous:2021caj} also makes predictions for OPE statistics using higher-point crossing.

It would be nice to check these predictions in free theories by explicitly constructing and analyzing three-point coefficients. It is also interesting to ask whether one can make connections with holography. Here the situation in higher dimensions seems somewhat different from the situation for Virasoro primaries in 2d. In the 2d context, the statistics of Virasoro primaries seems to be well captured by holographic wormhole calculations \cite{Collier:2019weq,Abajian:2023jye}. On the other hand, the statistics of OPE coefficients of global primaries and Virasoro primaries are quite different \cite{Benjamin:2023qsc}. Naively, we might expect wormhole calculations in higher dimensions to capture something more like the statistics of ``Virasoro-primary-like" states rather than global primaries. We do not know the correct definition of ``Virasoro-primary-like," but perhaps it means states that do not include a gas of gravitons around them (which would not modify the density of states, but could have nontrivial effects on three-point coefficients).

Let us mention some other possible directions for future exploration.

\subsection{Kinematics of the angular twist}

In the thermal flat limit studied in this work, all the angular fugacities in  
$g_L$,$g_R$,$g_M$ were set to zero. An immediate next step is to generalize 
to geometries with nontrivial $h_L,h_R$ and $h_M$.
In this case, the $SO(d-1)$ symmetry that fixes $\langle J_i\rangle$ to be zero 
will be broken and we expect the sum over 
$\left\{(\Delta_1,\lambda_1),(\Delta_2,\lambda_2),(\Delta_3,\lambda_3)\right\}$
to be dominated by a representation involving parametrically large spin. 
What would the saddle points of the shadow integral look like in this case? How 
do they depend on the three-point tensor structures? What would be the asymptotic behavior of
the corresponding OPE coefficients? 

As noted in section \ref{sec:thermal flat limit}, when $h_L,h_R$ are not infinitesimally close to the identity, thermal EFT typically breaks down. However, in \cite{Benjamin:2024kdg}, it was 
pointed out that in the case of rational angular fugacity, thermal EFT can still produce the correct 
high temperature asymptotics using the ``folding trick,'' which relates rational angular twists to physics at a lower temperature. Does the ``folding trick''
provide an obvious relation between $\text{tr}_{\mathcal{H}}(\mathcal{O}(\tau,\hat{n})e^{-\beta D})$ and 
$\text{tr}_{\mathcal{H}}(\mathcal{O}(\tau,\hat{n})e^{-\beta D}e^{i \pi  M_{ij}})$, or 
their normalized versions? 
For generic angles, it is natural to expect that 
$\langle\mathcal{O}(x)\rangle_{h}$ only depends on the geometry in the vicinity of the operator insertion. 
The presence of angular twists single out special subspaces in $\mathbb{R}^d$. The basis vectors of these subspaces,
when combined with the unit vector along the thermal circle, provide extra kinematic structures compatible with the symmetry of the 
geometric background. Can we systematically count and enumerate them? In dimensions greater than three, what 
are the normalized thermal one-point functions of mixed-symmetry tensors in the presence of angular twists?

\subsection{Understanding the genus-2 block}
In Appendix \ref{appendix:Casimir}, we discussed the Casimir equations solved by the
genus-2 block.
It would be nice to solve the Casimir equations 
recursively and to work out weight-shifting relations for the genus-2 blocks. It might also be interesting to derive Zamolodchikov-type recursion relations for genus-2 blocks, analogous to those derived for four-point blocks \cite{Hogervorst:2013sma,Kos:2013tga,Kos:2014bka,Penedones:2015aga,Erramilli:2019aa}.

\subsection{Comments on the thermal bootstrap}
In the thermal flat limit, the genus-2 crossing equation simplifies to:
\begin{equation}
    \sum_{\mathcal{O},J}b^2_\mathcal{O} q_J z^{\Delta_{\mathcal{O}}} 
    = 
    Z^{-1}(S^1_{\beta_{13}}\times S^{d-1})Z^{-1}(S^1_{\beta_{23}}\times S^{d-1}) \sum_{123}c_{123}^s c_{123}^{s'} B^{s,s'}_{123}.
\end{equation}
One might hope to use this crossing equation to ``complete the square'' of the thermal bootstrap \cite{Iliesiu:2018fao}, which on its own lacks the positivity conditions required for numerical bootstrap techniques. Here we offer a non-rigorous argument 
against the practicality of this idea: 
from our analysis in section \ref{sec:OPE asymptotic}, we see that when $r$ is very large, the product 
$B^{s,s'}_{\tilde{1}^{\dagger}\tilde{2}^{\dagger}\tilde{3}^{\dagger}} Z(\beta_{13})Z(\beta_{23})$
develops a saddle point with the corresponding one-loop factor scaling like $r^{-5}$. In the $r\rightarrow \infty$
limit, this saddle point will become a $\delta$-distribution. For this reason, we expect that the quotient $B_{123}^{s,s'}/Z(\beta_{13})Z_(\beta_{23})$,
as a function (more precisely: distribution) over the thermal flat locus, will be proportional to
 $\delta\left(z-\left(\frac{\Delta_1^2\Delta_2^2}{\pi f \Delta_3^3}\right)^{2/3}\right)$.
Thus, when $r=\infty$ the right hand side of the crossing equation will be proportional to:
\begin{equation}\label{eqn: sunrise_flat_limit}
\int d\Delta_1 d\Delta_2 c^2_{\Delta_1,\Delta_2,\Delta_3(z,\Delta_1,\Delta_2)}
\mu(\Delta_1,\Delta_2,\Delta_3(z,\Delta_1,\Delta_2)),
\end{equation}
with $\mu(\Delta_1,\Delta_2,\Delta_3(z,\Delta_1,\Delta_2))$
being a positive function and 
$\Delta_3(z,\Delta_1,\Delta_2)=\frac{\Delta _1^{2/3} \Delta _2^{2/3}}{(\pi f)^{1/3} \sqrt{z}}$.
Not only does the measure $\mu$ depend on $z$, but $z$ also determines which combinations
of $(\Delta_1,\Delta_2,\Delta_3)$ contribute in the sum.
It is not clear what one can learn from this crossing equation except for the fact that the left hand side should be 
positive for any given value of $z$. 
It is interesting to explore what happens if we turn on angular fugacities in the $r\rightarrow \infty$ limit.

\subsection{Towards the genus-2 numerical bootstrap}

For finite $r$, the genus-2 crossing equation (\ref{eq:ourequation}) has all the necessary positivity conditions on both sides to make the standard numerical bootstrap approach \cite{Rattazzi:2008pe} possible.
An advantage compared to the usual numerical bootstrap for four-point functions is that this equation ``knows" about all of the CFT data at once, as as opposed to just the data appearing in a fixed set of four-point functions. On the other hand, it may be technically challenging to numerically bootstrap this equation due to the large number of cross-ratios, and the multitude of quantum numbers conjugate to those cross ratios. Perhaps, to begin, one could study a lower-dimensional slice of the genus 2 moduli space. A numerical study of genus-2 crossing in 2d was performed in \cite{Cho:2017fzo}, where they expanded around a $\Z_3$ symmetric point in the genus-2 moduli space. We do not know of an analogous expansion point for the equation (\ref{eq:ourequation}), so it is not immediately obvious what the optimal expansion locus might be. This is an interesting question for future work.

\subsection{Genus-2 lightcone bootstrap and Lorentzian physics}

 In the context of the four-point bootstrap, a particularly interesting limit is the lightcone limit, which leads to the lightcone bootstrap and large-spin perturbation theory \cite{Komargodski:2012ek,Fitzpatrick:2012yx}. What is the analog of the lightcone limit for the genus-2 geometry? More generally, what kind of physics do we encounter on the genus-2 moduli space after analytic continuation away from Euclidean signature? Are there analogs of dispersive functionals \cite{Mazac:2016qev,Mazac:2018mdx,Mazac:2018ycv,Mazac:2019shk,Penedones:2019tng,Caron-Huot:2020adz,Carmi:2020ekr} for the genus-2 crossing equation?

\section*{Acknowledgements}
We thank Ilija Buric for initial collaboration on this work, and for helpful discussions. We thank Tom Hartman, Murat Kolo\u{g}lu, Sridip Pal and Francesco Russo for discussions, and Nathan Benjamin, Sridip Pal, and Yifan Wang for comments on the draft.
This material is based upon work supported by the U.S.
Department of Energy, Office of Science, Office of High Energy Physics, under Award Number DE-SC0011632. 
DSD and YX are supported in part by Simons Foundation grant 488657
(Simons Collaboration on the Nonperturbative Bootstrap).

\newpage

\appendix

\section{The angular expansion coefficients $C_{i,j}$}\label{app:coefficientsCij}

In this appendix, we record the angular expansion coefficients $C_{i,j}$ appearing in~(\ref{eq:equationwithangularcoefficients}).
\small
\begin{equation}
    \begin{split}
   C_{1,1}=&-\frac{\left(a_1-1\right){}^2 \gamma _1 \gamma _3 \Delta _3}{32 \beta _3 \gamma _2^3}+
   \frac{\Delta _3}{256 \left(a_1+1\right){}^2 \gamma _1^3 \gamma _2^7}
   \Big(
    \left(a_1-1\right) \left(a_1+1\right) \left(\gamma _2^2-\gamma _3\right) \gamma _1^6 \left(2
   \left(a_1^2+8 a_1+1\right) \gamma _2^2-5 \left(a_1+1\right){}^2 \gamma _3\right)\\
   &+\left(a_1-1\right)
   \gamma _2^2 \gamma _1^4 \left(2 \left(a_1^3+9 a_1^2+a_1-7\right) \gamma _2^4+6 \left(a_1-1\right)
   \left(a_1^2-1\right) \gamma _3 \gamma _2^2-3 \left(a_1-1\right){}^2 \left(a_1+1\right) \gamma
   _3^2\right)\\
   &+\left(a_1-1\right) \gamma _2^4 \gamma _3 \gamma _1^2 \left(\left(a_1+1\right)
   \left(a_1-1\right){}^2 \gamma _3+\left(a_1^3+7 a_1^2+23 a_1+17\right) \gamma
   _2^2\right)+\left(a_1+1\right){}^2 \left(a_1^2-1\right) \gamma _2^6 \gamma _3^2
\Big)\\&+O(r^d).
\end{split}
\end{equation}
\begin{equation}
    \begin{split}
   C_{2,2}=&
   -\frac{\left(a_1-1\right){}^2 \gamma _2 \gamma _3 \Delta _3}{32 \beta _3 \gamma _1^3}
   +
   \frac{\Delta _3}{256 \left(a_1+1\right){}^2 \gamma _1^7 \gamma _2^3}
   \Big(
    \left(a_1^2-1\right) \gamma _1^6 \left(2 \left(a_1^2+8 a_1-7\right) \gamma _2^4+\left(a_1^2+6
   a_1+17\right) \gamma _3 \gamma _2^2+\left(a_1+1\right){}^2 \gamma _3^2\right)\\
   &+\left(a_1-1\right)
   \gamma _2^2 \gamma _1^4 \left(2 \left(a_1+1\right) \left(a_1^2+8 a_1+1\right) \gamma _2^4+6
   \left(a_1-1\right) \left(a_1^2-1\right) \gamma _3 \gamma _2^2+\left(a_1-1\right){}^2
   \left(a_1+1\right) \gamma _3^2\right)\\
   &-\left(a_1-1\right) \left(a_1+1\right) \gamma _2^4 \gamma _3
   \gamma _1^2 \left(3 \left(a_1-1\right){}^2 \gamma _3+\left(7 a_1^2+26 a_1+7\right) \gamma
   _2^2\right)+5 \left(a_1+1\right){}^2 \left(a_1^2-1\right) \gamma _2^6 \gamma _3^2
\Big)\\&+O(r^d).
\end{split}
\end{equation}

\begin{equation}
    \begin{split}
   C_{3,3}=&
   -\frac{\left(\gamma _1^2-\gamma _2^2\right){}^2 \gamma _3 \Delta _3}{8 \beta _3 \gamma _1^3 \gamma _2^3}
   +
   \frac{\left(\gamma _1^2-\gamma _2^2\right){}^2 \Delta _3}{64 \left(a_1^2-1\right) \gamma _1^7 \gamma
   _2^7}
   \Big(
    \left(\gamma _2^2-\gamma _3\right) \gamma _1^6 \left(2 \left(a_1^2+8 a_1+1\right) \gamma _2^2-5
   \left(a_1+1\right){}^2 \gamma _3\right)\\
   &+\gamma _1^4 \left(2 \left(a_1^2+8 a_1+1\right) \gamma _2^6-2
   \left(a_1^2+14 a_1+1\right) \gamma _3 \gamma _2^4+\left(a_1^2+14 a_1+1\right) \gamma _3^2 \gamma
   _2^2\right)\\
   &+\gamma _2^4 \gamma _3 \gamma _1^2 \left(\left(a_1^2+14 a_1+1\right) \gamma _3-\left(7a_1^2+26 a_1+7\right) \gamma _2^2\right)+5 \left(a_1+1\right){}^2 \gamma _2^6 \gamma _3^2
   \Big)\\&+O(r^d).
\end{split}
\end{equation}
%C_{1,2}
\begin{equation}
    \begin{split}
   C_{1,2}=&
   \frac{\left(a_1-1\right){}^2 \gamma _3 \Delta _3}{16 \beta _3 \gamma _1 \gamma _2}
   +
   \frac{\Delta _3}{128 \left(a_1+1\right){}^2 \gamma _1^5 \gamma _2^5}
   \\ & \times\Big(
    \left(a_1-1\right) \gamma _1^6 \left(2 \left(a_1^3-3 a_1^2+a_1+5\right) \gamma _2^4+\left(3 a_1^3+13
   a_1^2+5 a_1-5\right) \gamma _3 \gamma _2^2-3 \left(a_1+1\right){}^3 \gamma
   _3^2\right)\\
   &+\left(a_1-1\right) \gamma _2^2 \gamma _1^4 \left(2 \left(a_1^3-3 a_1^2+a_1+5\right)
   \gamma _2^4-6 \left(a_1-1\right){}^2 \left(a_1+1\right) \gamma _3 \gamma _2^2+\left(a_1-1\right){}^2
   \left(a_1+1\right) \gamma _3^2\right)\\
   &+\left(a_1-1\right) \gamma _2^4 \gamma _3 \gamma _1^2
   \left(\left(a_1+1\right) \left(a_1-1\right){}^2 \gamma _3+\left(3 a_1^3+13 a_1^2+5 a_1-5\right)
   \gamma _2^2\right)-3 \left(a_1+1\right){}^2 \left(a_1^2-1\right) \gamma _2^6 \gamma _3^2
  \Big)\\&+O(r^d).
\end{split}
\end{equation}
%C_{1,3}
\begin{equation}
    \begin{split}
   C_{1,3}=&
   -\frac{\left(a_1-1\right) \left(\gamma _1^2-\gamma _2^2\right) \gamma _3 \Delta _3}{8 \beta _3 \gamma _1
   \gamma _2^3}
   +
   \frac{\left(\gamma _1^2-\gamma _2^2\right) \Delta _3}{64 \left(a_1+1\right) \gamma _1^5 \gamma _2^7}
   \Big(
    \left(\gamma _2^2-\gamma _3\right) \gamma _1^6 \left(2 \left(a_1^2+8 a_1+1\right) \gamma _2^2-5
   \left(a_1+1\right){}^2 \gamma _3\right)\\
   &-\gamma _1^4 \left(2 \left(a_1^2-4 a_1+5\right) \gamma _2^6-2
   \left(a_1^2-10 a_1+1\right) \gamma _3 \gamma _2^4+\left(a_1^2-10 a_1+1\right) \gamma _3^2 \gamma
   _2^2\right)\\
   &+\gamma _2^4 \gamma _3 \gamma _1^2 \left(\left(-3 a_1^2-10 a_1+5\right) \gamma
   _2^2+\left(a_1^2+6 a_1+1\right) \gamma _3\right)+3 \left(a_1+1\right){}^2 \gamma _2^6 \gamma _3^2
  \Big)\\&+O(r^d).
    \end{split}
\end{equation}
%C_{2,3}
\begin{equation}
    \begin{split}
   C_{2,3}=&
   \frac{\left(a_1-1\right) \left(\gamma _1^2-\gamma _2^2\right) \gamma _3 \Delta _3}{8 \beta _3 \gamma
   _1^3 \gamma _2}
   +
   \frac{\left(\gamma _1^2-\gamma _2^2\right) \Delta _3}{64 \left(a_1+1\right) \gamma _1^7 \gamma _2^5}
   \Big(
    \gamma _1^6 \left(2 \left(a_1^2-4 a_1+5\right) \gamma _2^4+\left(3 a_1^2+10 a_1-5\right) \gamma _3
   \gamma _2^2-3 \left(a_1+1\right){}^2 \gamma _3^2\right)\\
   &-\gamma _1^4 \left(2 \left(a_1^2+8
   a_1+1\right) \gamma _2^6+2 \left(a_1^2-10 a_1+1\right) \gamma _3 \gamma _2^4+\left(a_1^2+6
   a_1+1\right) \gamma _3^2 \gamma _2^2\right)\\
   &+\gamma _2^4 \gamma _3 \gamma _1^2 \left(\left(7 a_1^2+26
   a_1+7\right) \gamma _2^2+\left(a_1^2-10 a_1+1\right) \gamma _3\right)-5 \left(a_1+1\right){}^2 \gamma
   _2^6 \gamma _3^2
  \Big)\\&+O(r^d).
\end{split}
\end{equation}
\normalsize

\section{Details on  Casimir equations}\label{appendix:Casimir}

\paragraph{Notation:} For a string $I=(i_1,j_1,i_2,j_2,\cdots)$, 
the symbol $\tau^\rho_{I}$ is defined as:
\begin{equation}
\tau^\rho_I =\Tr(\rho(g_{i_1})^{-1} \rho(g_{j_1})\rho(g_{i_2})^{-1}\rho(g_{j_2})\cdots),
\end{equation}
where $\rho$ is a representation of $G$. For example,$\tau^\rho_{1213}\equiv \Tr(\rho_{g_1}^{-1}\rho(g_2)\rho_{g_1}^{-1}\rho(g_3))$. The choice of $\rho$ is usually obvious from context so to
lighten the notation, we will often times just use $\tau_I$ and write:
\begin{equation}
    \tau_I = \text{Tr}(g_{i_1}^{-1} g_{j_1} g_{i_2}^{-1} g_{j_2}\cdots).
\end{equation}
\subsection{$d=1$, vector representation}
In 1d, the moduli space of the flat conformal structure is three dimensional and we will choose the following set of 
invariants as coordinates on this space:
\begin{equation}
    \left\{ \tau_{12},\tau_{23},\tau_{13} \right\}.
\end{equation}
For functions on this moduli space, infinitesimal left actions are realized as a 
differential operator in terms of $\tau_{12},\tau_{23},\tau_{31}$. For example,
\begin{equation}
    \begin{split}
    &\mathcal{L}_{AB}^{(1)} B_{123}(g_1,g_2,g_3)\equiv \frac{d}{d t}\Bigg|_{t=0} B_{123}(e^{tL_{AB}}g_1,g_2,g_3)\\
    &=\frac{d}{d t}\Bigg|_{t=0}\Tr((e^{t L_{AB}}g_1)^{-1} g_2)\frac{\partial B_{123}}{\partial \tau_{12}}+\frac{d}{d t}\Bigg|_{t=0}\Tr((e^{t L_{AB}}g_1)^{-1} g_3)\frac{\partial B_{123}}{\partial \tau_{13}}
    \\
    &= -\Tr(g_1^{-1}L_{AB}g_2)\frac{\partial B_{123}}{\partial \tau_{12}}-\Tr(g_1^{-1}L_{AB} g_3)\frac{\partial B_{123}}{\partial \tau_{13}},
    \end{split}
\end{equation}
where it is understood that in expressions such as $\Tr(g_1^{-1}L_{AB} g_2)$, $g_1,L_{AB},g_2$ denote
the corresponding three dimensional matrices in the vector representation of $SO(1,2)$ .
\\
We can then combine the generators $\mathcal{L}_{AB}$ into the quadratic Casimir:
\begin{equation}
    \begin{split}
    &\mathcal{C}_2^{(1)} \equiv -\frac{1}{2}\mathcal{L}_{AB}^{(1)}  \mathcal{L}^{AB,(1)} 
    = -\frac{1}{2}\Tr(g_1^{-1}L_{AB}g_2)\Tr(g_1^{-1}L^{AB}g_2)\frac{\partial^2}{\partial 
    \tau_{12}^2}-\frac{1}{2}\Tr(g_1^{-1}L_{AB}g_3)\Tr(g_1^{-1}L^{AB}g_3)\frac{\partial^2}{\partial 
    \tau_{13}^2}\\
    &-\Tr(g_1^{-1} L_{AB}g_2)\Tr(g_1^{-1}L^{AB} g_3)\frac{\partial^2}{\partial \tau_{12}\partial \tau_{13}}
    -\frac{1}{2}\Tr(g_1^{-1}L_{AB}L^{AB}  g_2)\frac{\partial}{\partial \tau_{12}}
    -\frac{1}{2}\Tr(g_1^{-1}L_{AB}L^{AB} g_3 )\frac{\partial}{\partial \tau_{13}}.
    \end{split}
\end{equation}
In vector representations, matrix elements of the generators are given by:
\begin{equation}
    (L_{AB})^{I}_{J}=\delta^I_A \eta_{BJ}-\delta^I_B \eta_{AJ},
\end{equation}
therefore,
\begin{equation}
\begin{split}
\Tr{(L_{AB} h_1 L^{AB}h_2)}&=2 \Tr(h_1^{-1}h_2)-2\Tr(h_1)\Tr(h_2),\\
\Tr(L_{AB} h_1)\Tr(L^{AB}h_2)& =2 \Tr(h_1^{-1}h_2)-2\Tr(h_1 h_2),
\end{split}
\end{equation}
thus,
\begin{equation}
    \Tr(g_1^{-1} L_{AB} g_2)\Tr(g_{1}^{-1}L^{AB}g_2)=6-2\tau_{1212},
    \quad
    \Tr(g_1^{-1} L_{AB} g_2)\Tr(g_{1}^{-1}L^{AB}g_3)=2 \tau_{23}-2\tau_{1213},
\end{equation}
\begin{equation}
    \Tr(g_1^{-1}L_{AB}L^{AB} g_2) = -4\tau_{12},
\end{equation}
where we've used the fact that the Casimir operator of $SO(1,d+1)$ vector representations 
is always $(d+1) \mathbb{I}$. Invariants such as $\tau_{1212},\tau_{1313},\tau_{1213}$ are 
related to $\tau_{12},\tau_{23},\tau_{13}$ as:
\begin{equation}\label{eqn:trace_relation_1d}
    \begin{split}
\tau_{1212}&=\tau_{12}^2-2\tau_{12},\quad
\tau_{1313}=\tau_{13}^2-2\tau_{13},\\
\tau_{1213}&=\tau_{12}\tau_{13}+\tau_{12}+\tau_{13}+\tau_{23}
+1\pm \sqrt{(\tau_{12}+1)(\tau_{13}+1)(\tau_{23}+1)}.
    \end{split}
\end{equation}
The first two identities above follow from the Cayley-Hamilton theorem, that is, the matrices
$g_{i}^{-1} g_j$ satisfy the following matrix identity:
\begin{equation}
    (g_i^{-1}g_j)^3-\Tr(g_i^{-1}g_j)(g_i^{-1}g_j)^2+\Tr(g_i^{-1}g_j)(g_i^{-1}g_j)-\mathbb{I}=0,
\end{equation} 
the last identity in  (\ref{eqn:trace_relation_1d}) looks complicated and 
the $\pm$ sign introduces ambiguity. To understand it, we need to consider the 
spinor representation of $SO(1,2)$. Since the vector representation is just the symmetric product of 
the spinor representation, we have:
\begin{equation}
        \tau^{\text{vector}}_I=(\tau^{\text{spinior}}_I)^2-1.
\end{equation}
The $SL(2,\mathbb{R})$ matrices satisfy the following identities:
\begin{equation}
    g^2 - \Tr(g) g+\mathbb{I}=0,
\end{equation}
therefore:
\begin{equation}
    \tau_{1213}^{\text{spinor}}=  \tau_{12}^{\text{spinor}}\tau_{13}^{\text{spinor}}-\tau_{23}^{\text{spinor}}.
\end{equation}
This explains the last identity in (\ref{eqn:trace_relation_1d}) and the sign in
front of $\sqrt{\cdots}$ is determined by $\text{sgn}(\tau^{\text{spinor}}_{12}\tau^{\text{spinor}}_{13}\tau^{\text{spinor}}_{23})$.
\subsection{$d=1$, spinor representation}
From the discussion above, it is clear that characters in the spinor representation provide a better set of coordinates.
We write down the complete expression of the Casimir operator in terms of $\tau^{\text{spinor}}_I$ in this subsection:
\begin{equation}\label{eqn:1dCasimir}
\begin{split}
\mathcal{C}_2^{(1)}&=\left(\frac{1}{4}\tau_{12}^2-1\right)\partial_{\tau_{12}}^2+\left(\frac{1}{4}\tau_{13}^2-1\right)\partial_{\tau_{13}}^2+\left(\frac{1}{2}\tau_{12}\tau_{13}-\tau_{23}\right)\partial_{\tau_{13}}\partial_{\tau_{12}}\\
&\quad +\frac{3}{4}\tau_{12}\partial_{\tau_{12}}+\frac{3}{4}\tau_{13}\partial_{\tau_{13}},\\
\mathcal{C}_2^{(2)}&=\left(\frac{1}{4}\tau_{12}^2-1\right)\partial_{\tau_{12}}^2+\left(\frac{1}{4}\tau_{23}^2-1\right)\partial_{\tau_{23}}^2+\left(\frac{1}{2}\tau_{12}\tau_{23}-\tau_{13}\right)\partial_{\tau_{13}}\partial_{\tau_{23}}\\
&\quad +\frac{3}{4}\tau_{12}\partial_{\tau_{12}}+\frac{3}{4}\tau_{23}\partial_{\tau_{23}},\\
\mathcal{C}_2^{(3)}&=\left(\frac{1}{4}\tau_{13}^2-1\right)\partial_{\tau_{13}}^2+\left(\frac{1}{4}\tau_{23}^2-1\right)\partial_{\tau_{23}}^2+\left(\frac{1}{2}\tau_{13}\tau_{23}-\tau_{12}\right)\partial_{\tau_{13}}\partial_{\tau_{23}}\\
&\quad +\frac{3}{4}\tau_{23}\partial_{\tau_{23}}+\frac{3}{4}\tau_{13}\partial_{\tau_{13}}.\\
\end{split}
\end{equation}
Solving the 1d Casimir differential equations order by order in the 
low temperature limit\footnote{By ``low temperature limit'', we mean when $e^{-\beta_{ij}}\sim \epsilon$
and $\epsilon\rightarrow 0$. Here $\beta_{ij} = 2\cosh^{-1} (\tau_{ij}/2)$.}, we find
the following expansion for the 1d genus-2 sunrise block:
\be\label{eq:low_temperature_expansion}
&B_{\Delta_1,\Delta_2,\Delta_3} \nn\\
&=y_{12}^{\Delta _1+\Delta _2-\Delta _3} y_{13}^{\Delta _1-\Delta _2+\Delta _3} y_{23}^{-\Delta _1+\Delta _2+\Delta _3} 
\nn
\\
&\quad \times \Bigg(   
    1
    +\frac{(\Delta _1+\Delta _2-\Delta_3)(-\Delta _1+\Delta _2+\Delta _3) }{2 \Delta _2}\frac{y_{12}y_{23}}{y_{13}}
  \nn\\
&\quad \quad \quad +\frac{(\Delta _1-\Delta _2+\Delta _3) (-\Delta _1+\Delta
   _2+\Delta _3)}{2 \Delta _3} \frac{y_{13} y_{23}}{y_{12}}
\nn     
\\
&\quad \quad \quad+\frac{(\Delta _1+\Delta _2-\Delta _3) (\Delta _1-\Delta _2+\Delta
   _3)}{2 \Delta _1} \frac{y_{13} y_{12}}{y_{23}}
\nn 
\\ 
&\quad \quad \quad+\frac{(\Delta _1-\Delta _2-\Delta _3){}^2 (\Delta _1+\Delta _2-\Delta
   _3-1) (\Delta _1-\Delta _2+\Delta _3-1)}{4 \Delta _2 \Delta
   _3} y_{23}^2
\nn 
\\
&\quad \quad \quad-\frac{(\Delta _1-\Delta _2-\Delta _3+1) (\Delta _1+\Delta _2-\Delta
   _3-1) (\Delta _1-\Delta _2+\Delta _3)^2 }{4 \Delta _1
   \Delta _3} y_{13}^2
\nn  
\\ 
&\quad \quad \quad-\frac{(\Delta _1-\Delta _2-\Delta _3+1) (\Delta _1+\Delta _2-\Delta
   _3){}^2 (\Delta _1-\Delta _2+\Delta _3-1)}{4 \Delta _1
   \Delta _2} y_{12}^2
\nn. 
\\
&\quad \quad \quad+\frac{(\Delta _1-\Delta _2-\Delta _3-1) (\Delta _1-\Delta _2-\Delta
   _3) (\Delta _1-\Delta _2+\Delta _3) (\Delta _1-\Delta _2+\Delta
   _3+1)}{4 \Delta _3 (2 \Delta _3+1)}\frac{y_{13}^2 y_{23}^2}{y_{12}^2}
\nn  
\\ 
&\quad \quad \quad+\frac{(\Delta _1-\Delta _2-\Delta _3-1) (\Delta _1-\Delta _2-\Delta
   _3) (\Delta _1+\Delta _2-\Delta _3) (\Delta _1+\Delta _2-\Delta
   _3+1)}{4 \Delta _2 (2 \Delta _2+1) } \frac{y_{12}^2 y_{23}^2}{y_{13}^2}
\nn 
\\ 
&\quad \quad \quad+\frac{(\Delta _1+\Delta _2-\Delta _3) (\Delta _1+\Delta _2-\Delta
   _3+1) (\Delta _1-\Delta _2+\Delta _3) (\Delta _1-\Delta
   _2+\Delta _3+1)}{4 \Delta _1 (2 \Delta _1+1)
   } \frac{y_{12}^2 y_{13}^2}{y_{23}^2}
\nn 
\\ &\quad \quad \quad+\cdots
\Bigg),
\ee 
where 
\be
y_{ij} = e^{-\beta_{ij}/2},\quad \tau_{ij} = 2 \cosh\left(\frac{\beta_{ij}}{2}\right).
\ee

Each term in this expansion comes from a descendant state in the multiplet $\cR_{\Delta_1}\otimes \cR_{\Delta_2}\otimes \cR_{\Delta_3}$:
\be
y_{12}^{n_1+n_2-n_3} y_{13}^{n_1+n_3-n_2} y_{23}^{n_2+n_3-n_1}   
\leftrightarrow 
(P^{n_1}|\Delta_1\rangle)\otimes (P^{n_2}|\Delta_2\rangle) \otimes (P^{n_3}|\Delta_3\rangle)
\ee
Observe that:
\be
\lim_{\Delta_3\rightarrow 0} \lim_{\Delta_1,\Delta_2\rightarrow \Delta}B_{\Delta_1,\Delta_2,\Delta_3}(y_{12},y_{13},y_{23})
 = y_{12}^{2\Delta}(1+y_{12}^2+\cdots),
\ee
which is the low-temperature expansion of the character $e^{-\beta_{12}\Delta}/(1-e^{-\beta_{12}})$.
Similarly:
\be 
\lim_{\Delta_1\rightarrow 0}\lim_{\Delta_2,\Delta_3\rightarrow \Delta} 
   B_{\Delta_1,\Delta_2,\Delta_3}(y_{12},y_{13},y_{23}) = y_{23}^{2\Delta}(1+y_{23}^2+\cdots),
\\
\lim_{\Delta_2\rightarrow 0}\lim_{\Delta_1,\Delta_3\rightarrow \Delta} 
   B_{\Delta_1,\Delta_2,\Delta_3}(y_{12},y_{13},y_{23})= y_{13}^{2\Delta}(1+y_{13}^2+\cdots).
\ee 
\subsection{$d=3$, spinor representation}
Given our experience with 1d Casimir equation, we focus on the spinor representation. We will be using the following set of invariants as coordinates on the moduli space:
\begin{equation}\label{eqn:3d_coordinates}
    \mathcal{S} = \{ \tau_{12},\tau_{23},\tau_{13},\tau_{1212},\tau_{1313},\tau_{2323},\tau_{1213},\tau_{1232},
\tau_{1323},\tau_{121323} \}. 
\end{equation}
Following a similar derivation as in the 1d case, we find:
\begin{equation}\label{eqn:3dCas}
    \begin{split}
  \mathcal{C}_2^{(1)} &=
  -2\left(\tau _{12}-\tau _{121212}\right) \partial _{\tau _{12}} \partial _{\tau
   _{1212}}
   +\frac{1}{2} \left(\tau _{1212}-4\right) \partial _{\tau _{12}}^2
   +2\left(\tau _{121213}-\tau _{13}\right) \partial_{\tau _{12}} \partial _{\tau _{1213}}\\
   %-----------1-----------%
   &+2\left(\tau _{12121323}-\tau _{1323}\right) \partial_{\tau _{12}}  \partial _{\tau
   _{121323}}
   + \left(\tau _{121232}-\tau _{23}\right) \partial _{\tau _{12}} \partial _{\tau
   _{1232}}
   -2\left(\tau _{12}-\tau _{121313}\right) \partial _{\tau _{1213}} \partial_{\tau _{13}}\\
   %-----------2------------%
   &+\left(\tau _{1213}-\tau _{23}\right) \partial _{\tau _{12}} \partial _{\tau_{13}}
   +2\left(\tau _{121313}-\tau _{1323}\right)\partial _{\tau _{12}}  \partial _{\tau
   _{1313}}
   -2\left(\tau _{12}-\tau _{1213231323}\right) \partial _{\tau _{121323}} \partial
   _{\tau _{1323}}\\
   %------------3---------------%
   &+\left(\tau _{121323}-\tau _{2323}\right) \partial _{\tau _{12}}  \partial_{\tau _{1323}}
   -2\left(\tau _{12}-\tau _{13131323}\right) \partial _{\tau _{1313}}\partial _{\tau _{1323}}
   +2\left(\tau _{12121212}-4\right)\partial _{\tau _{1212}}^2\\
   %----------4-----------------%
   &+4 \left(\tau _{12121213}-\tau _{23}\right)\partial _{\tau _{1212}} \partial _{\tau_{1213}}
   +4  \left(\tau _{1212121323}-\tau _{2323}\right)\partial _{\tau _{1212}} \partial_{\tau _{121323}}
   +2\left(\tau _{12121232}-\tau _{13}\right)\partial _{\tau _{1212}}\partial _{\tau _{1232}}\\
   %--------------5----------%
   &+2  \left(\tau _{121213}-\tau_{1232}\right) \partial _{\tau _{1212}}\partial _{\tau _{13}}
   +4  \left(\tau_{12121313}-\tau _{123132}\right) \partial _{\tau _{1212}}\partial _{\tau _{1313}}
   +2 \left(\tau _{12121323}-\tau _{123232}\right) \partial _{\tau _{1212}} \partial _{\tau _{1323}}\\
   %--------------6---------------%
   &+\left(\tau _{12121313}+\tau _{12131213}-\tau _{123123}-4\right)\partial_{\tau_{1213}}^2 \\
   %---------------7--------------------%
   &+\left(\tau _{1212131323}+\tau _{1212132313}+\tau _{1213121323}-\tau _{12312323}-\tau _{12323123}-2 \tau _{23}\right)\partial_{\tau _{1213}} \partial_{\tau _{121323}}\\
   %---------------8-------------------------%
   &+\left(\tau _{121213231323}+\tau _{121323121323}-\tau_{1232312323}-4\right)\partial _{\tau_{121323}}^2 
    +\left(\tau_{12123213}+\tau _{12131232}-2 \tau _{1323}\right)\partial _{\tau _{1213}} \partial _{\tau _{1232}} \\
    %--------------9--------------------%
   &+\left(\tau _{1212321323}+\tau _{1213231232}-2 \tau _{132323}\right)\partial _{\tau _{121323}} \partial_{\tau _{1232}} 
   +4 \left(\tau _{12131313}-\tau _{23}\right) \partial _{\tau _{1213}}\partial _{\tau_{1313}}\\
   %----------------10--------------------%
   &+ \left(\tau _{12131323}+\tau
   _{12132313}-2 \tau _{1232}\right)\partial _{\tau _{1213}} \partial _{\tau _{1323}}
  +2\left(\tau _{1213131323}+\tau _{1213231313}-2 \tau _{123123}\right)\partial _{\tau _{121323}} \partial _{\tau _{1313}}\\
   %---------------------11--------------------%
   &+\left(\tau _{12131323}+\tau _{12132313}-2 \tau_{13}\right)\partial_{\tau_{121323}} \partial _{\tau _{13}} 
   +\frac{1}{2} \left(\tau _{12321232}-4\right) \partial _{\tau_{1232}}^2
   +\left(\tau _{123213}-\tau _{2323}\right) \partial_{\tau _{13}}\partial _{\tau _{1232}} \\
   %--------------------12------------------------%
   &+2\left(\tau _{12321313}-\tau _{132323}\right)\partial _{\tau _{1313}}\partial_{\tau _{1232}}
   + \left(\tau _{12321323}-\tau_{232323}\right) \partial _{\tau _{1323}}\partial _{\tau _{1232}}
   -2\left(\tau _{13}-\tau _{131313}\right)\partial _{\tau _{13}} \partial _{\tau _{1313}}\\
   %------------------13--------------------------%
   & +\frac{1}{2} \left(\tau _{1313}-4\right)\partial _{\tau _{13}}^2
   +\left(\tau _{131323}-\tau _{23}\right)\partial _{\tau _{1323}}\partial _{\tau _{13}}
   +2\left(\tau _{13131313}-4\right)\partial _{\tau_{1313}}^2
   +\frac{1}{2} \left(\tau _{13231323}-4\right) \partial _{\tau _{1323}}^2\\
   %------------linear order terms -----------------%
   & + \frac{5}{2}\tau_{12}\partial_{\tau_{12}}+\frac{1}{2}(5 \tau_{1212}+\tau_{12}^2+4)\partial_{\tau_{1212}}
     + \frac{5}{2}\tau_{31}\partial_{\tau_{31}}+\frac{1}{2}(5 \tau_{3131}+\tau_{31}^2+4)\partial_{\tau_{3131}}\\
    &+(5 \tau_{1213}+\tau_{12}\tau_{31}+\tau_{23})\partial_{\tau_{1213}}
    +\frac{5}{2}\tau_{1323}\partial_{\tau_{1323}}
    +\frac{5}{2}\tau_{1232}\partial_{\tau_{1232}}
    +(5 \tau_{121323}+\tau_{12}\tau_{1323}+\tau_{2323})\partial_{\tau_{121323}}.
    \end{split}
\end{equation}
The expressions for $\cC_2^{(2)},\cC_2^{(3)}$ take analogous forms and we omit them for brevity. The derivation of the 3d Casimir operator is almost identical as the 1d version except that one needs to be 
careful about multiple occurrence of the same letter. Let's look at an example:
\begin{equation}
    \begin{split}
    &-\frac{1}{2}\left(\mathcal{L}_{AB} \tau_{1213}\right) \left(\mathcal{L}^{AB} \tau_{121323}\right)\\
    &=-\frac{1}{2}\left(\Tr(g_1^{-1}L_{AB} g_2 g_1^{-1}g_3)+\Tr(g_1^{-1} g_2 g_1^{-1} L_{AB} g_3) \right)
    \left(\Tr(g_1^{-1}L^{AB}g_2 g_1^{-1} g_3g_2^{-1}g_3)+\Tr(g_1^{-1}g_2 g_1^{-1} L^{AB} g_3g_2^{-1}g_3)\right)\\
    &= \left(\tau _{1212131323}+\tau _{1212132313}+2
   \tau _{1213121323}-\tau _{12312323}-\tau _{12323123}-2 \tau _{23}\right),
    \end{split}
\end{equation}
where we've used the fact that in spinor representation:
\begin{equation}
    \Tr(L_{AB}h_1) \Tr(L^{AB}h_2)=\Tr(h_2^{-1}h_1)-\Tr(h_1 h_2).
\end{equation}

In  (\ref{eqn:3dCas}), there are a few invariants which do not belong to the set of 10 coordintas:
\begin{equation}
    \begin{split}
    & \tau_{121212},\tau _{12121212},\tau _{121313},\tau _{1213231323},\tau
   _{1213131323},\tau _{1213231313}, \tau_{123123},\tau _{12121313},\tau _{12131213},\tau_{123132},\tau _{121213},
   \\
   & \tau _{12131323},\tau _{12132313},\tau _{12321232},\tau
   _{121213231323},\tau _{121323121323},\tau _{1232312323},\tau _{12121323},\tau
   _{123232},\tau _{12121232},\tau _{131313},\\ 
   &\tau _{13131313},\tau _{13131323}, \tau_{12123213},\tau _{12131232},\tau _{13231323},\tau _{12321313},\tau_{132323},\tau
   _{1212321323},\tau _{1213231232},\tau _{12121213},\tau _{12131313},
   \\
   & \tau _{121232},\tau
   _{131323},\tau _{1212131323},\tau _{1212132313},\tau _{1213121323},\tau _{12312323},\tau
   _{12323123},\tau _{1212121323},\tau _{123213},\tau _{12321323},\tau_{232323}.
    \end{split}
\end{equation}
Many of them can be reduced to a polynomial of the 10 coordinates using the following Cayley-Hamilton identity:
\begin{equation}
g^4-\Tr(g) g^3 + \frac{1}{2}(\Tr(g)^2-\Tr(g^2)) g^2 - \Tr(g) g^3+\mathbb{I}=0,
\end{equation}
for example:
\begin{equation}
    \begin{split}
\tau_{12(13)^2}&=-\frac{1}{2} \tau _{12} \left(\tau _{13}^2-\tau _{1313}\right)+\tau _{1213} \tau
   _{13}+\tau _{13} \tau _{23}-\tau _{1323},\\
  \tau_{(12)^3} &=\frac{1}{2} \left(3 \tau _{12} \tau _{1212}-\tau _{12}^3+6 \tau _{12}\right),\\
  \tau_{(12)^4} &= \frac{1}{2} \left(2 \tau _{12}^2 \left(\tau _{1212}+4\right)-\tau _{12}^4+\tau
   _{1212}^2-8\right),\\ 
   \tau_{(12)^3 32}&=\frac{1}{2} \left(\tau _{12} \left(\left(\tau _{1212}+2\right) \tau _{23}-2 \tau
   _{1213}\right)+\tau _{12}^2 \left(\tau _{1232}+2 \tau _{13}\right)-\tau _{12}^3
   \tau _{23}+\tau _{1212} \tau _{1232}-2 \tau _{13}\right).
  \end{split}
\end{equation}
However, there are some other invariants which we don't know how to express in terms of the ten coordinates.
One such example is $\tau_{123123}$. 
By numerically computing the Jacobian between the invariants and group parameters, we verify that $\tau_{123123}$
is not independent from the ones listed in \ref{eqn:3d_coordinates}. From experience in 1d, one expect to find 
a polynomial relation between $\tau_{123123}$ and those in $\mathcal{S}$. We numerically searched over the following set of monomials:
\begin{equation}
   \left \{ \prod_{I\in \mathcal{S'}}\tau_{I}^{n_I} \Big| \sum_{I} n_I \leq 4 \right \},\quad \mathcal{S'}=\mathcal{S}\cup \{\tau_{123123}\}.      
\end{equation}
but no linear relation was found. 
As a first sanity check, note that when acting on a function which only depends on
$\tau_{12}$ and $\tau_{1212}$, $\cC_2^{(1)}$ reduces to:
\be 
\cC_2^{(1)}f(\tau_{12},\tau_{1212})&=
4 \left(\tau
   _{12121212}-4\right) \partial _{\tau _{1212}}^2 f+2 \left(\tau _{121212}-\tau _{12}\right) \partial _{\tau _{12}} \partial _{\tau
   _{1212}}f
   +2 \left(\frac{\tau _{1212}}{4}-1\right) \partial _{\tau _{12}}^2f 
    \nn \\ 
    &+2\left(\frac{\tau _{12}^2}{2}+\frac{5 \tau _{1212}}{2}+2\right) \partial _{\tau
   _{1212}}f+\frac{5}{2} \tau _{12} \partial _{\tau _{12}}f
\ee
Switching to the coordiantes $y_{12},x_{12}$ defined using the relation:
\be
y_{12} = \exp(-\beta_{12}/2),&\quad x_{12} = \exp(i\theta_{12}/2)
\\ 
\tau_{12} = x_{12} y_{12} + \frac{1}{x_{12} y_{12}}+\frac{x_{12}}{y_{12}}+\frac{y_{12}}{x_{12}},
&\quad 
\tau_{1212} = x_{12}^2 y_{12}^2 + \frac{1}{x_{12}^2 y_{12}^2}+\frac{x_{12}^2}{y_{12}^2}+\frac{y_{12}^2}{x_{12}^2}
\ee
we get:
\be
&\cC_2^{(1)}f(x_{12},y_{12}) = 
\frac{1}{4}\Bigg( x_{12}\partial^2_{x_{12}} +y_{12}\partial^2_{y_{12}}
\nn \\ &+\frac{y_{12} \left(7 x_{12}^2 y_{12}^6-3 x_{12}^4 y_{12}^4-3 x_{12}^2 y_{12}^4-x_{12}^4
   y_{12}^2-x_{12}^2 y_{12}^2+5 x_{12}^2-3
   y_{12}^4-y_{12}^2\right)}{\left(y_{12}-1\right) \left(y_{12}+1\right)
   \left(y_{12}-x_{12}\right) \left(x_{12}+y_{12}\right) \left(x_{12} y_{12}-1\right)
   \left(x_{12} y_{12}+1\right)}\partial_{y_{12}}
\nn \\ &+\frac{x_{12} \left(7 x_{12}^6 y_{12}^2-3 x_{12}^4 y_{12}^4-3 x_{12}^4 y_{12}^2-x_{12}^2
   y_{12}^4-x_{12}^2 y_{12}^2-3 x_{12}^4-x_{12}^2+5 y_{12}^2\right)}{
   \left(x_{12}-1\right) \left(x_{12}+1\right) \left(x_{12}-y_{12}\right)
   \left(x_{12}+y_{12}\right) \left(x_{12} y_{12}-1\right) \left(x_{12} y_{12}+1\right)}\partial_{x_{12}}
\Bigg) f(x_{12},y_{12}).
\ee
It is then easy to check that the conformal character is indeed an eigenfunction of $\cC_2^{(1)}$:
\be 
\cC_2^{(1)}\chi_{\Delta,j}(x_{12},y_{12})
&=\left[\Delta(\Delta-3)+j(j+1)\right]\chi_{\Delta,j}(x_{12},y_{12}),
\nn \\
\chi_{\Delta,j}(x_{12},y_{12})&= \frac{y_{12}^{2\Delta}}{(1-y_{12}^2) (1-x_{12}^2y_{12}^2) (1-x_{12}^{-2}y_{12}^2)} \frac{x_{12}^{2j+1}-x_{12}^{-2j-1}}{x_{12}^{2}-x_{12}^{-2}}.
\ee
\subsection{Casimir equation for the dumbbell block}
For completeness, we include Casimir equations for the dumbbell block in. The dumbbell block 
$B_{L,R,M}$ should satisfy the follwoing three Casimir equations simultaneously:
\be
\begin{split}
\cC_2^{(L)} B_{L,R,M} &= C_2(\pi_L) B_{L,R,M},\\
\cC_2^{(R)} B_{L,R,M} &= C_2(\pi_R)B_{L,R,M},\\
\cC_2^{(M)} B_{L,R,M} &= C_2(\pi_M) B_{L,R,M}.
\end{split}
\ee
In 1d, following our convention in section \ref{sec:thermal flat limit}, $\tau_{12},\tau_{13},\tau_{23}$
can be written as:
\be
\tau_{13} = \Tr(g_L^{-1}), \quad \tau_{23} = \Tr(g_R),\quad \tau_{12} = \Tr(g_M^{-1} g_L^{-1} g_M g_R).
\ee
Working with the spinor representatio, we find that $\cC_2^{(L)}=\cC_2^{(1)}$, $\cC_2^{(R)}=\cC_2^{(2)}$ but $\cC^{(M)}_2$ 
is different from $\cC^{(3)}_2$. Since $\tau_{12}$ is the only variable sensitive to $g_M$, we have:
\be
\begin{split}
\cC^{(M)}_{2} 
    &= -\frac{1}{2}\Tr(g_M^{-1}[g_L^{-1},L_{AB}]g_M g_R)\Tr(g_M^{-1}[g_L^{-1},L^{AB}]g_M g_R) \frac{\ptl^2}{\ptl \tau_{12}^2}\\
    & -\frac{1}{2}\Tr(g_M^{-1}\left[[g_L^{-1},L_{AB}],L^{AB}\right] g_M g_R) \frac{\ptl}{\ptl \tau_{12}},
\end{split}
\ee
with
\be
\begin{split}
-\frac{1}{2}\Tr(g_M^{-1}[g_L^{-1},L_{AB}]g_M g_R)\Tr(g_M^{-1}[g_L^{-1},L^{AB}]g_M g_R)  
& =-\tau_{12} \tau_{13} \tau_{23}+\tau_{12}^2+\tau_{13}^2+\tau_{23}^2-4, \\
-\frac{1}{2}\Tr(g_M^{-1}\left[[g_L^{-1},L_{AB}],L^{AB}\right] g_M g_R) 
&=  2 \tau_{12}-\tau_{13}\tau_{23},
\end{split}
\ee
where we've used the following identities for matrices in the 2-dimensional 
representation of $Spin(2,1)$:
\be
\begin{split}
\Tr(L_{AB}h_1)\Tr(L^{AB}h_2)&=\frac{1}{2}\Tr(h_1)\Tr(h_2)-\Tr(h_1 h_2),\\
\Tr(L_{AB}h_1 L^{AB}h_2)&=\frac{1}{2}\Tr(h_1 h_2)-\Tr(h_1) \Tr(h_2).
\end{split}
\ee
The 1d dumbbell block has the following low-temperature expansion:
\be
&B_{\Delta_L,\Delta_R,\Delta_M}(y_{12},y_{13},y_{23})
=y_{13}^{2\Delta_L}y_{23}^{2\Delta_R}
\Bigg(
1-\frac{y_{13} y_{23} \left(\Delta _M-1\right) \Delta _M}{y_{12}}
\nn  
\\ 
&+\frac{y_{13}^2 \left(2 \Delta _L \left(\left(\Delta _M-1\right) \Delta _M+1\right)+\left(\Delta _M-1\right) \Delta _M\right)}{2 \Delta
   _L}
+\frac{y_{23}^2 \left(\Delta _M^2 \left(2 \Delta _R+1\right)-\Delta _M \left(2 \Delta _R+1\right)+2 \Delta _R\right)}{2 \Delta _R}
\nn  
\\
&+\frac{y_{13}^2 y_{23}^2 \left(\Delta _M-2\right) \left(\Delta _M-1\right) \Delta _M \left(\Delta _M+1\right)}{4 y_{12}^2}
+\cdots \Bigg).
\ee
Similarly in 3d, the ten coordinates in $\cS$ can be written as:
\be
   &\tau_{13} = \Tr(g_L^{-1}),\ \tau_{23}= \Tr(g_R),\ \tau_{12} = \Tr(g_M^{-1} g_L^{-1} g_M g_R),
   \nn 
   \\
   &\tau_{1313} = \Tr(g_L^{-2}),\ \tau_{2323}= \Tr(g_R^2), \ \tau_{1212} = \Tr((g_M^{-1} g_L^{-1} g_M g_R)^2),
   \nn \\
   &\tau_{1213} = \Tr(g_M^{-1} g_L^{-2} g_M g_R),
   \ \tau_{1232}=\Tr(g_M^{-1} g_L^{-1} g_M g_R^2),
   \ \tau_{1323} =\Tr(g_M^{-1} g_L^{-1} g_M g_R^{-1})
   \nn \\ 
  & \tau_{121323} = \Tr( g_M^{-1} g_L^{-1} g_M g_R^{-1} g_M^{-1} g_L^{-1} g_M g_R).
\ee 
Working with the spinor representation, we can again check that $\cC_2^{(L)}=\cC_2^{(1)}$, $\cC_2^{(R)}=\cC_2^{(2)}$
while $\cC_2^{(M)}$ is given by: 
\be 
&\cC_2^{(M)} 
=
\left(\tau _{1212}+\tau _{123123}-\tau_{123213}-4\right)
\partial _{\tau _{12}}^2    
\nn
\\ 
&+
    2\left(-2 \tau _{12}+2 \tau _{121212}+\tau _{12123123}-\tau_{12123213}-\tau _{12131232}+\tau _{12132132}\right)
        \partial _{\tau _{12}} \partial _{\tau _{1212}}
\nn
\\ 
&+
    4\left( \tau _{12121212}+ \tau _{1212312123}- \tau _{1212321213}-4\right)
        \partial _{\tau _{1212}}^2
\nn
\\ 
&+
    \left(2\tau _{121213}+\tau _{12312313}+\tau_{12313123}-2\tau _{12321313}-2\tau _{13}\right)
        \partial _{\tau _{12}} \partial _{\tau _{1213}}
\nn  
\\
&+ 2\left(\tau _{121323}+\tau _{123132}-\tau _{1313}-\tau _{2323}\right)
    \partial _{\tau _{12}} \partial _{\tau _{1323}}
\nn 
\\ 
&+
    \left(4 \tau _{12121323}-2\tau _{121313}+2\tau _{12132123}+\tau_{1231231323}+\tau_{1231323123}-\tau_{1232131323}-\tau _{1232132313}-4 \tau _{1323}\right)
        \partial _{\tau _{12}} \partial _{\tau _{121323}}
\nn 
\\ 
&+
    \left(4 \tau _{12121213}+2\tau_{1212313123}-2\tau _{1212321313}-2\tau_{1213131232}+2\tau _{1213213132}-4 \tau _{23}\right)
        \partial _{\tau _{1212}} \partial _{\tau _{1213}}
\nn     
\\
&+ 
    4\left(\tau _{12121323}+ \tau _{12123132}- \tau _{121313}- \tau_{123232}\right)
        \partial _{\tau _{1212}} \partial _{\tau _{1323}}
\nn 
\\ 
&+  
    2\left(4 \tau _{1212121323}+2 \tau _{1212312132}+\tau_{121231323123}-\tau_{121232132313}-2 \tau _{12131213}-\tau_{121313231232}+\tau _{121321323132}-4 \tau _{2323}\right)
\nn \\  &\quad \quad \quad \quad\times \partial _{\tau _{1212}} \partial _{\tau _{121323}}
\nn 
\\ 
&+
    \left(\tau _{12131213}+\tau _{1231312313}-\tau_{1232131313}-4\right)
        \partial _{\tau _{1213}}^2
\nn  
\\ 
&+
    \left(\tau _{12131323}+\tau _{12132313}+2\tau_{12313213}-2\tau _{1232}-2\tau _{131313}\right)
        \partial _{\tau _{1213}} \partial _{\tau _{1323}}
\nn 
\\ 
&+
    \Big(\tau _{1212131323}+\tau _{1212132313}+2\tau_{1213121323}-2\tau_{12131313}+2\tau_{1213213123}-\tau_{12312323}  
\nn \\ 
&\quad\quad\quad\quad+\tau _{123131231323}+\tau_{123132312313}-\tau _{123213131323}-\tau_{123213231313}-\tau_{12323123}-2\tau _{23}\Big)
    \partial_{\tau _{1213}} \partial_{\tau _{121323}}
\nn 
\\ 
&+ 
    \left(\tau _{123123}-\tau _{13132323}+\tau_{13231323}-4 \right)
        \partial _{\tau _{1323}}^2
\nn 
\\ 
&+ 
    \left(-4 \tau _{12}+\tau _{12123123}-\tau_{1213132323}+\tau_{12132132}+4 \tau_{1213231323}-\tau _{1213232313}+2\tau _{1231321323}-2\tau_{13131323}\right)
    \nn \\ &\quad\quad\quad\quad \times   \partial _{\tau _{121323}} \partial _{\tau _{1323}}
\nn 
\\ 
&+ 
    \Big(2\tau _{121213231323}+\tau _{1212312123}-\tau_{121312132323}-2\tau _{1213132313}+2\tau _{121321323123}+2\tau_{121323121323}
    \nn \\ &\quad\quad\quad\quad  +\tau _{12313231231323}-\tau_{12321323131323}-2\tau _{1232312323}-8\Big)
        \partial _{\tau _{121323}}^2
\nn 
\\
&+
    \left(2\tau _{121232}+\tau _{12312323}+\tau_{12323123}-2\tau _{12323213}-2\tau _{23}\right)
        \partial _{\tau _{12}} \partial _{\tau _{1232}}
\nn 
\\
&+ 
     \left(4 \tau _{12121232}+2\tau _{1212312323}-2\tau _{1212323213}-2\tau_{1213123232}+2\tau _{1213232132}-4\tau _{13}\right)
      \partial _{\tau _{1212}} \partial _{\tau _{1232}}
\nn 
\\ 
&+
    \left(-2\tau _{1213}+\tau _{12313232}+2\tau_{12321323}+\tau _{12323132}-2\tau _{232323}\right)
        \partial _{\tau _{1232}} \partial _{\tau _{1323}}
\nn 
\\ 
&+ 
     \Big(-2\tau _{121213}+2\tau _{1212321323}+\tau_{1213212323}+2\tau_{1213231232}+\tau_{1213232123}+\tau _{123123231323}+\tau_{123231323123}
     \nn \\ &\quad\quad\quad\quad -\tau _{123232131323}-\tau_{123232132313}-4 \tau _{132323}\Big)
    \partial _{\tau _{121323}} \partial _{\tau _{1232}}
\nn 
\\ 
&+ 
    \left(\tau _{12321232}+\tau _{1232312323}-\tau_{1232323213}-4\right)
        \partial _{\tau _{1232}}^2  
%-----------first order terms---------------
\nn 
\\ 
&+   \left(5 \tau _{12}-\tau _{13} \tau _{23}-\tau_{1323}\right)
        \partial _{\tau _{12}}
    +\left(2\tau _{12}^2+10 \tau _{1212}-2\tau _{1213} \tau _{23}-2\tau_{121323}-2\tau _{123132}-2\tau _{1232} \tau _{13}+8\right)
        \partial _{\tau _{1212}}
\nn 
\\ 
&+  \left(5 \tau _{1213}-\tau _{1313} \tau_{23}-\tau _{131323}\right)
        \partial _{\tau _{1213}} 
    + \left(-\tau _{12}-\tau _{13} \tau _{23}+5 \tau _{1323}\right)
        \partial _{\tau _{1323}}
 \nn 
 \\  
 &+
    \left(5 \tau _{1232}-\tau _{13} \tau_{2323}-\tau _{132323}\right)
        \partial _{\tau_{1232}}
\nn 
\\ 
&+  \left(2\tau _{12} \tau _{1323}-\tau _{1212}-\tau_{1213} \tau_{23}+10 \tau _{121323}-2\tau _{123123}-2\tau
_{13}^2-\tau _{131323} \tau _{23}-\tau_{13231323}+2\tau _{2323}\right)
    \partial _{\tau_{121323}}
\ee 
\bibliographystyle{JHEP}
\bibliography{refs}

\end{document}